\documentclass[a4paper,11pt]{article}
\pdfoutput=1 

\usepackage{jheppub} 

\usepackage{epstopdf} 
\usepackage[T1]{fontenc} 
\usepackage[dvipsnames]{xcolor}

\usepackage{graphics}
\usepackage{inputenc}
\usepackage{xspace}
\usepackage{amsmath}
\usepackage{amssymb}
\usepackage{url}
\usepackage{rotating}
\usepackage{enumerate}
\usepackage{graphicx}
\usepackage{pdflscape, afterpage, capt-of} 
\usepackage{slashed} 
\usepackage{subfig} 
\usepackage[normalem]{ulem}
\usepackage{nicefrac}

\usepackage{def} 

\newcommand{\private}[1]{} 

\title{\boldmath The chirality-flow formalism}

\author[a]{Andrew Lifson,}
\author[a]{Christian Reuschle}
\author[a]{and Malin Sjodahl}

\affiliation[a]{Department of Astronomy and Theoretical Physics, Lund
  University, S{\"o}lvegatan 14A, 223\,62 Lund, Sweden}

\emailAdd{andrew.lifson@thep.lu.se}
\emailAdd{christian.reuschle@thep.lu.se}
\emailAdd{malin.sjodahl@thep.lu.se}

\abstract{
  We take a fresh look at Feynman diagrams in the spinor-helicity formalism.
  Focusing on tree-level massless QED and QCD, we 
  develop a new and conceptually simple graphical method for their calculation.
  In this pictorial method, 
  which we dub the chirality-flow formalism,
  Feynman diagrams are directly represented in terms of chirality-flow lines
  corresponding to spinor inner products,
  without the need to resort to intermediate algebraic manipulations.

}

\begin{document} 
\preprint{LU-TP 20-16, MCNET-20-10}
\maketitle
\flushbottom

\section{Introduction}
\label{sec:introduction}

During the past decades, 
various techniques for calculating scattering amplitudes have emerged, 
resulting in both compact analytic formulae and efficient numerical approaches. 
In this context, strategies for managing quantum numbers,
such as color and helicity have played a major role.
Both for color and helicity, factorized approaches,
where amplitudes have a particular color structure or helicity assignment, 
have been used.

In color decomposition, 
amplitudes are decomposed into color factors multiplied by smaller, 
gauge-invariant pieces, 
so-called partial amplitudes, 
and various approaches exist, 
differing in the way 
of choosing the set of vectors in which the color decomposition is obtained 
\cite{MacFarlane:1968vc,Paton:1969je,tHooft:1973alw,cvi76,Butera:1979na,Cvitanovic:1980bu,
Berends:1987cv,Berends:1987me,Mangano:1987xk,Mangano:1988kk,Zeppenfeld:1988bz,Kosower:1988kh,Berends:1988yn,Berends:1988zn,Berends:1989hf,Bern:1990ux,Bern:1994fz,DelDuca:1999rs,Maltoni:2002mq,Weinzierl:2005dd,Ellis:2008qc,Sjodahl:2008fz,Sjodahl:2009wx,Ita:2011ar,Keppeler:2012ih,PhDReuschle,Schuster:2013aya,Reuschle:2013qna, Du:2015apa,Sjodahl:2015qoa,Sjodahl:2018cca,Kilian:2012pz}.
For helicity amplitudes,
i.e. amplitudes with assigned helicities, 
and in particular for partial amplitudes,
the spinor-helicity formalism 
in which diagrams and amplitudes are expressed in terms of two-component Weyl spinors,
has been very successful
\cite{DeCausmaecker:1981jtq,Berends:1981rb,Berends:1981uq,DeCausmaecker:1981wzb,Berends:1983ez,Kleiss:1984dp,Berends:1984gf,Gunion:1985bp,Gunion:1985vca,Kleiss:1985yh,Hagiwara:1985yu,Kleiss:1986ct,Kleiss:1986qc,Xu:1986xb,Gastmans:1987qz,Schwinn:2005pi},
especially after realizing that also the polarization vectors of external vector bosons can be expressed in terms of two-component Weyl spinors.
In the Weyl-van-der-Waerden formalism 
\cite{Farrar:1983wk,Berends:1987cv,Berends:1987me,Berends:1988yn,Berends:1988zn,Berends:1989hf,Dittmaier:1993bj,Dittmaier:1998nn,Weinzierl:2005dd},
diagrams and amplitudes can even be expressed in such a way as to
avoid Lorentz four-vectors entirely, using that Dirac spinors and
Lorentz four-vectors transform under the $(1/2,0)\oplus(0,1/2)$
and $(1/2,1/2)=(1/2,0)\otimes(0,1/2)$ representations of the Lorentz
group respectively. This is a fact which we also rely on in this paper.

For diagrams or amplitudes in the spinor-helicity formalism, 
particularly compact analytic expressions exist in the form of the Parke-Taylor formula 
and other maximally-helicity-violating (MHV) amplitudes 
\cite{Grisaru:1977px,Parke:1985pn,Parke:1986gb,Berends:1987me,Schwinn:2006ca}.
However, also helicity amplitudes with more complicated next-to-maximally-helicity-violating (NMHV) configurations, etc., have been studied.

In the spinor-helicity formalism, 
calculations of single Feynman diagrams, 
as well as complete scattering amplitudes, 
are significantly simplified by expressing them in terms of spinor inner products. 
In this paper, 
we take a fresh look at Feynman diagrams in the spinor-helicity formalism in massless QED and QCD. 
In an attempt to further simplify their calculations, 
we extend the spinor-helicity formalism with an intuitive pictorial representation,
reminiscent of the pictorial representations often used in the treatment of color degrees of freedom. 

In the context of the color-flow picture,
a graphical representation 
which provides an intuitive approach in terms of the flow of color is used
\cite{tHooft:1973alw,Maltoni:2002mq,Weinzierl:2005dd,Reuschle:2013qna,Kilian:2012pz}.
Here, indices in the adjoint representation are converted to (pairs of) indices in the fundamental representation of SU($N$) color,
and the color degrees-of-freedom are accounted for by considering all possible connections in the space of fundamental color indices, 
i.e. all possible color flows. 
As an example, in a condensed notation,  
the four-gluon vertex can be expressed as 
\begin{align} \label{eq:4gluon_colour}
\scriptV_{gggg}
\;\;\propto
\sum\limits_{S(2,3,4)}
\;\raisebox{-0.9cm}{\includegraphics[scale=0.325]{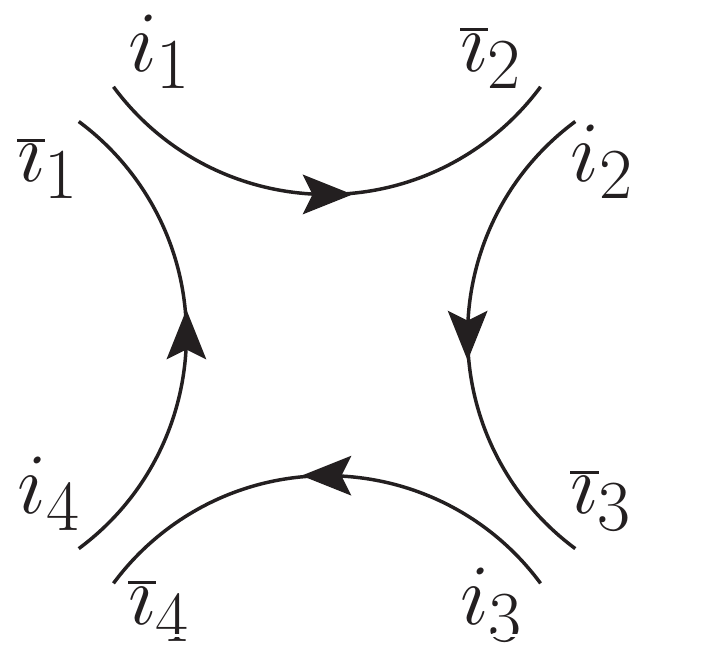}}
\;V_{gggg}(1,2,3,4)\;,
\end{align}
where $V_{gggg}(1,2,3,4)$ denotes the color-stripped four-gluon vertex in the color-flow picture, 
$S(2,3,4)$ denotes the set of permutations of the integers $2,3,4$,
and the lines represent the color flows.

Noting that the Lorentz group algebra consists of two copies of the (complexified)
su(2) algebra, it can be anticipated that an analogous graphical approach should
be applicable for the Lorentz group\footnote{\label{ft:note2}
  Based on the Weyl-van-der-Waerden formalism, 
  a double-line notation analogous to the double-line notation in the
  color-flow picture for spinor-helicity diagrams has been observed before
  \cite{Weinzierl:2005dd}.
  A birdtrack \cite{cvi76,Cvi08} based graphical formalism has
  also been noted \cite{Kennedy:1981ei}.
  However, 
  to the best of our knowledge, 
  a corresponding pictorial representation, 
  with a directed continuous flow has not previously been formulated.
}.
Unlike the single su(3) color algebra,
this would require two different types of flow lines 
--- one dotted and one undotted ---
which can never be contracted into each other,
since the corresponding object would not be Lorentz invariant.
We dub this graphical approach the chirality-flow formalism.

More concretely, 
the chirality-flow formalism relies on the fact that objects carrying Lorentz indices can be converted to objects carrying spinor indices instead.
Feynman diagrams can then be rewritten in terms of contractions of spinor indices
in a flow-like picture, 
similar to the color-flow picture.
For example, the four-gluon vertex can be expressed 
in a similarly condensed notation 
as
\begin{eqnarray} \label{eq:4gIntro}
  \scriptV_{gggg}
  \;\;\propto
  \sum\limits_{Z(2,3,4)}
  \raisebox{-0.425\height}{\includegraphics[scale=0.3]{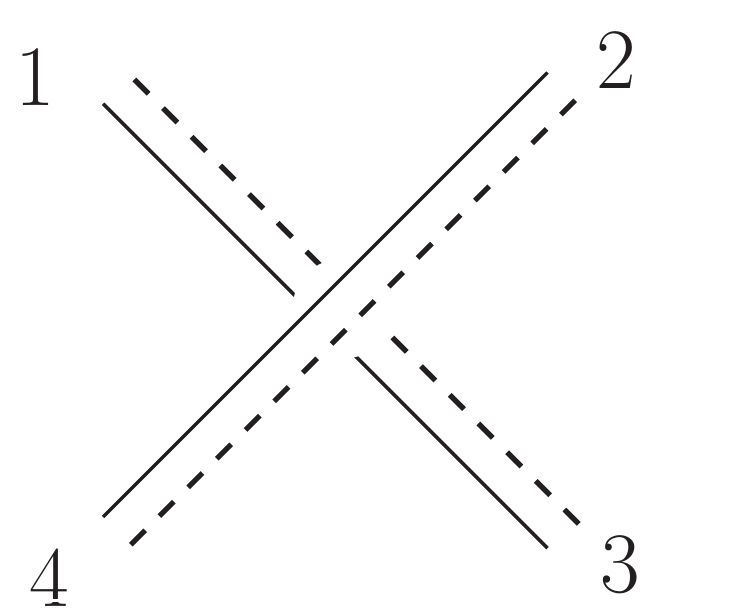}}
  \Big(f^{a_1a_2b}f^{ba_3a_4}+f^{a_4a_1b}f^{ba_2a_3}\Big)\;,
\end{eqnarray}
where $Z(2,3,4)$ denotes the set of cyclic permutations of the integers $2,3,4$,
and the lines represent the chirality flows.

In this new graphical formalism,
the contractions of the spinor indices in Feynman diagrams are more obvious,
allowing for a more transparent, shorter journey from Feynman diagrams to spinor inner products.
The usual algebraic identities needed to transform a Feynman diagram into a set of spinor inner products
are directly built into the pictorial representation.

We intend for this paper to serve as a readable introduction for beginners to the spinor-helicity formalism, 
and in this spirit we have written it in a self-contained manner, 
although some basic identities are only given in \appref{sec:conventions}.
Further introductions and overviews can be found in 
\cite{Mangano:1990by,Gastmans:1990xh,Dixon:1996wi,Dittmaier:1998nn,Weinzierl:2005dd,Weinzierl:2007vk,Dreiner:2008tw,Ellis:2011cr,Peskin:2011in,Elvang:2013cua,Dixon:2013uaa}.

The rest of the paper is organized as follows. 
In \secref{sec:Color} we warm up by reviewing the color-flow idea.
The traditional spinor-helicity formalism, 
as well as chirality-flow representations of external particles, 
inner products and slashed momenta are introduced in \secref{sec:HelicityBasics}. 
Section \ref{sec:HelicityFlow} paves the way for the chirality-flow Feynman rules, 
as we complete and prove the validity of the chirality-flow picture for Feynman
diagrams in massless QED and QCD.
The Feynman rules are then collected in \secref{sec:FeynmanRules}.
In \secref{sec:Examples} we give examples, and in \secref{sec:Conclusion} we summarize and conclude.

\section{Color flow}
\label{sec:Color}

As a warm-up, 
let us start with considering a well-known example of a flow-like representation in the context of SU($N$) scattering amplitudes 
--- color flow (with $N$ colors).

In the color-flow formalism 
\cite{tHooft:1973alw,Maltoni:2002mq,Weinzierl:2005dd,Reuschle:2013qna,Kilian:2012pz} 
the color factors of Feynman rules are converted into color-flow rules. 
Color indices in the adjoint representation of SU($N$) are thereby converted to pairs of color indices, 
one in the fundamental representation and one in the  antifundamental representation,
and color factors are given by Kronecker $\delta$'s,
connecting the fundamental index of one parton to the antifundamental index of another parton. 
In other words, 
we can write the Fierz identity for the  SU($N$) generators
\begin{equation}
t^a_{ij}t^a_{kl}
=
\delta_{il}\delta_{kj}-\frac{1}{N}\delta_{ij}\delta_{kl}\,,
\label{eq:color_fierz0}
\end{equation}
in a graphical representation
\begin{align}
\raisebox{-0.3625\height}{\includegraphics[scale=0.45]{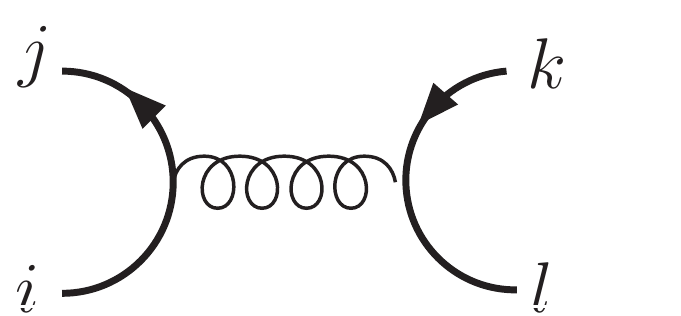}}
\!\!\!\!=\;\; 
\left[\,\, 
 \raisebox{-0.3625\height}{\includegraphics[scale=0.425]{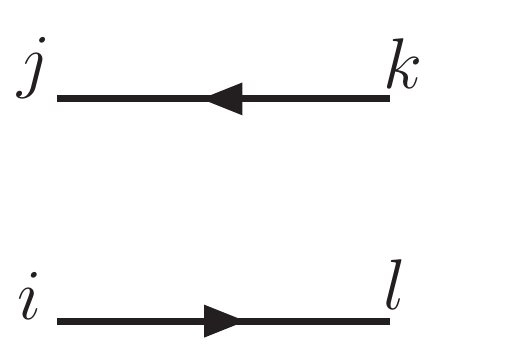}}\!\!\!\!
 -\,\frac{1}{N}\, 
 \raisebox{-0.3325\height}{\includegraphics[scale=0.45]{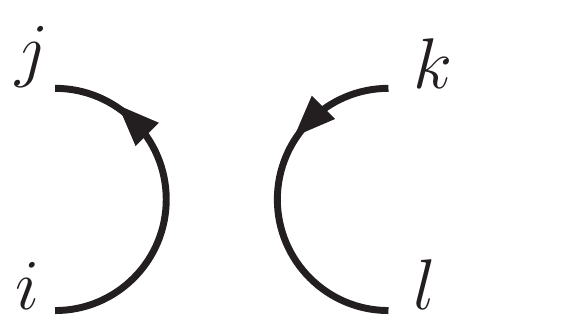}}\!\! 
\right]\;,
\label{eq:color_fierz2}
\end{align}
where the adjoint representation is replaced by (a linear combination of) flows of color. 

\Eqref{eq:color_fierz2} can be seen as an example of color flow in the case of a gluon exchanged between two quark lines,
in which case one also talks about the exchange of a U($N$) gluon in the first term, 
and a U(1) gluon in the second, 
color-suppressed term. 

For gluon exchange between a quark line and a gluon,
or in pure gluon interactions,
the color-suppressed terms drop out, 
and it is possible to express gluons as U($N$) gluons only.
For example, the color flows of the four-gluon vertex are given in \eqref{eq:4gluon_colour}.

Given a diagram or amplitude one can write down all possible flows of color,
and from these infer the corresponding color-stripped amplitudes or diagrams.
The color-stripped Feynman rules in the color-flow formalism, 
sometimes also referred to as the double-line formalism, 
come with the benefit that diagrams and amplitudes are easier to calculate and can be given a more intuitive interpretation\footnote{
  Various approaches for color decomposition of QCD amplitudes, or SU($N$) amplitudes in general, exist, 
  differing in the way of choosing the set of vectors in which the color decomposition is obtained. 
  The trace 
  \cite{Paton:1969je,Berends:1987cv,
  Berends:1987me,Mangano:1987xk,Mangano:1988kk,Kosower:1988kh,Berends:1988yn,Berends:1988zn,Berends:1989hf,
  Bern:1990ux,Bern:1994fz,
  Ellis:2008qc,Ita:2011ar,
  PhDReuschle,Schuster:2013aya,Reuschle:2013qna,
  Nagy:2007ty,Sjodahl:2009wx,Platzer:2012np,
  Sjodahl:2012nk,Sjodahl:2014opa} 
  and color-flow 
  \cite{tHooft:1973alw,Maltoni:2002mq,Weinzierl:2005dd,Reuschle:2013qna,Kilian:2012pz}
  representations have been very useful for deriving
  amplitudes, and analytic formulae exist for many cases
  \cite{Berends:1987me,Mangano:1987xk,Mangano:1988kk,Kosower:1988kh,Berends:1988yn,Berends:1988zn,Berends:1989hf,
  Bern:1990ux,Bern:1994fz,
  PhDReuschle,Schuster:2013aya,Reuschle:2013qna}.
  On the other hand, orthogonal, group-theory based multiplet bases
  \cite{cvi76,Butera:1979na,Cvitanovic:1980bu,
  Zeppenfeld:1988bz,
  MacFarlane:1968vc,Keppeler:2012ih, Du:2015apa, Sjodahl:2015qoa,Sjodahl:2018cca,
  Cvi08,Sjodahl:2012nk,Alcock-Zeilinger:2016cva,
  Kyrieleis:2005dt,Dokshitzer:2005ig,Sjodahl:2008fz}
  are superior for squaring amplitudes, and adjoint (DDM) bases \cite{DelDuca:1999rs}
  may be advantageous for the pure gluon case.
}.
In this paper we define a similarly intuitive picture,
which we dub the chirality-flow formalism, 
for Feynman diagrams in the spinor-helicity formalism.

For future reference we define the generators $t$ and structure constants $f$ of the SU($N$) color algebra by
\begin{align}
\trace\big(t^at^b\big)
&=\delta^{ab}\,,\label{eq:color_norm}\\
if^{abc}&=
\trace\big(t^a[t^b,t^c]\big)\,,\label{eq:f_expr}
\end{align}
with the indices $a,b,c=1,...,N^2-1$ in the adjoint representation of SU($N$),
and the indices $i,j,k,l=1,...,N$ in the (anti)fundamental representation.
We normalize the generators to unity in
\eqref{eq:color_norm}
to avoid carrying around unnecessary powers of $2$ in QCD algebra relations.
This fixes the constant on the right-hand side of \eqref{eq:color_fierz0}
to unity. 
Combining \eqref{eq:color_norm} with the algebra relation
$[t^a,t^b]=i f^{abc} t^c$, \eqref{eq:f_expr} follows.

\section{The basics of the spinor-helicity formalism}
\label{sec:HelicityBasics}

In this section we review some basics of the spinor-helicity formalism, 
focusing on massless fermions and vector bosons (some additional relations are given in \appref{sec:conventions}, 
and more detailed introductions can  be found in \cite{Mangano:1990by,Gastmans:1990xh,Dixon:1996wi,Dittmaier:1998nn,Weinzierl:2005dd,Weinzierl:2007vk,Dreiner:2008tw,Ellis:2011cr,Peskin:2011in,Elvang:2013cua,Dixon:2013uaa}). 
We also introduce the chirality-flow representations of spinors, 
spinor inner products and bispinors.  

\subsection{Spinors and spinor inner products}
\label{sec:spinors}

Let us first consider an incoming fermion or outgoing anti-fermion of momentum $p$ in the chiral, or Weyl, representation.
For massless fermions and in a condensed notation we can write the corresponding four-component spinors in momentum space as
\begin{equation} 
  \label{eq:DiracSpinor}
  u(p)
  =
  \begin{pmatrix}
    u_L\\
    u_R 
  \end{pmatrix}
  =
  \begin{pmatrix}
    \tla_p^{\da}\\
    \la_{p,\be} 
  \end{pmatrix}
  ~,
  \quad
  v(p)
  =
  \begin{pmatrix}
    v_L\\
    v_R
  \end{pmatrix}
  =
  \begin{pmatrix}
    \tla_p^{\da}\\
    \la_{p,\be}
  \end{pmatrix}
  ~,
\end{equation}
where we have introduced the two-component Weyl spinors\footnote{
  Explicit representations of the Weyl spinors are given in 
  \eqsrefa{eq:la}{eq:tla}.
} 
$\la_{p,\be}$ and $\tla_p^{\da}$.
The state $\tla_p^{\da}$, with a dotted index, transforms under the left-chiral $(\frac{1}{2},0)$-representation of the Lorentz group,
while the state $\la_{p,\be}$, with an undotted index, transforms under the right-chiral $(0,\frac{1}{2})$-representation (see \appref{sec:appspinors}).
These states are projected out from the four-component spinors by the chiral projection operators $P_{\nicefrac{R}{L}}=\frac{1}{2}(1\nicefrac{+}{-}\gamma^5)$,
such that for example $P_Lv(p)=\big(\begin{smallmatrix}v_L\\0\end{smallmatrix}\big)=\big(\begin{smallmatrix}\tla_p^{\da}\\0\end{smallmatrix}\big)$ and $P_Rv(p)=\big(\begin{smallmatrix}0\\v_R\end{smallmatrix}\big)=\big(\begin{smallmatrix}0\\\la_{p,\be}\end{smallmatrix}\big)$,
where we use the Dirac matrices in the chiral basis, 
\begin{align}
    \gamma^{\mu} 
    = 
    \begin{pmatrix}
      0 & \sigma^{\mu,\da\be} \\
      \Bar{\sigma}^{\mu}_{\be\da} & 0
    \end{pmatrix}
    =
    \begin{pmatrix}
      0 & \sqrt{2}\tau^{\mu,\da\be} \\
      \sqrt{2}\Bar{\tau}^{\mu}_{\be\da} & 0
    \end{pmatrix}
    \;,
    \quad
    \gamma^5 
    = 
    i \gamma^0 \gamma^1 \gamma^2\gamma^3 
    =
    \begin{pmatrix}
      -1_{2\times2} & 0 \\
      0 & 1_{2\times2}
    \end{pmatrix}
    \;.
    \label{eq:gamma}
\end{align}
Here we have introduced normalized versions of the Pauli matrices\footnote{
  They are also known in the literature as Infeld-van-der-Waerden symbols (see e.g.\ \cite{Dreiner:2008tw}).
  Explicit representations of the Pauli matrices are given in \eqref{eq:sigma}.
}
such that, 
analogous to \eqref{eq:color_norm},
we have
\begin{equation}
  \trace\big(\tau^\mu \tau^\nu\big)
  = 
  \delta^{\mu \nu}
  \quad \Leftrightarrow \quad
  \trace\big(\tau^\mu \bar{\tau}^\nu\big)
  = 
  g^{\mu \nu}\;.
\end{equation}
The spinors for outgoing fermions and incoming anti-fermions are then given by
\begin{equation}
  \Bar{u}(p) 
  \equiv 
  u^{\dagger}(p)\gamma^0
  = 
  \left( (u_R)^{\dagger}, (u_L)^{\dagger}\right)
  = 
  \left(\tla_{p,\db}, \la_p^{\al}\right)
  \;,
  \quad
  \Bar{v}(p)
  \equiv 
  v(p)^{\dagger}\gamma^0
  = 
  \left( (v_R)^{\dagger}, (v_L)^{\dagger}\right)
  = 
  \left(\tla_{p,\db}, \la_p^{\al}\right)
  \;,
\label{eq:DiracSpinor2}
\end{equation}
such that for example $\bar{u}(p)P_L=\big((u_R)^{\dagger},0\big)=\big(\tla_{p,\db},0\big)$ and $\bar{u}(p)P_R=\big(0,(u_L)^{\dagger}\big)=\big(0,\la_p^{\al}\big)$,
where we have used the Hermitian conjugate relations for massless spinors,
\begin{equation}
  (\la_{p,\be})^{\dagger}=\tla_{p,\db}\quad\mbox{and}\quad
  \left( \tla_p^{\da} \right)^{\dagger}= \la_p^{\al}\;,
  \label{eq:spinorhermitianmain}
\end{equation}
which implies for the components $(\la_{p,\be})^*=(\tla_{p,\db})$ and $(\tla_p^{\da})^*=(\la_p^{\al})$,
for $\be=\dbe$ and $\dal=\al$.

In this paper we will use the convention of counting all particles
in a scattering process as outgoing. 
The four types of {\it outgoing} spinors we need to consider are thus
\begin{eqnarray}
    \mbox{Right-chiral fermion} & \lambda_i^{\alpha}  \leftrightarrow  \langle i| 
    & = \raisebox{-0.2\height}{\includegraphics[scale=0.4]{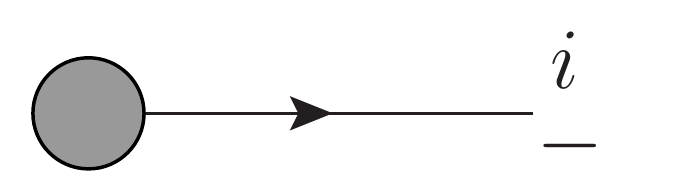}} 
      = \raisebox{-0.2\height}{\includegraphics[scale=0.40]{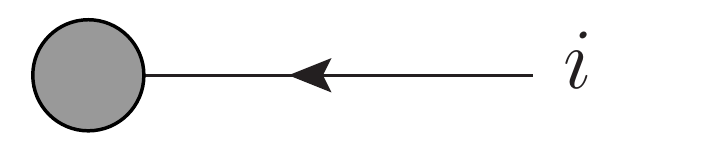}}  
    \label{eq:lambdaupper}
    \\
    \mbox{Right-chiral anti-fermion} & \la_{j,\alpha}  \leftrightarrow | j \rangle  
    & = \raisebox{-0.2\height}{\includegraphics[scale=0.4]{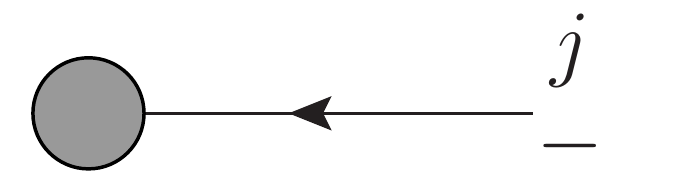}} 
      = \raisebox{-0.2\height}{\includegraphics[scale=0.40]{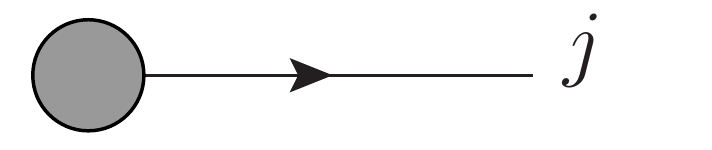}}  
    \label{eq:lambdalower}
    \\
    \mbox{Left-chiral fermion} & \tilde{\lambda}_{i,\da}  \leftrightarrow  [i|  
    & = \raisebox{-0.2\height}{\includegraphics[scale=0.4]{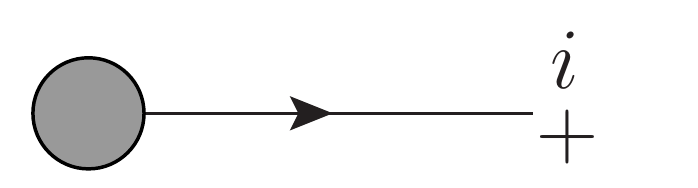}} 
      = \raisebox{-0.2\height}{\includegraphics[scale=0.40]{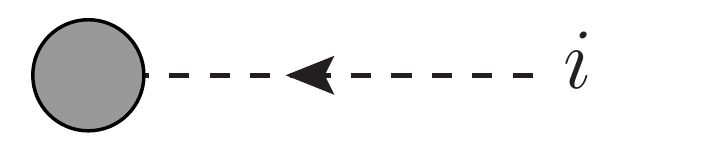}}  
    \label{eq:lambdatildeupper}
    \\
    \mbox{Left-chiral anti-fermion} & \tilde{\lambda}_j^{\dot{\alpha}}  \leftrightarrow  |j] 
    & = \raisebox{-0.2\height}{\includegraphics[scale=0.4]{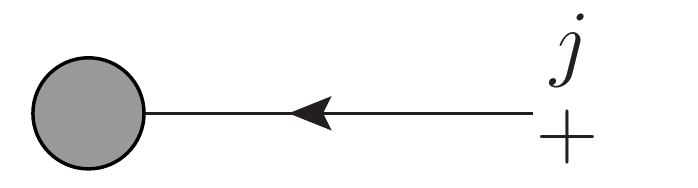}} 
      = \raisebox{-0.2\height}{\includegraphics[scale=0.4]{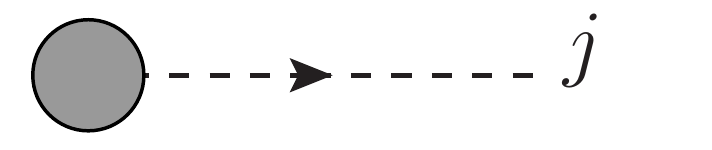}}  
    \label{eq:lambdatildelower}
    \;,
\end{eqnarray}
where we further introduce the common bra-ket notation for Weyl spinors, 
and where the left graphical rules correspond to the conventional Feynman rules 
(showing fermion-flow arrow, momentum label and helicity label)
whereas in the right pictorial rules,
we use dotted lines to denote particles with dotted indices
(outgoing positive helicity)
and solid lines to denote particles with undotted indices (outgoing negative helicity)\footnote{\label{ft:helicityvschirality}
  For massless, outgoing fermions as well as anti-fermions, 
  the positive and negative helicity states are described by the corresponding left- and right-chiral Weyl spinors respectively,
  i.e. $\bar{u}^+=\big((\bar{u}^+)_L,(\bar{u}^+)_R\big)=\big((\bar{u}^+)_L,0\big)$ and $\bar{u}^-=\big((\bar{u}^-)_L,(\bar{u}^-)_R\big)=\big(0,(\bar{u}^-)_R\big)$ as well as $v^+=\Big(\begin{smallmatrix}(v^+)_L\\(v^+)_R\end{smallmatrix}\Big)=\big(\begin{smallmatrix}(v^+)_L\\0\end{smallmatrix}\big)$ and $v^-=\Big(\begin{smallmatrix}(v^-)_L\\(v^-)_R\end{smallmatrix}\Big)=\big(\begin{smallmatrix}0\\(v^-)_R\end{smallmatrix}\big)$,
  where we have made the helicity labels explicit.
}. 
Also note the arrow direction in the rightmost pictorial rules, which
--- at this point --- goes against the fermion-flow arrow direction.
This arrow direction follows what we dub a chirality flow,
i.e.\ a flow between Weyl spinors in the $(1/2,0)$-representation
(for dotted lines) or the $(0,1/2)$-representation (for solid lines).

We will find situations where the chirality-flow arrow cannot be 
kept anti-aligned with respect to the fermion-flow arrow.
Nevertheless, 
we adopt the convention of reading all chirality-flow expressions,
such as spinor inner products,
{\it following} the chirality-flow arrow.
For convenience we collect our conventions 
in \tabsref{tab:Flow rules comparison QED} and \ref{tab:Flow rules comparison QCD} in \appref{sec:Rosetta stone}.

Spinor indices are raised and lowered by the Levi-Civita tensor (sometimes called the spinor metric) 
$\epsilon^{\alpha \beta}/\epsilon^{\da \db}/\epsilon_{\alpha \beta}/\epsilon_{\da \db}$,
which we define as
\begin{equation}
  \epsilon^{12} = -\epsilon^{21} = \epsilon_{21} = -\epsilon_{12} = 1~,
  \label{eq:epsilon}
\end{equation}
such that $\eps_{\al\be}\eps^{\be\ga}\!=\!\delta_{\al}^{\ \ga}$ and 
$\eps^{\dal\dbe}\eps_{\dbe\dga}\!=\!\delta^{\dal}_{\ \dga}$.
With our definition of the $\epsilon$-tensor above, 
the operations for lowering and raising spinor indices are\footnote{
  Note that $\la_{p,\al}=-\epsilon_{\be\al}\la_p^{\be}=-\la_p^{\be}\epsilon_{\be\al}$.
}
\begin{equation}
\la_{p,\al} = \epsilon_{\al\be}\la_p^{\be}~, 
\quad 
\tla_{p,\dal} = \epsilon_{\dal\dbe}\tla_p^{\dbe}~,
\quad
\la_p^{\al} = \epsilon^{\al\be}\la_{p,\be}~, 
\quad 
\tla_p^{\dal} = \epsilon^{\dal\dbe}\tla_{p,\dbe}~.
\label{eq:epsilon_action2}  
\end{equation}
Considering that $\epsilon$ is the SL(2,C) invariant object,
the definitions for the (antisymmetric,  Lorentz invariant) spinor inner products follow as
\begin{align}
  \langle i j \rangle &
  \defequal \langle i || j \rangle
  = \la_i^{\al}\la_{j,\al}
  = \epsilon^{\al \be}\la_{i,\be}\la_{j,\al}
  = -\epsilon^{\be \al}\la_{i,\be}\la_{j,\al} = -\la_{i,\be}\la_j^{\be} = -\langle j i \rangle~, 
  \label{eq:spinorproduct1}
  \\
  [ i j ] &
  \defequal [ i || j ]
  = \tla_{i,\da}\tla_j^{\da}
  = \epsilon_{\da \db}\tla_i^{\db}\tla_j^{\da}
  = -\epsilon_{\db \da}\tla_i^{\db}\tla_j^{\da}
  = - \tla_i^{\db}\tla_{j,\db} = -[ji]~,
  \label{eq:spinorproduct2}
\end{align}
implying in particular $\langle i i \rangle=[ii]=0$.
Having defined spinors and their graphical representations in the chirality-flow picture, 
we can represent the spinor inner products pictorially as well,
\begin{align} \label{eq:Angle Prod Flow}
  \langle i j \rangle\; 
  & = \; \raisebox{-0.2\height}{\includegraphics[scale=0.40]{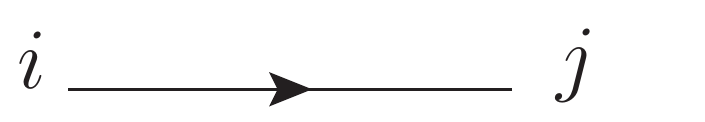}} 
  \;\; , \quad \quad 
  \langle j i \rangle\; 
  = \; \raisebox{-0.2\height}{\includegraphics[scale=0.40]{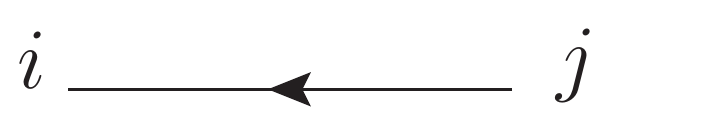}}
  \;\; , \\
  [ i j ] \;
  & = \; \raisebox{-0.2\height}{\includegraphics[scale=0.40]{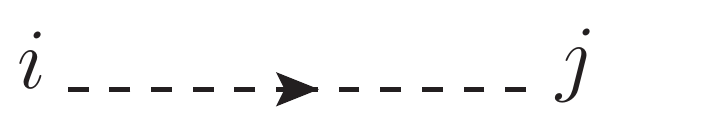}} 
  \;\; , \quad \quad 
  [ j i ] \;
  \,= \; \raisebox{-0.2\height}{\includegraphics[scale=0.40]{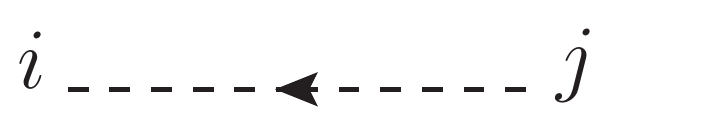}}
  \;\; ,
  \label{eq:Square Prod Flow}
\end{align}
where we read \textit{following} the chirality-flow arrow.

Any scattering amplitude can be written in terms of Lorentz-invariant spinor inner products, 
and in the following sections we will see that we can always find corresponding graphical representations for its Feynman diagrams in terms of chirality-flow lines.

\subsection{Four-vectors}
\label{sec:fourvectors}

Next, we consider Lorentz four-vectors.
A four-vector $p_\mu$ can be mapped to Hermitian $2\times2$-matrices,
or bispinors\footnote{
  Explicit representations of momentum bispinors are given in 
  \eqsrefa{eq:p_contr2_app}{eq:p_contr3_app}.
},
\begin{align}
  \label{eq:p_contr0}
  p^{\dal\be}
  &\defequal 
  p_{\mu}\tau^{\mu,\dal\be} 
  =
  \frac{1}{\sqrt{2}}
  p_{\mu}\si^{\mu,\dal\be} 
  ~, \qquad \quad
  \slashed{p}
  \defequal   
  p_{\mu}\si^{\mu} 
  ~,
  \\ 
  \bar{p}_{\al\dbe}
  &\defequal 
  p_{\mu} \Bar{\tau}^{\mu}_{\al\dbe} 
  =
  \frac{1}{\sqrt{2}}
  p_{\mu} \Bar{\si}^{\mu}_{\al\dbe}
    ~, \qquad \quad
  \quad \ \, \bar{\slashed{p}}
  \defequal   
  p_{\mu}\bar{\si}^{\mu} ~,
  \label{eq:p_contr}
\end{align}
where we have introduced a slash notation for the momentum bispinors,
not to be confused with the Feynman slash.
The ordinary Lorentz four-vector transformation rules are recovered by boosting and rotating the bispinors as indicated by the spinor index structure,
which translates to the statement that Lorentz four-vectors transform under the $(1/2,1/2)=(1/2,0)\otimes(0,1/2)$ representation of the (restricted) Lorentz group.

Translating the index structure into the chirality-flow formalism allows us to define a convenient momentum-dot notation
\begin{equation}
  \label{eq:p_contr2}
  \slashed{p} 
  \;\;\leftrightarrow\;\;
  \sqrt{2} p^{\da\be} 
  \defequal
  \raisebox{-0.15\height}{\includegraphics[scale=0.55]{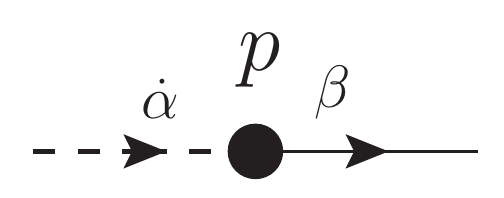}}
  \quad\quad\text{and}\quad\quad
  \bar{\slashed{p}} 
  \;\;\leftrightarrow\;\;
  \sqrt{2}\bar{p}_{\al\db} 
  \defequal
  \raisebox{-0.15\height}{\includegraphics[scale=0.55]{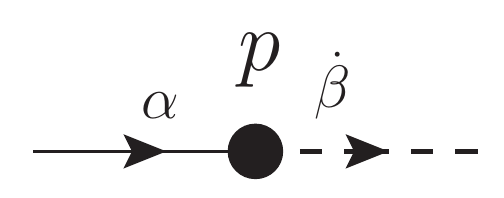}}
  \;\;.
\end{equation}
If a momentum $p$ is massless,
i.e. light-like,
the corresponding momentum bispinors can be expressed as outer products, 
or dyads, 
of Weyl spinors,
\begin{align}
    \slashed{p} 
    = 
    |p]\langle p| 
      & \;\;\leftrightarrow\;\;
    \sqrt{2}p^{\da\be} 
    =
    \tla_p^{\da}\la_p^{\be}~,
    \quad\;\;\;\,\, \text{for}\;\;
    p^2=0~,
    \label{eq:Mom two-spinors}
    \\
    \bar{\slashed{p}} 
    = 
    |p\rangle [p|
      & \;\;\leftrightarrow\;\;
    \sqrt{2}\bar{p}_{\al\db} 
    = 
    \la_{p,\al}\tla_{p,\db}~,
    \quad\text{for}\;\; p^2=0~,
    \label{eq:Mombar two-spinors}
\end{align}
or in the chirality-flow picture, 
\begin{eqnarray}
  \quad \raisebox{-0.15\height}{\includegraphics[scale=0.55]{./Jaxodraw/FermionPropFlowg}}
  \;
  &=
  \;\; \raisebox{-0.20\height}{\includegraphics[scale=0.55]{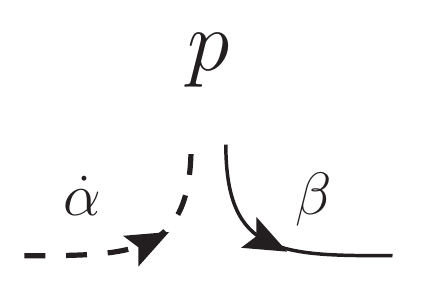}}\;,
  \quad\mbox{for}\quad p^2=0\;,
  \\
  \quad \raisebox{-0.15\height}{\includegraphics[scale=0.55]{./Jaxodraw/FermionPropFlowh}}
  \;
  &=
  \;\; \raisebox{-0.20\height}{\includegraphics[scale=0.55]{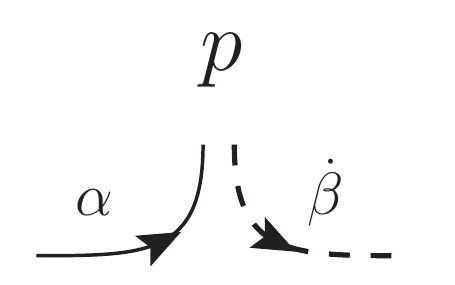}},
  \quad\mbox{for}\quad p^2=0\;,
\end{eqnarray}
where $p$ in the right graphical rules denotes that the line ends correspond to spinors with momentum $p$.
If a momentum $p$ is a linear combination of several external momenta 
$p_i, ~ p=\sum_ic_ip_i$ with $p_i^2=0$
--- which is always the case at tree level ---
we may use the linearity of 
\eqsrefa{eq:Mom two-spinors}{eq:Mombar two-spinors} 
in $p$ to write
\begin{align}
    \label{eq:Mom two-spinors_pi}
    \slashed{p} 
    = 
    \sum\limits_i
    c_i
    |p_i]\langle p_i|
    & \;\;\leftrightarrow\;\;
    \sqrt{2}p^{\da\be} 
    =
    \sum\limits_i
    c_i
    \tla_{p_i}^{\da}\la_{p_i}^{\be}~,
       \\
    \label{eq:Mombar two-spinors_pi}
    \bar{\slashed{p}} 
    = 
    \sum\limits_i
    c_i
    |p_i\rangle [p_i|
      & \;\;\leftrightarrow\;\;
    \sqrt{2}\bar{p}_{\al\db} 
    = 
    \sum\limits_i
    c_i
    \la_{p_i,\al}\tla_{p_i,\db}~,
\end{align}
or in the chirality-flow picture, 
\begin{eqnarray}
  \raisebox{-0.15\height}{\includegraphics[scale=0.55]{./Jaxodraw/FermionPropFlowg}}
  \;
  &=
  \;\; 
  \displaystyle\sum\limits_ic_i\raisebox{-0.20\height}{\includegraphics[scale=0.55]{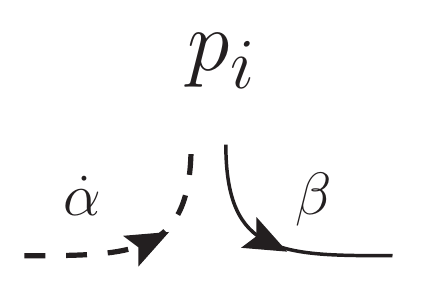}},
  \label{eq:p_sum0}
  \\
  \raisebox{-0.15\height}{\includegraphics[scale=0.55]{./Jaxodraw/FermionPropFlowh}}
  \;
  &=
  \;\; 
  \displaystyle\sum\limits_ic_i\raisebox{-0.20\height}{\includegraphics[scale=0.55]{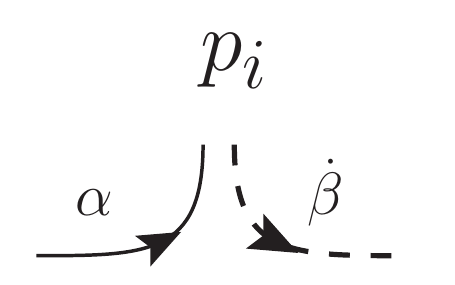}}.
  \label{eq:p_sum}
\end{eqnarray}

It is sometimes useful to write a four-vector $p^{\mu}$ in terms of a spinor contraction,
rather than as a bispinor.
We can do so by using 
\eqref{eq:tautrace0} with \eqref{eq:p_contr0},
then using \eqref{eq:Mom two-spinors},
\begin{align} \label{eq:Gordona}
p^{\mu} 
= p^{\da\be}\taubar^{\mu}_{\be\da}
= \frac{1}{\sqrt{2}} \la_p^{\be}\taubar^{\mu}_{\be\da}\tla_p^{\da}
= \frac{1}{\sqrt{2}} \langle p | \taubar^{\mu} | p]~,
\end{align}
or similarly, 
by using \eqref{eq:tautrace0} with \eqref{eq:p_contr},
then using \eqref{eq:Mombar two-spinors}, 
\begin{align} \label{eq:Gordonb}
p^{\mu} 
= \bar{p}_{\al\db}\tau^{\mu,\db\al}
= \frac{1}{\sqrt{2}} \tla_{p,\db}\tau^{\mu\db\al}\la_{p,\al}
= \frac{1}{\sqrt{2}} [p | \tau^{\mu} | p\rangle~.
\end{align}
The above relations will be utilized when proving the chirality-flow picture in \secref{sec:HelicityFlow},
and in the chirality-flow Feynman rules of the fermion propagator and the triple-gluon vertex in \secref{sec:FeynmanRules}.

As a final comment, 
we remind that for massless spinors the Dirac equation separates into two massless Weyl equations, 
which in the spinor-helicity formalism take on a particularly simple form.
For example, 
using \eqref{eq:Mom two-spinors} we have
\begin{align}
  \slashed{p}|p\rangle
  \overset{\scriptscriptstyle p^2=0}{=}
  \big(|p]\langle p|\big)|p\rangle
  = |p]\langle pp\rangle
  = 0~,
\end{align}
which is easily confirmed to be true, 
as $\langle pp\rangle=0$ due to the antisymmetry of the spinor inner products
(for the other three Weyl equations see \appref{sec:appfourvectors}).

\subsection{Polarization vectors}
\label{sec:polarizationvectors}

Aside from external spin-$1/2$ fermions and momentum four-vectors
we also need to treat external vector bosons, 
for which the outgoing polarization vectors can also be written in terms of Weyl spinors 
\cite{Berends:1981rb,DeCausmaecker:1981jtq,
Gunion:1985bp,Gunion:1985vca,Kleiss:1986qc,Xu:1986xb,
Gastmans:1990xh},
\begin{eqnarray}
  \label{eq:polvec1}
  \epsilon_-^{\mu}(p_i,r)
  &=& \frac{\la_i^{\al}\taubar^{\mu}_{\al\db}\tla_r^{\db}}{\tla_{i,\dga} \tla_{r}^{\dga}}
  = \frac{\langle i|\taubar^{\mu}|r]}{[ir]}
  = \raisebox{-0.25\height}{\includegraphics[scale=0.5]{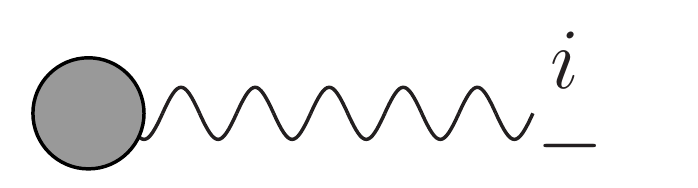}}\,,
  \\
  \epsilon_+^{\mu}(p_i,r)
  &=&
  \frac{\la_r^{\al}\taubar^{\mu}_{\al\db}\tla_i^{\db}}{\la_r^{\ga} \la_{i,\ga}}
  = 
  \frac{\langle r|\taubar^{\mu}|i]}{\langle ri\rangle}
  = 
  \raisebox{-0.25\height}{\includegraphics[scale=0.5]{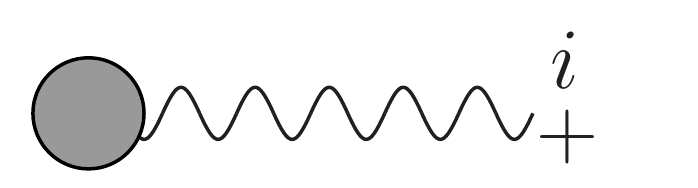}}\,,
  \label{eq:polvec2}
\end{eqnarray}
where $p_i$ is the vector boson momentum and $r$ is an arbitrary light-like reference momentum satisfying $p_i\cdot r \neq 0$.
The subscripts in $\eps_{\pm}$ denote helicity labels,
and the two polarizations are related by complex conjugation,
i.e. $\big(\epsilon_-^{\mu}(p_i,r)\big)^*=\epsilon_+^{\mu}(p_i,r)\,$.

As for any four-vector, 
we may trade the Lorentz index in $\epsilon_\pm^\mu$ for spinor indices
using \eqsrefd{eq:p_contr0}{eq:p_contr}
and \eqref{eq:taufierz0}
and write them as bispinors, 
\begin{eqnarray}
  \label{eq:epsilon0}
  \eps_{-}^{\dbe\al}(p_i,r)
  &=&
  \frac{\tla_r^{\db}\la_i^{\al}}{\tla_{i,\dga} \tla_{r}^{\dga}}
  = 
  \frac{|r]\langle i|}{[ir]}
  =
  \frac{1}{[ir]}
  \raisebox{-0.25\height}{\includegraphics[scale=0.5]{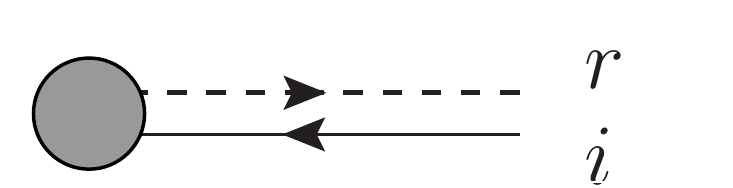}}\;,
  \\
  \eps_{+}^{\dbe\al}(p_i,r)
  &=&
  \frac{\tla_i^{\db}\la_r^{\al}}{\la_r^{\ga} \la_{i,\ga}}
  =
  \frac{|i]\langle r|}{\langle ri\rangle}
  = 
  \frac{1}{\langle ri\rangle}
  \raisebox{-0.25\height}{\includegraphics[scale=0.5]{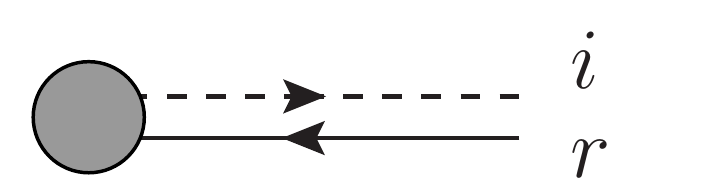}} \;.
  \label{eq:epsilon1}
\end{eqnarray}
From the bispinor representations, we note that an external vector boson has the same numerator structure as an external fermion-antifermion pair, 
giving a natural graphical interpretation in terms of chirality-flow lines.
Also, the two polarization-bispinors are not directly related by complex conjugation.
Rather, viewed as matrices we have 
$\big(\eps_{-}(p_i,r)\big)^*=\big(\eps_{+}(p_i,r)\big)^\intercal$,
which follows from the Hermitian conjugation relations between Weyl spinors, \eqref{eq:spinorhermitianmain}.

As we will see in the next section, 
we can equally well write the polarization vectors as
\begin{eqnarray}
  \label{eq:polvec3}
  \epsilon_-^{\mu}(p_i,r)
  &=& \frac{\tla_{r,\da}\tau^{\mu,\da\be}\la_{i,\be}}{\tla_{i,\dga} \tla_r^{\dga}}
  = \frac{[r|\tau^{\mu}|i\rangle}{[ir]}
  =\raisebox{-0.25\height}{\includegraphics[scale=0.5]{./Jaxodraw/PhotonExtMi}}\,, 
  \\
  \epsilon_+^{\mu}(p_i,r)
  &=&\frac{\tla_{i,\da}\tau^{\mu,\da\be}\la_{r,\be}}{\la_r^\ga \la_{i,\ga}}
  =\frac{[i|\tau^{\mu}|r\rangle}{\langle ri\rangle}
  = \raisebox{-0.25\height}{\includegraphics[scale=0.5]{./Jaxodraw/PhotonExtPl}}\,,
  \label{eq:polvec4}
\end{eqnarray}
or again as bispinors, 
\begin{eqnarray}
  \label{eq:epsilon2}
  \bar{\eps}_{-,\be\dal}(p_i,r)
  &=&
  \frac{\la_{i,\be}\tla_{r,\da}}{\tla_{i,\dga} \tla_{r}^{\dga}}
  = 
  \frac{|i\rangle[r|}{[ir]}
  =
  \frac{1}{[ir]}
  \raisebox{-0.25\height}{\includegraphics[scale=0.5]{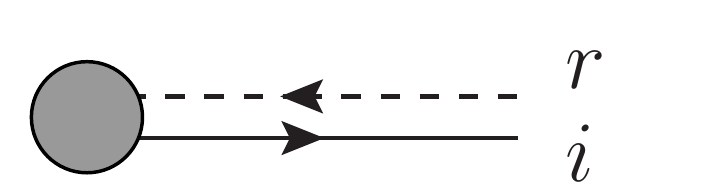}}\;,
  \\
  \bar{\eps}_{+,\be\dal}(p_i,r)
  &=&
  \frac{\la_{r,\be}\tla_{i,\da}}{\la_r^{\ga} \la_{i,\ga}}
  =
  \frac{|r\rangle[i|}{\langle ri\rangle}
  = 
  \frac{1}{\langle ri\rangle}
  \raisebox{-0.25\height}{\includegraphics[scale=0.5]{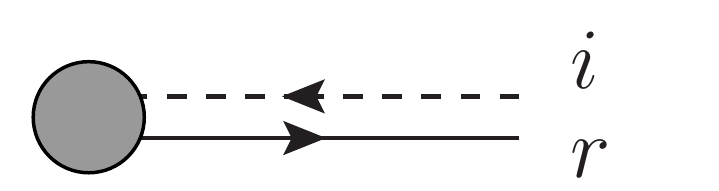}}\,.
  \label{eq:epsilon3}
\end{eqnarray}
Some more details are provided at the end of \appref{sec:appfourvectors}.

\section{Building the chirality-flow picture}
\label{sec:HelicityFlow}

\private{\commentMS{
Should still be checked with care.
}}

In this section we will see that we can always cast any tree-level Feynman diagram into a linear combination of products of chirality flows, 
which 
--- when contracted with external spinors --- 
result in spinor inner products.
We will also be able to express these chirality flows diagrammatically.
This sets the scene for 
the tree-level chirality-flow Feynman rules in the next section.
Here and in the following we work in Feynman gauge.

\subsection{A simple QED example}

Let us first consider the example of
single photon exchange between two distinct fermions. 
We note that in the chiral representation, the Lorentz structure of the 
fermion-photon vertex may be separated into two parts,
corresponding to two different vertices
\begin{equation}
  \bar{u}(p_1)\gamma^\mu v(p_2)
  =\left(\tla_{1,\db}, \la_1^{\al}\right)
  \begin{pmatrix}
    0 & \sqrt{2}\tau^{\mu,\db \eta}  \\
    \sqrt{2}\Bar{\tau}^{\mu}_{\al \Dot{\gamma}} & 0
  \end{pmatrix}
  \begin{pmatrix}
    \tla_2^{\Dot{\gamma}} \\
    \la_{2,\eta} 
  \end{pmatrix}
  =\sqrt{2}\tla_{1,\db} \tau^{\mu,\db \eta}\la_{2,\eta}
  +\sqrt{2}\la_1^{\al}\Bar{\tau}^{\mu}_{\al \Dot{\gamma}}\tla_2^{\Dot{\gamma}}\;.
  \label{eq:ubar_gamma_u}
\end{equation}
For photon exchange,
i.e. two fermion-photon vertices with an intermediate photon propagator,
there are four possible terms, 
corresponding to the four possible helicity combinations. 
Focusing only on the 
Lorentz 
structure we have
\begin{align}
&\!\Big(\bar{u}(p_1)\gamma^\mu v(p_2)\Big)
g_{\mu\nu}
\Big(\bar{u}(p_3)\gamma^\nu v(p_4)\Big)
\sim
\label{eq:photon_exchange}
\\
&(\tla_{1,\dal} \tau^{\mu,\dal \be}\la_{2,\be})\! (\tla_{3,\dga} \tau_{\mu}^{\dga \eta}\la_{4,\eta})
+(\tla_{1,\dal} \tau^{\mu,\dal \be}\la_{2,\be})\! (\la_3^{\ga}\Bar{\tau}_{\mu,\ga \deta}\tla_4^{\deta})
+(\la_1^{\al}\Bar{\tau}^{\mu}_{\al \dbe}\tla_2^{\dbe})\! (\tla_{3,\dga} \tau_{\mu}^{\dga \eta}\la_{4,\eta})
+(\la_1^{\al}\Bar{\tau}^{\mu}_{\al \dbe}\tla_2^{\dbe})\! (\la_3^{\ga}\Bar{\tau}_{\mu,\ga \deta}\tla_4^{\deta})\nonumber \\
&\sim 
\hspace{2ex}
\raisebox{-0.4\height}{\includegraphics[scale=0.55]{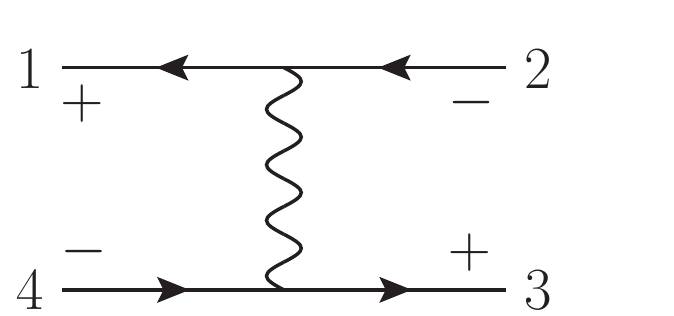}}
\hspace{-0.25ex}+\hspace{3.25ex}
\raisebox{-0.4\height}{\includegraphics[scale=0.55]{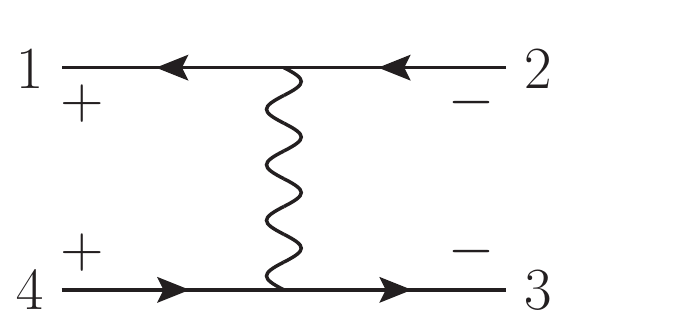}}
\hspace{-0.1ex}+\hspace{1.5ex} 
\raisebox{-0.4\height}{\includegraphics[scale=0.55]{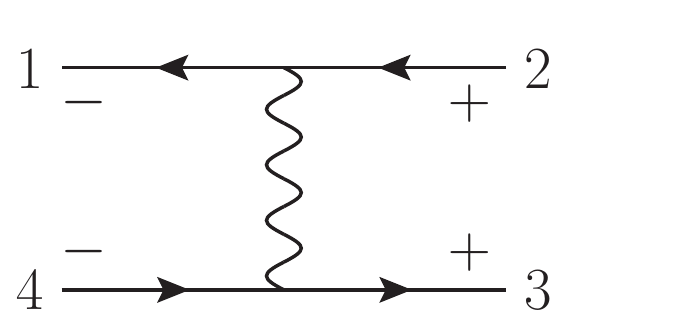}}
\hspace{-1.125ex}+\hspace{0.875ex} 
\raisebox{-0.4\height}{\includegraphics[scale=0.55]{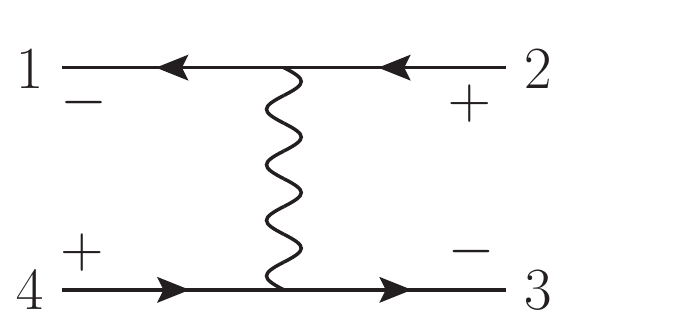}}
\hspace{-1.75ex}\nonumber.
\end{align}

Considering the second or third term in \eqref{eq:photon_exchange}, 
having the structure $\tau^{\mu}\Bar{\tau}_{\mu}$, 
we note that the Lorentz indices can be contracted using the Fierz identity. 
Graphically, for example for the third term, 
we have in the chirality-flow picture
\begin{equation} 
  \label{eq:Fierz Rearrangement Tau}
  \Bar{\tau}^{\mu}_{\al \dbe} \tau_{\mu}^{\dga \eta}
  \;\;= \raisebox{-0.42\height}{\includegraphics[scale=0.40]{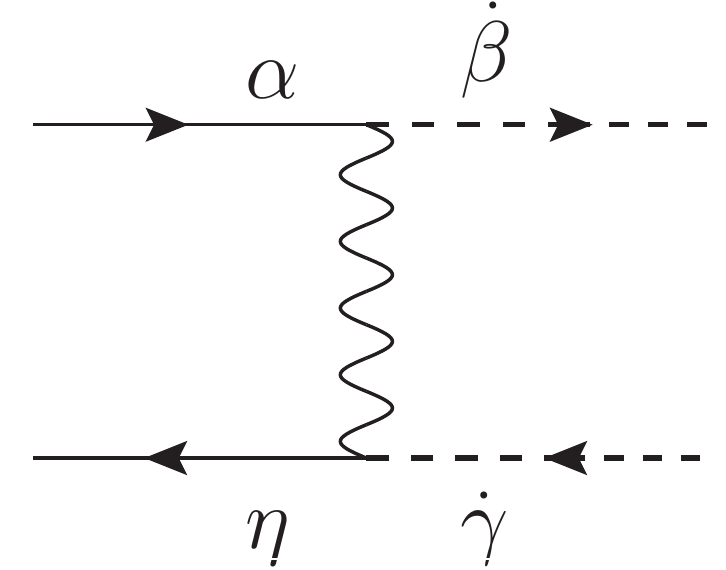}} =
  \raisebox{-0.42\height}{ \includegraphics[scale=0.40]{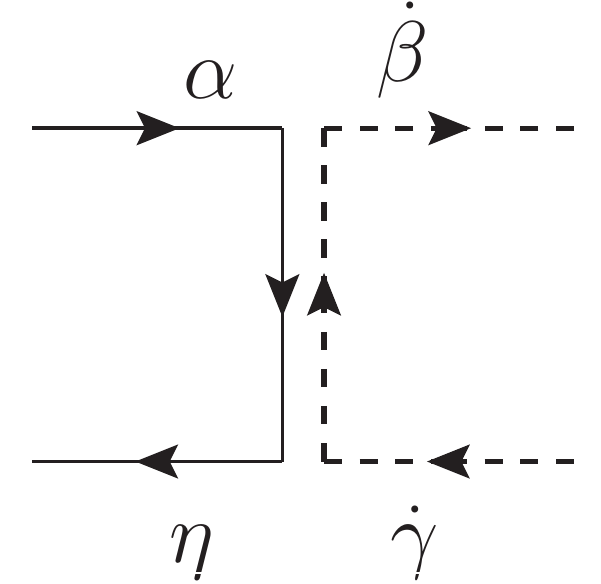}} 
  = \;\;\delta_{\alpha}^{~\eta} \delta^{\Dot{\gamma}}_{~\Dot{\beta}} \;,
\end{equation}
where we have not yet applied the external spinors.
This is a graphical embodiment of the Fierz identity,
and can be compared to the corresponding identity for SU(3) color,
\eqref{eq:color_fierz0}.
The left diagram in \eqref{eq:Fierz Rearrangement Tau} is a hybrid representation between a conventional Feynman diagram and a diagram in the chirality-flow picture.
Here
we let the photon line denote the group theory structure only, 
in analogy with the gluon line in 
\eqref{eq:color_fierz2}\footnote{\label{ft:hybrid}
  The absence of the $1/N$-suppressed term in \eqref{eq:Fierz Rearrangement Tau},
  compared to 
  \eqsref{eq:color_fierz0} 
  or (\ref{eq:color_fierz2}), 
  can be understood by noting that this term is canceled against the contribution from
  $\taubar^{0}_{\al\dbe}\tau_{0}^{\dga\eta}$.
  An equivalent way of viewing this is that we are summing over the generators of U(2) in \eqref{eq:Fierz Rearrangement Tau}, 
  meaning that we should not expect an additional term.
}.
We see that, at least in this case, 
the Lorentz structure of the photon propagator in the chirality-flow picture 
may be represented by a double line
\begin{equation}
  g^{\mu\nu}\quad\rightarrow\quad
  \raisebox{-0.25\height}{\includegraphics[scale=0.5]{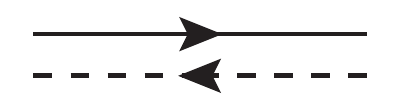}}\quad\mbox{or}\quad \raisebox{-0.25\height}{\includegraphics[scale=0.5]{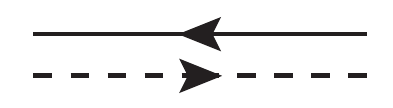}}\;\;,
  \label{eq:g_munuPic}
\end{equation}
where we defined a double line to be two parallel lines with opposing arrows, 
where one line is dotted and the other is undotted.

Note that \eqref{eq:Fierz Rearrangement Tau}, 
when applied to external spinors, 
results directly in the spinor inner products 
\begin{equation}
  \taubar^{\mu}_{\al\dbe}\tau_{\mu}^{\dga\eta}
  \la_1^\al\tla_2^{\dbe}\tla_{3,\dga}\la_{4,\eta}
  =\langle 1\,4 \rangle [3\,2]\,,
\label{eq:bla}
\end{equation}
or pictorially\footnote{
  We remark that the graphical appearance of \eqref{eq:Fierz Rearrangement applied} is very similar to the one of \eqref{eq:Fierz Rearrangement Tau}, 
  the difference being only in the labels of the external lines.
  We will usually supply the external lines of a chirality-flow diagram with particle labels, 
  i.e. labels of spinor momenta, 
  but the external lines may in principle also be kept ``free'',
  i.e. with spinor indices to act as placeholders, 
  to be supplied with particle labels at some later stage. 
},
\begin{equation} \label{eq:Fierz Rearrangement applied}
  \raisebox{-0.42\height}{\includegraphics[scale=0.40]{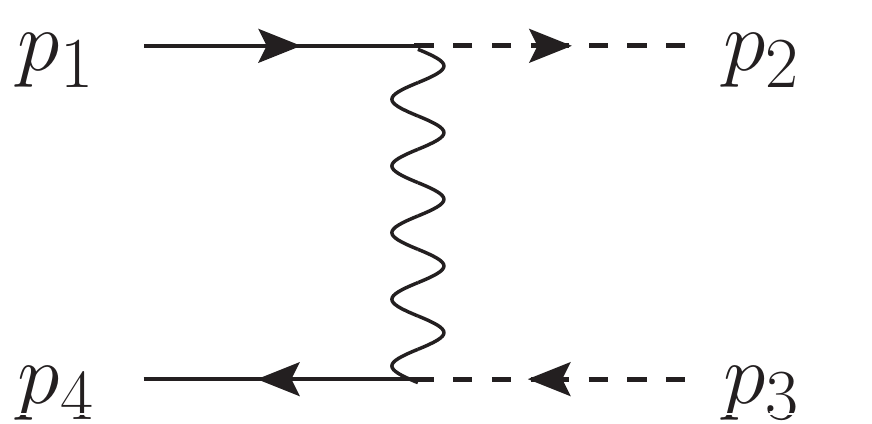}}
  =
  \raisebox{-0.42\height}{ \includegraphics[scale=0.40]{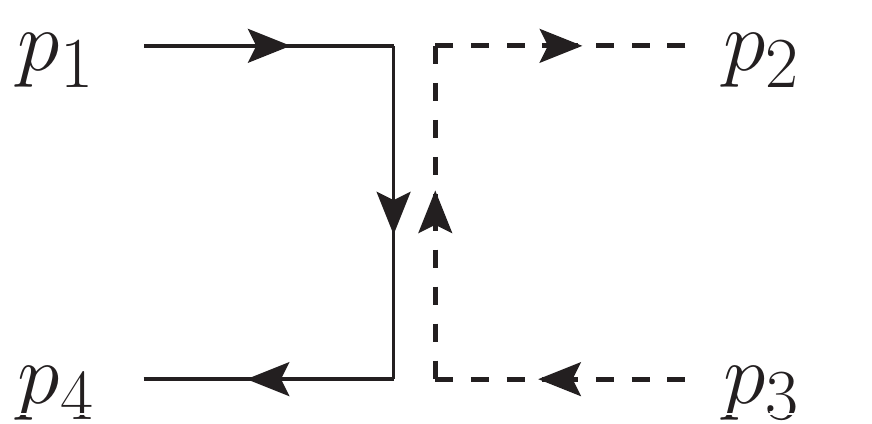}}\;.
\end{equation}

In the first and fourth terms in \eqref{eq:photon_exchange}, 
of the forms $\tau^{\mu} \tau_\mu$ and $\taubar^{\mu} \Bar{\tau}_\mu$ respectively,
the arrow directions in the chirality-flow picture would 
--- at this point ---
not match. 
For example, 
for the fourth term we have (using \eqref{eq:taufierz1})
\begin{equation} \label{eq:Fierz Rearrangement Wrong dir}
  \taubar^{\mu}_{\alpha \dbe} \Bar{\tau}_{\mu,\gamma \deta}
  \;\;= 
  \raisebox{-0.42\height}{\includegraphics[scale=0.40]{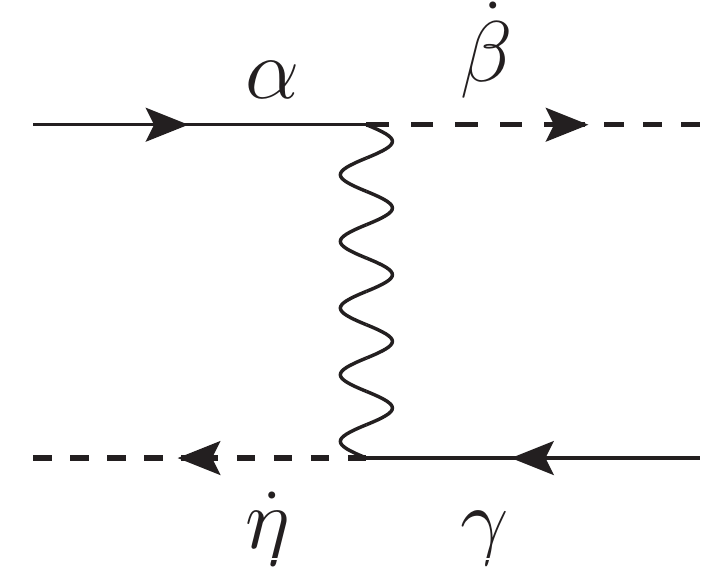}}
  =\;\;
  \epsilon_{\al\ga}\epsilon_{\dbe\deta}\;,
\end{equation}
where we note that the dotted (undotted) lines point away from (towards) each other.
Here it is less obvious how to proceed,
but we will show that the flow picture as 
applied in \eqref{eq:Fierz Rearrangement Tau}, the case of matching arrows,
can actually be applied here as well.

To see this we use the identity for charge conjugation of a current\footnote{
  In the last step we use \eqref{eq:tau taubar}, the relation between $\tau$ and $\Bar{\tau}$, 
  which can also be seen from the explicit matrix representation for the $\tau$ matrices, 
  \eqref{eq:sigma}.
} 
\begin{equation} 
  \la_i^{\al}\taubar^{\mu}_{\al\dbe}\tla_j^{\dbe}
  \defequal 
  \epsilon^{\al\ga}\epsilon^{\dbe\dde}\la_{i,\ga}\taubar^{\mu}_{\al\dbe}\tla_{j,\dde}
  = 
  \tla_{j,\dde}(-\epsilon^{\ga\al})(-\epsilon^{\dde\dbe})\taubar^{\mu}_{\al\dbe}\la_{i,\ga}
  = 
  \tla_{j,\dde}\tau^{\mu,\dde\ga}\la_{i,\ga}\;,
  \label{eq:charge conj current indices}
\end{equation}
or equivalently 
\begin{equation} 
\langle i|\taubar^{\mu}|j]=[j|\tau^{\mu}|i\rangle\;,
\label{eq:charge conj current matrix}
\end{equation}
which we have already used to relate the polarization vectors in 
\eqsrefa{eq:polvec1}{eq:polvec2} to the ones in \eqsrefa{eq:polvec3}{eq:polvec4} respectively. 
This means that whenever a $\bar{\tau}$ is squeezed between external spinors 
we can trade it for a $\tau$, or vice versa, 
\textit{if} we also perform the corresponding index lowering 
or raising 
operations on the spinors.

The pictorial representation of 
\eqsrefa{eq:charge conj current indices}{eq:charge conj current matrix}
in the hybrid representation is
\begin{equation} \label{eq:charge conj current easy}
  \raisebox{-0.4\height}{\includegraphics[scale=0.40]{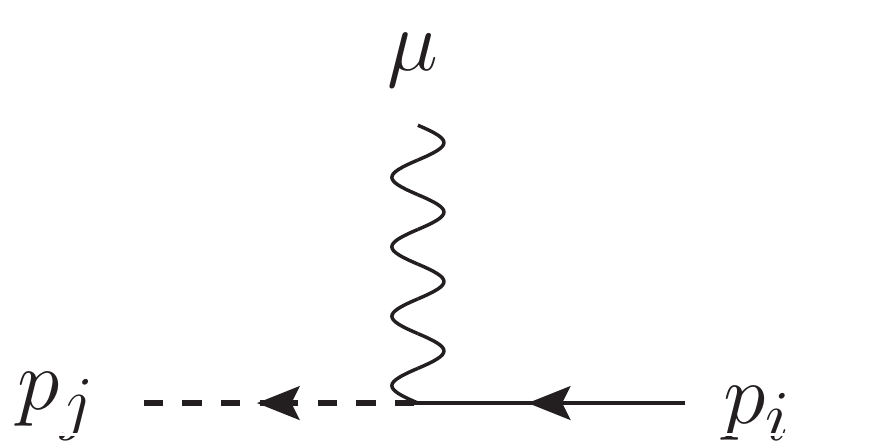}} 
  =\;\;
  \raisebox{-0.4\height}{\includegraphics[scale=0.40]{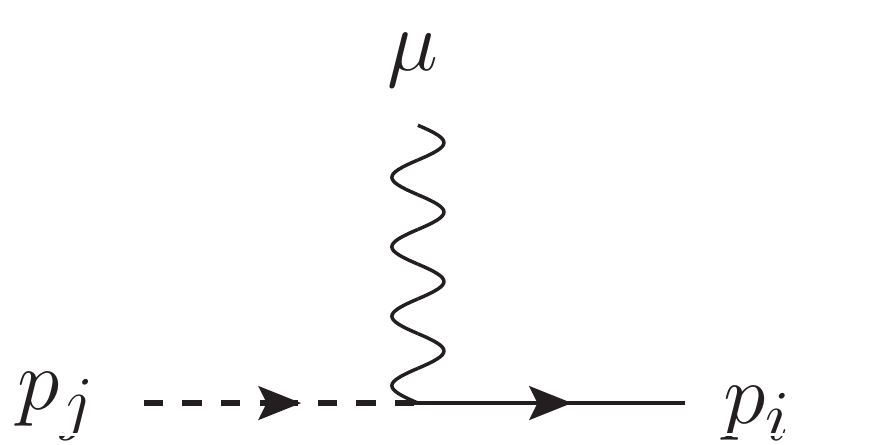}}\;,
\end{equation}
from which we conclude that, 
for gauge boson exchange between free fermions, 
the directions of the chirality-flow arrows at a fermion-photon vertex may be flipped
such that the resulting diagram is one with matching chirality-flow arrows.
Once flipped, the Fierz identity \eqref{eq:Fierz Rearrangement Tau} may be used such that,
for example, 
the fourth term in \eqref{eq:photon_exchange} can be expressed as
\begin{equation} \label{eq:Fierz dir changed}
  \raisebox{-0.42\height}{\includegraphics[scale=0.40]{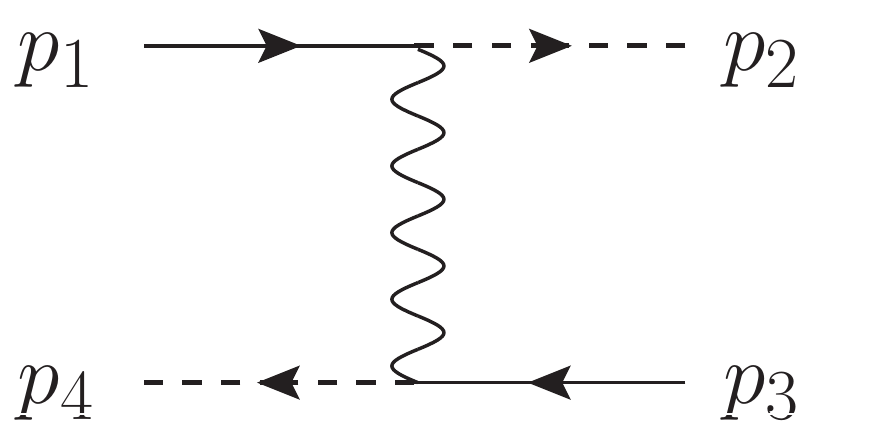}}\;
  =\raisebox{-0.42\height}{\includegraphics[scale=0.40]{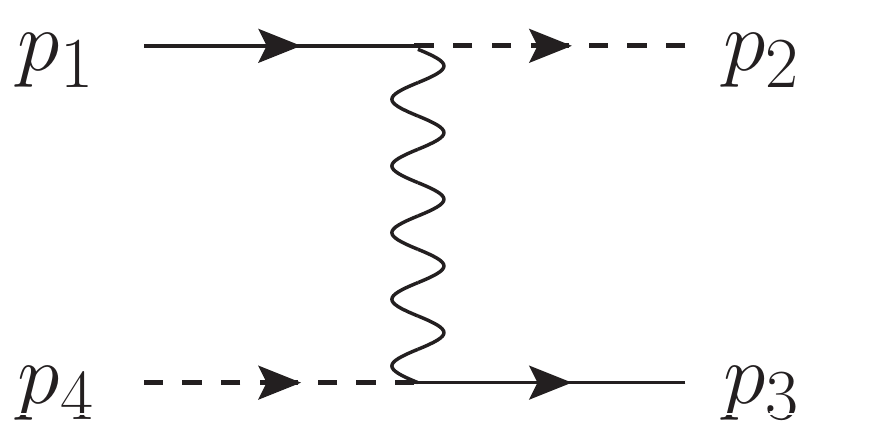}}
  = \raisebox{-0.42\height}{ \includegraphics[scale=0.40]{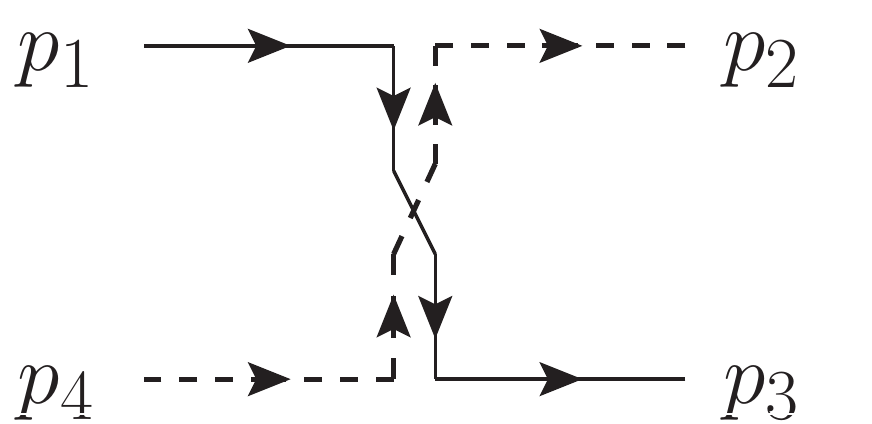}}\;,
\end{equation}
where we have performed the arrow swap at the lower vertex.
Note that we could have alternatively performed the arrow swap at the upper vertex. 
Graphically we see that also in this diagram,
the Lorentz structure of the photon propagator in the chirality-flow picture
may be represented by a double line with arrows opposing each other, as in \eqref{eq:g_munuPic}.

Therefore, in the example of single photon exchange between fermions, 
we see that for every allowed combination of external helicities we can turn the Feynman diagram into a chirality-flow diagram.

\subsection{Proof for QED}

We will now show that this can be done for \textit{all} QED tree-level Feynman diagrams with explicit external helicities. 
To do so, 
we must show that it is always possible to swap the chirality-flow arrows such that we can use the Fierz identity, 
\eqref{eq:Fierz Rearrangement Tau}, for each contraction of vector indices. 
While proving this, we will also show that we can always write $g^{\mu\nu}$ (or $g_{\mu\nu}$) as a double line with opposing arrows,
i.e. that \eqref{eq:g_munuPic} always holds.

We have just seen that we can always swap the chirality-flow arrows on a fermion line which emits a single photon, 
\eqref{eq:charge conj current easy}. 
Similarly,
it is possible to show that we can swap the chirality-flow arrows for a fermion line which emits an arbitrary number of photons,
i.e.\
\begin{align}
    \label{eq:charge conj current pic}
    \raisebox{-0.5\height}{ \includegraphics[scale=0.40]{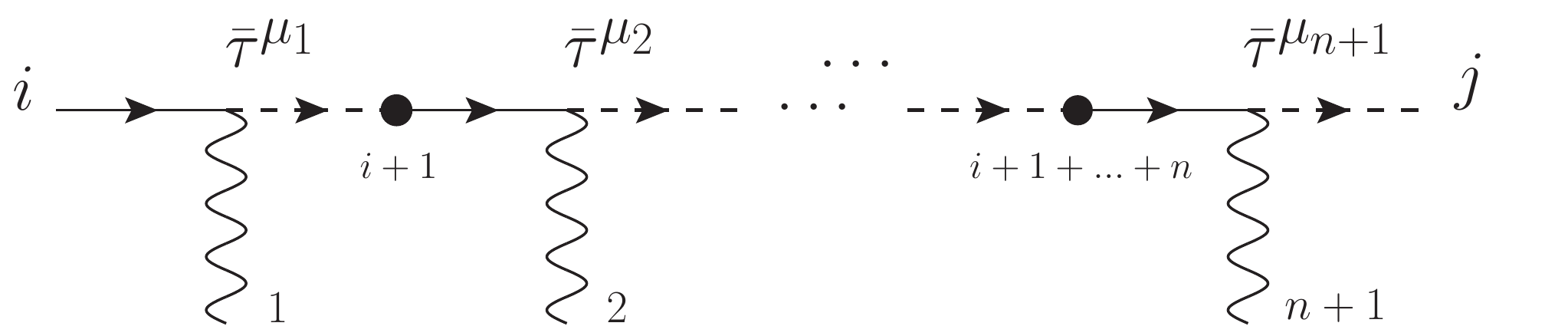}} 
    = 
    \raisebox{-0.5\height}{ \includegraphics[scale=0.40]{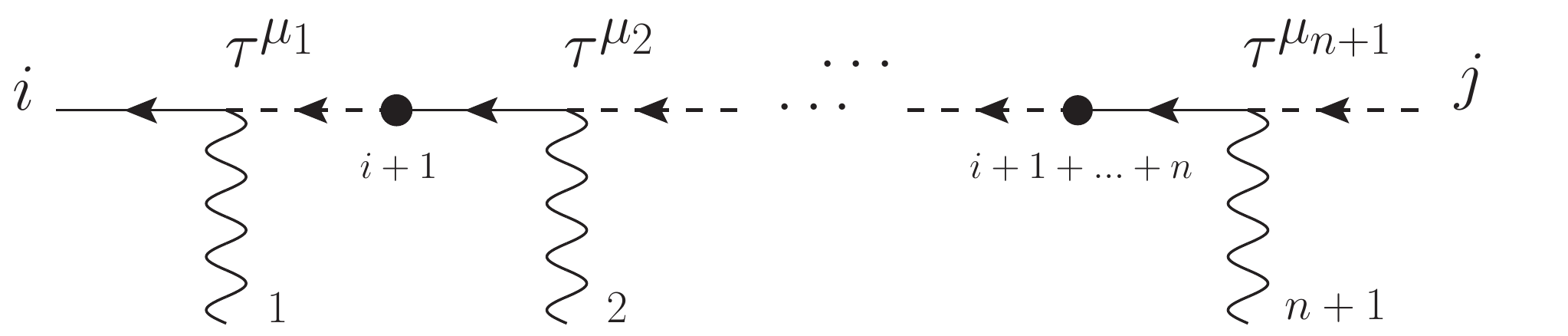}}\;,
\end{align} 
where we use the momentum-dot notation from \eqref{eq:p_contr2} for the fermion propagators. 
We can prove \eqref{eq:charge conj current pic} by using
\eqsrefd{eq:Mom two-spinors_pi}{eq:p_sum} to write it as
\begin{align} \label{eq:charge conj long bra-ket}
& \ \langle i|
\taubar^{\mu_1}
\Big(\;
|i]\langle i| 
\;+\; 
|1]\langle 1|
\;\Big)
\taubar^{\mu_2}
\,\dots\;
\taubar^{\mu_{n}}
\Big(\;
|i]\langle i| 
\;+\; 
|1]\langle 1| 
\;+\; 
... 
\;+\; 
|n]\langle n|
\;\Big) 
\taubar^{\mu_{n+1}}
|j] \nonumber \\
=& \ [i|
\tau^{\mu_1}
\Big(\;
|i\rangle [i| 
\;+\; 
|1\rangle [1|
\;\Big)
\tau^{\mu_2}
\,\dots\;
\tau^{\mu_{n}}
\Big(\;
|i\rangle [i| 
\;+\; 
|1\rangle [1| 
\;+\; 
... 
\;+\; 
|n\rangle [n|
\;\Big) 
\tau^{\mu_{n+1}}
|j\rangle
\;,
\end{align}
which is seen by applying
\eqref{eq:charge conj current matrix} 
to transform each $\langle k|\taubar^\mu|l] \leftrightarrow [l|\tau^\mu|k\rangle$.

The photons attached to the fermion line in \eqref{eq:charge conj current pic} may be either internal or external.
If they are external,
we use that the Lorentz structure of an external photon is the same as that of a fermion-antifermion pair connected to the fermion line by that photon, 
i.e.\ $\eps^{\mu} \sim \langle i | \taubar^{\mu} | j ]$ as noted at the end of \secref{sec:polarizationvectors}.
We call such a structure a pseudo-vertex, and for each
pseudo-vertex we can always adjust the chirality-flow arrows
such that the Fierz identity can be applied.

If a photon in \eqref{eq:charge conj current pic} is internal,
it must be attached to another fermion line. 
In QED, 
we can build any Feynman diagram by using internal photons to iteratively stitch together such fermion lines.
At tree level, each new fermion line attaches to only one already existing fermion line,
so we can freely adjust the chirality-flow arrows on the new line such that the Fierz identity holds at the attachment.
Remembering that the external photons are equivalent to fermion lines, 
we can therefore write any Feynman diagram in such a way that the chirality-flow arrows match,
and \eqref{eq:g_munuPic} holds.
We conclude that any QED tree-level Feynman diagram can be written as a chirality-flow diagram
multiplied by scalar factors from vertices, propagators, and external photons.

\subsection{Proof for QCD}

It remains to show that the flow picture can be applied to QCD tree-level Feynman diagrams as well.
We begin by remarking that the fermion-boson QCD vertex
and the external polarization vectors have the same Lorentz structure as in the QED case.
What remains therefore is to understand the Lorentz structures of the triple-gluon vertex, 
made up of terms 
$\sim p^{\mu}g^{\nu\rho}$,
and the four-gluon vertex, with terms of the form 
$\sim g^{\mu\nu}g^{\rho\sigma}$.

Let us first ignore factors of $p^{\mu}$ from the triple-gluon vertex.
In this case, 
the metric factors $g^{\mu\nu}$ from the non-abelian vertices
(possibly combined with metric factors from propagators)
will only act to contract indices from $\tau$ or $\taubar$ matrices in fermion lines,
as in QED.
Therefore, we can swap the chirality-flow arrow as required, 
and \eqref{eq:g_munuPic} still holds.

Next, we include the factors of $p^{\mu}$.
Using \eqsrefa{eq:Gordona}{eq:Gordonb},
\begin{equation} \label{eq:Gordon}
p^{\mu} = \frac{1}{\sqrt{2}} \langle p | \taubar^{\mu} | p] = \frac{1}{\sqrt{2}} [p | \tau^{\mu} |p\rangle~,
\end{equation}
we see that the momentum $p^{\mu}$ can be viewed as another type of pseudo-vertex,
and we can again proceed as in the QED case.
We thus conclude that \eqref{eq:g_munuPic} still holds,
and that any tree-level QCD Feynman diagram with explicit helicities can be cast as a
(sum of)  chirality-flow diagram(s)
multiplied by (Lorentz) scalar factors from vertices, propagators, and external gluons.

\subsection{QCD remarks} \label{sec:QCDremarks}

A novel feature of the QCD non-abelian vertices is the existence of disconnected Lorentz structures.
For example, in the four-gluon vertex 
consisting of terms of the form $g^{\mu\nu}g^{\rho\sigma}$,
the flow of Lorentz indices in $g^{\mu\nu}$
does not affect the flow of the Lorentz indices in $g^{\rho\sigma}$.
This leads to disconnected pieces of a chirality-flow diagram,
i.e.\ spinor lines not related to each other by either momentum dots 
or a shared double line.
We can flip the arrow direction of one disconnected piece 
without affecting the arrow direction of the other.

Additionally, 
a new chirality-flow rule for $p^{\mu}$ in the triple-gluon vertex must be found.
If $p^\mu$ is contracted with a fermion line or the polarization vector
of an external gluon, it will result in
$p^{\mu}\tau_{\mu} = \slashed{p}/\sqrt{2}$
(or $p^{\mu}\taubar_{\mu} =\bar{ \slashed{p}}/\sqrt{2}$),
allowing to use \eqref{eq:p_contr2} to identify 
\begin{equation} \label{eq:tripGlueP}
p^{\mu} \rightarrow \frac{1}{\sqrt{2}}\raisebox{-0.15\height}{\includegraphics[scale=0.55]{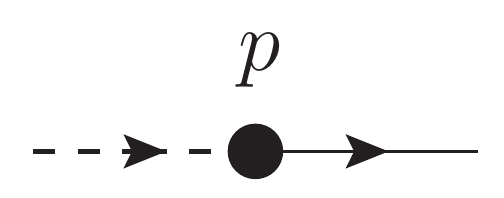}}
\quad \text{or} \quad
p^{\mu} \rightarrow \frac{1}{\sqrt{2}}\raisebox{-0.15\height}{\includegraphics[scale=0.55]{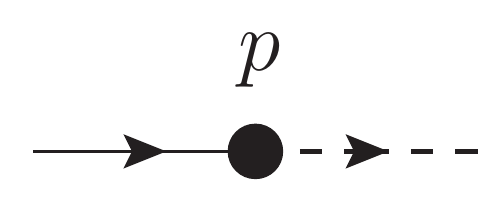}},
\end{equation}
in these cases. 

The remaining situation to consider is when the momentum $p_i^{\mu}$ is instead contracted with another momentum $p_j^{\nu}$.
In this case, we obtain
\begin{align}
  p_i\cdot p_j
  = 
  p_i^{\mu}p_j^{\nu}\trace\big(\tau_{\mu}\taubar_{\nu}\big)
  =
  \frac{1}{2}\trace\big(\slashed{p}_i\bar{\slashed{p}}_j\big)
  =
  \frac{1}{2}\raisebox{-0.4\height}{\includegraphics[scale=0.5]{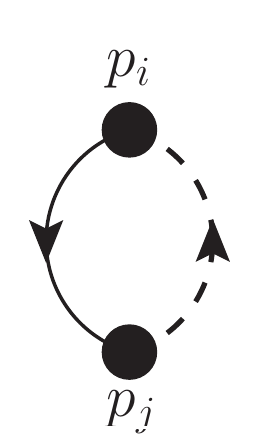}}\;\;,
  \label{eq:pi.pj pic main}
\end{align}
for which we can choose the chirality-flow arrow independently from the rest of the diagram.
Such a term is therefore another example of a disconnected piece of a chirality-flow diagram.

We thus conclude that the chirality-flow rule for $p^{\mu}$ in the triple-gluon vertex
can be taken to be \eqref{eq:tripGlueP}.
Note that due to \eqref{eq:Gordon}, 
we may always use the version of 
\eqref{eq:tripGlueP} 
required to obtain chirality-flow arrows which match the rest of the diagram.

\section{Chirality-flow Feynman rules}
\label{sec:FeynmanRules}

In this section we will collect the chirality-flow Feynman rules for massless QED and QCD,
using the result of the previous section
that any tree-level Feynman diagram can be cast as a (sum of) chirality-flow diagram(s).
The corresponding rules for external spinors and polarization vectors have been collected in \secsref{sec:spinors} and \ref{sec:polarizationvectors} already.  
For convenience we also collect the full set of Feynman rules, 
in various representations, 
in the `Rosetta stones' in \tabsref{tab:Flow rules comparison QED} and \ref{tab:Flow rules comparison QCD} in \appref{sec:Rosetta stone}.

\subsection{Vertices}
\label{sec:verticesqedqcd}

We start with the fermion-photon vertex. 
Suppressing spinor indices, 
it is given by 
\begin{eqnarray}
\raisebox{-0.4\height}{\includegraphics[scale=0.5]{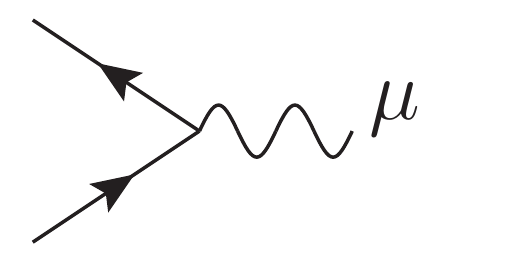}}
\!\!\!\!\!= ieQ_f\gamma^\mu\;,
\end{eqnarray}
with $e$ and $Q_f$ being the electromagnetic coupling constant and the charge respectively.
In the Weyl representation 
we can separate the vertex into two parts,
\begin{eqnarray}
  \label{eq:fermion_photon_vertex0}
  ieQ_f\sqrt{2}\tau^{\mu,\da\be}
  = \raisebox{-0.4\height}{\includegraphics[scale=0.5]{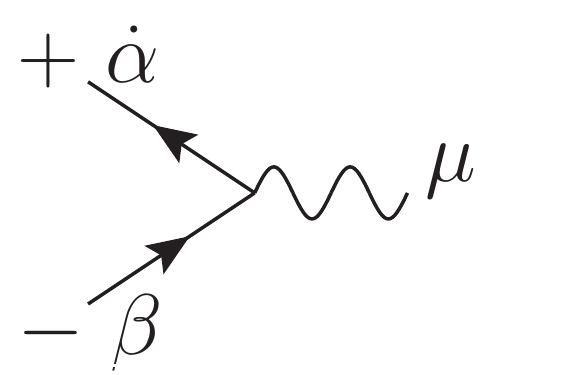}}
  &\rightarrow& \;\;\; 
  ieQ_f\sqrt{2}\raisebox{-0.35\height}{\includegraphics[scale=0.425]{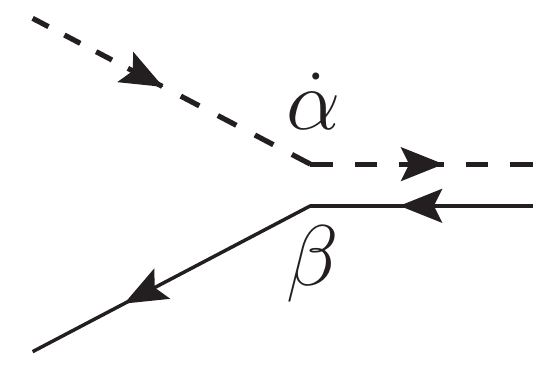}}\;,
  \\
  ieQ_f\sqrt{2}\taubar^{\mu}_{\al\db}
  = \raisebox{-0.4\height}{\includegraphics[scale=0.5]{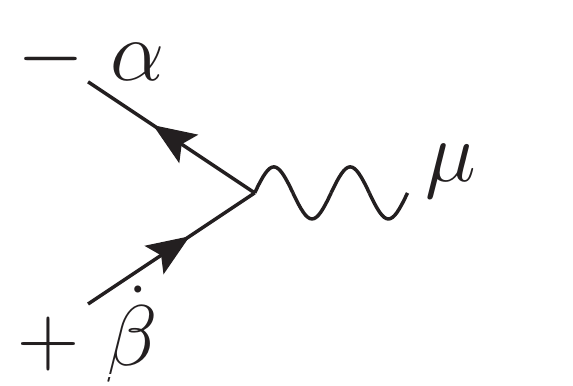}}
  &\rightarrow& \;\;\; 
  ieQ_f\sqrt{2}\raisebox{-0.4\height}{\includegraphics[scale=0.425]{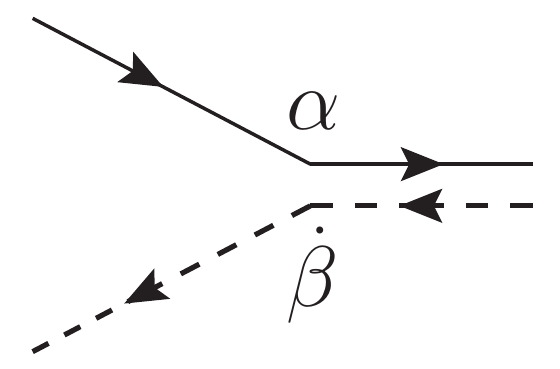}}\;,
  \label{eq:fermion_photon_vertex}
\end{eqnarray}
depending on the helicity configurations of the fermions. 
Here we make the Weyl spinor indices explicit,
and the right graphical rules represent the two parts in the chirality-flow picture. 
If the fermion-antifermion pair is a quark-antiquark pair we have to multiply by an additional $\delta_{i\jbar}$,
with $i$ and $\jbar$ being the color indices of the (outgoing) quark and antiquark respectively.

The quark-gluon vertex is similarly given by\footnote{
  Note that $i$ and $\jbar$ are the color indices of {\it outgoing} quarks and antiquarks respectively,
  meaning that indices on $t^a_{i\jbar}$, as well as color-flow arrows are
  read {\it against} the fermion flow. Correspondingly, the reading direction
  of $f^{abc}$ is clockwise. Opposite reading directions, for both quark color (along
  fermion arrow) and structure constants (counter clockwise) may be imposed
  without introducing additional signs.
  In the color-flow picture, 
  and with our conventions in \eqsrefa{eq:color_fierz0}{eq:color_norm},
  $t^a_{i\jbar}$ is translated to $\delta_{i\ibar_a}\delta_{i_a\jbar}$.
}
\begin{eqnarray}
\raisebox{-0.4\height}{\includegraphics[scale=0.5]{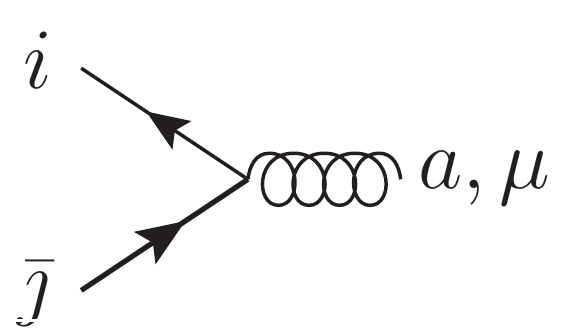}}
= i\frac{g_s}{\sqrt{2}}t^a_{i\jbar}\gamma^\mu
\;,
\end{eqnarray}
with $g_s$ denoting the strong coupling constant. 
Note that, 
comparing to the (perhaps most common) 
convention where $\trace(t^at^b)=\frac{1}{2}\delta^{ab}$,
our generators and structure constants are a factor $\sqrt{2}$ larger,
which affects the normalization of our QCD vertices.

In the Weyl representation we can again separate the vertex into two parts,
\begin{eqnarray}
  \label{eq:fermion_gluon_vertex0}
  ig_st^a_{i\jbar}\tau^{\mu,\da\be}
  = \raisebox{-0.4\height}{\includegraphics[scale=0.5]{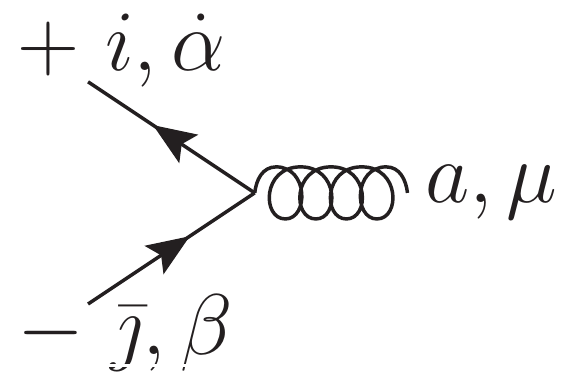}}
  \;\;\;&\rightarrow& \;\;\; 
  ig_s t^a_{i\jbar}\raisebox{-0.35\height}{\includegraphics[scale=0.425]{./Jaxodraw/TauVertexHighb}}\;,
  \\
  ig_st^a_{i\jbar}\taubar^{\mu}_{\al\db}
  = \raisebox{-0.4\height}{\includegraphics[scale=0.5]{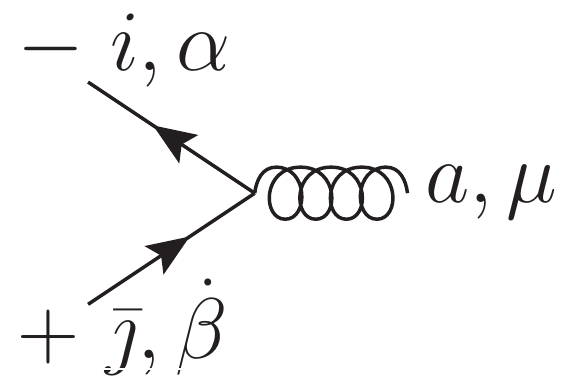}}
  \;\;\;&\rightarrow& \;\;\;
  ig_s t^a_{i\jbar}\raisebox{-0.4\height}{\includegraphics[scale=0.425]{./Jaxodraw/TauVertexLowa}}\;,
  \label{eq:fermion_gluon_vertex}
\end{eqnarray}
depending on the helicity configurations of the quarks.

We also need the three- and four-gluon vertices.
For the three-gluon vertex, 
the color and kinematic parts factorize trivially,\,
\begin{eqnarray}
  \raisebox{-1.0cm}{\includegraphics[scale=0.4]{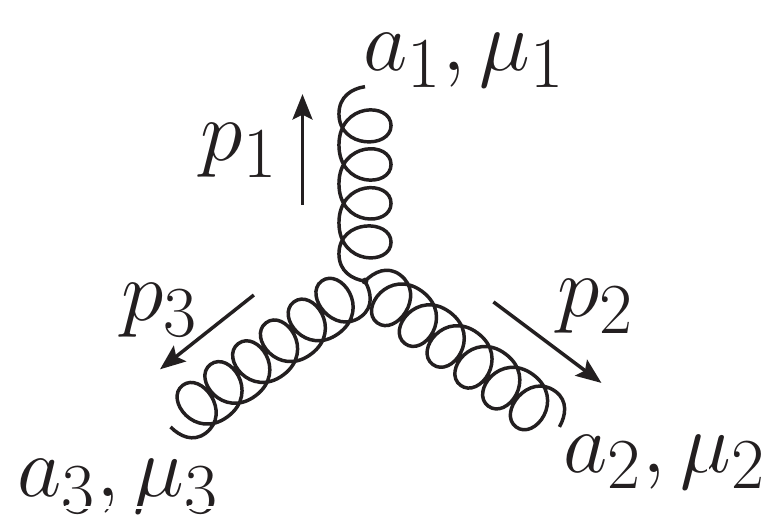}}
  \!\!
  &=&
  \;\;
  i\frac{g_s}{\sqrt{2}} 
  \!\!\!\!\!
  \underbrace{if^{a_1a_2a_3}}_{\includegraphics[scale=0.4]{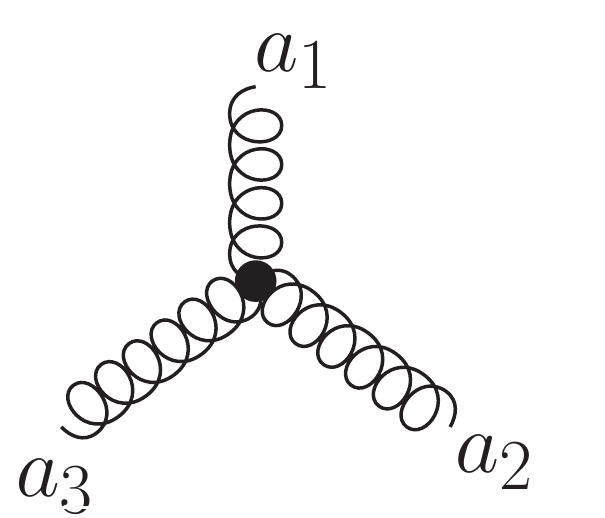}}
  \!\!\!\!\!  
  \underbrace{
  \Big(
    g^{\mu_1\mu_2}(p_1-p_2)^{\mu_3}
    +g^{\mu_2\mu_3}(p_2-p_3)^{\mu_1}
    +g^{\mu_3\mu_1}(p_3-p_1)^{\mu_2}
  \Big)
  }_{\equiv V_3^{\mu_1\mu_2\mu_3}(p_1,p_2,p_3)}\;,\quad
\label{eq:trpgluvrt}
\end{eqnarray}
where we take the graph with only color indices and a dot in the vertex to equal $if^{a_1 a_2 a_3}$. 
The kinematic part can also be written in a more condensed form as $V_3^{\mu_1\mu_2\mu_3}(p_1,p_2,p_3)=\sum_{Z(1,2,3)}g^{\mu_1\mu_2}(p_1-p_2)^{\mu_3}$,
where $Z(1,2,3)$ denotes the set of cyclic permutations of the integers $1,2,3$.
Using \eqsrefa{eq:g_munuPic}{eq:tripGlueP}
to translate $g^{\mu_1\mu_2}$ and $(p_1-p_2)^{\mu_3}$ etc. 
to the chirality-flow picture, 
we get\,
\begin{eqnarray}
  \raisebox{-1.0cm}{\includegraphics[scale=0.4]{Jaxodraw/CR/3GluonVertex_Full}} \!\!\!\!\!\! 
  \rightarrow 
  \;i\frac{g_s}{\sqrt{2}} 
  \raisebox{-1.0cm}{\includegraphics[scale=0.4]{Jaxodraw/MS/3GluonVertex_Color}}
  \!\!\!\frac{1}{\sqrt{2}}
  \left(
  \raisebox{-0.875cm}{\includegraphics[scale=0.375]{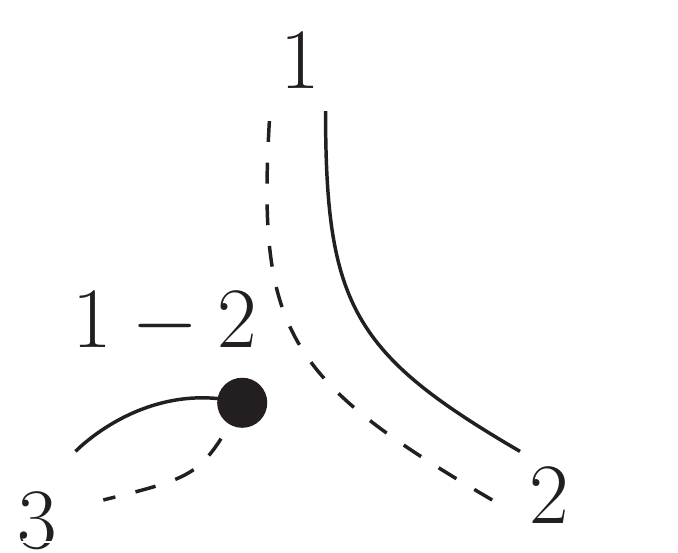}} \!\!\!
  + \raisebox{-0.825cm}{\includegraphics[scale=0.375]{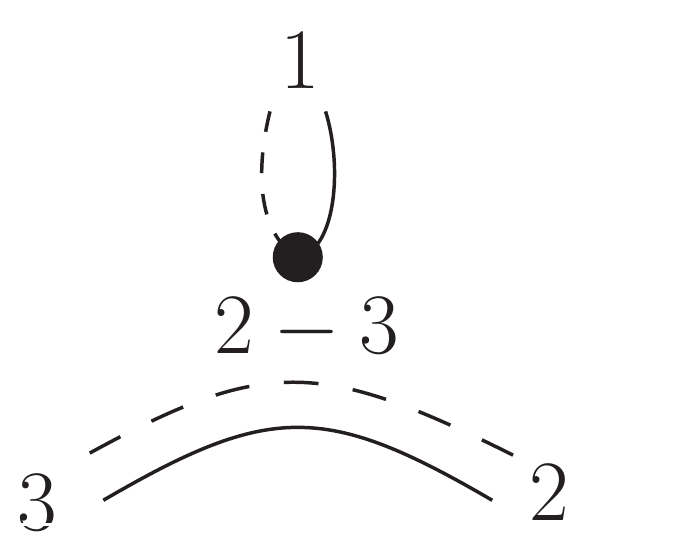}} \!\!\!
  + \raisebox{-0.875cm}{\includegraphics[scale=0.375]{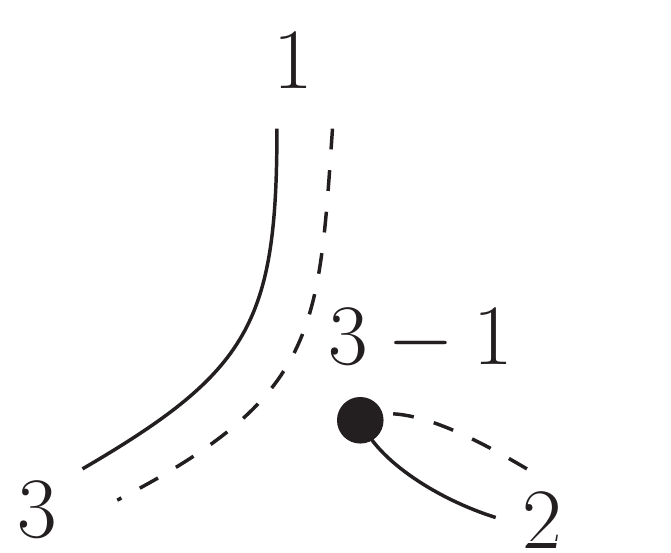}}\!\!\!
  \right),
\label{eq:trpgluvrt2}
\end{eqnarray}
where the chirality-flow arrows have been removed since they must flow in different directions for different terms in the sum.
However, the arrow directions of the dotted and undotted chirality-flow lines 
must always oppose each other in the double lines (from the metric), 
and form a continuous flow in the lines joined by a momentum-dot (from the momentum).
Note that 
since we only consider tree-level diagrams with massless particles, 
the momentum parts can always be written as linear combinations of external momenta $p_i$, 
with $p_i^2=0$, 
such that we can use \eqsrefd{eq:Mom two-spinors_pi}{eq:p_sum} to reduce the corresponding momentum bispinors to outer products, or dyads, of momentum spinors.
Using \eqsrefa{eq:f_expr}{eq:color_fierz0},
and realizing that $V_3^{\mu_1\mu_2\mu_3}(p_1,p_2,p_3)=-V_3^{\mu_1\mu_3\mu_2}(p_1,p_3,p_2)$, 
we can further translate the color part of the three-gluon vertex to the color-flow picture,
such that
\begin{eqnarray}
  \!\!\raisebox{-1.0cm}{\includegraphics[scale=0.4]{Jaxodraw/CR/3GluonVertex_Full}}
  \!\!\!\!\!\! \rightarrow \;
  i\frac{g_s}{\sqrt{2}}
  \sum\limits_{S(2,3)}
  \!\!\!\!\!
  \underbrace{\delta_{i_1\ibar_2} \delta_{i_2\ibar_3} \delta_{i_3\ibar_1}}_{\;\;\,\includegraphics[scale=0.35]{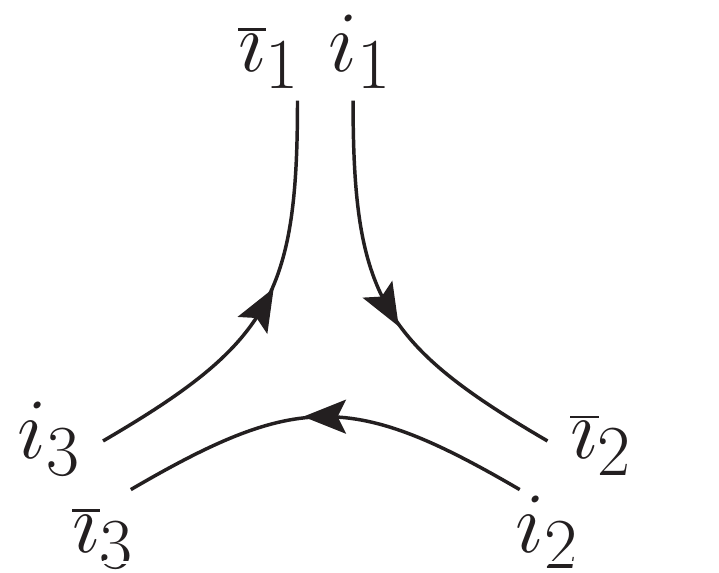}}
  \!\!\!\!
  \frac{1}{\sqrt{2}}
  \left(
  \raisebox{-0.875cm}{\includegraphics[scale=0.375]{Jaxodraw/MS/3GluonVertex_HelFlow1}} \!\!\!\!\!
  + \!\! \raisebox{-0.825cm}{\includegraphics[scale=0.375]{Jaxodraw/MS/3GluonVertex_HelFlow2}} \!\!\!\!\!
  + \!\! \raisebox{-0.875cm}{\includegraphics[scale=0.375]{Jaxodraw/MS/3GluonVertex_HelFlow3}}\!\!\!\!\!
  \right),
\label{eq:trpgluvrt3}
\end{eqnarray}
where $S(2,3)$ denotes the set of permutations of the integers $2,3$, and
where the lines in the lower graph represent the color flows\footnote{
  In the trace basis, 
  and with our conventions in \eqsref{eq:color_fierz0} and  (\refd{eq:color_norm}{eq:f_expr}),
  this vertex is\\ $ig_s/\sqrt{2}\sum_{S(2,3)}\trace\big(t^{a_1}t^{a_2}t^{a_3}\big)V_3^{\mu_1\mu_2\mu_3}(p_1,p_2,p_3)$.
}.

For the four-gluon vertex, 
we translate the metric to the chirality-flow picture in the same way as before,\,
\begin{eqnarray}
  \label{eq:four_gluon_vertex}
  \raisebox{-0.425\height}{\includegraphics[scale=0.325]{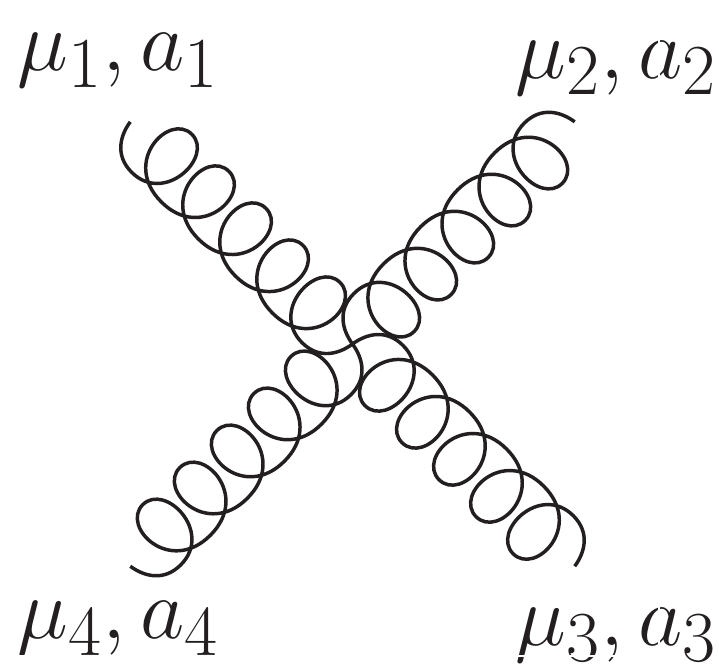}}
  =
  \;\;
  i\left(\!\frac{g_s}{\sqrt{2}}\right)^{\!\!2}\!\!\! 
  &&\bigg(\;\;
    \underbrace{i f^{a_1a_2b}i f^{ba_3a_4}}_{\includegraphics[scale=0.275]{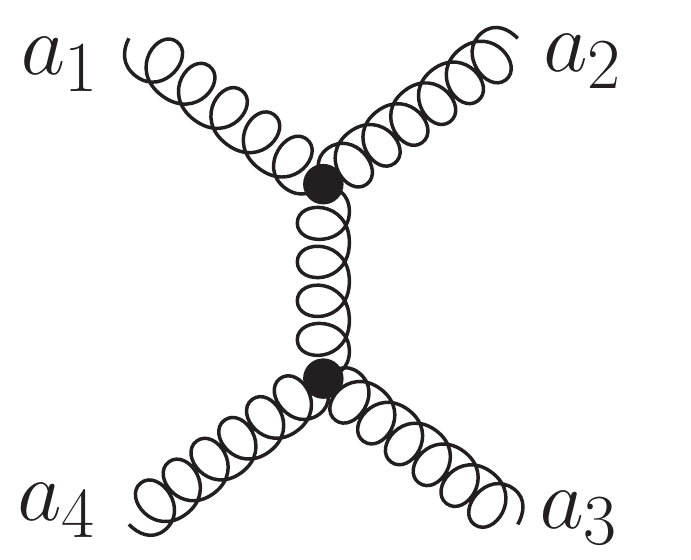}}\quad
    \underbrace{\big(g^{\mu_1\mu_3}g^{\mu_2\mu_4}-g^{\mu_1\mu_4}g^{\mu_2\mu_3}\big)}_{
      \includegraphics[scale=0.275]{Jaxodraw/MS/4GluonVertex_Kin13}
      \raisebox{+3.5\height}{$-$}\quad
      \includegraphics[scale=0.275]{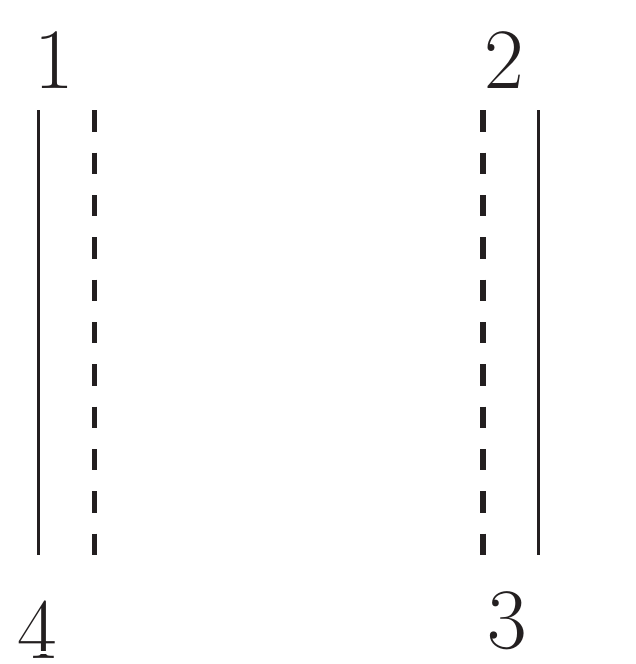}
    } \nonumber\\
    &&+\;\; 
    \underbrace{i f^{a_1a_3b}i f^{ba_2a_4}}_{\includegraphics[scale=0.275]{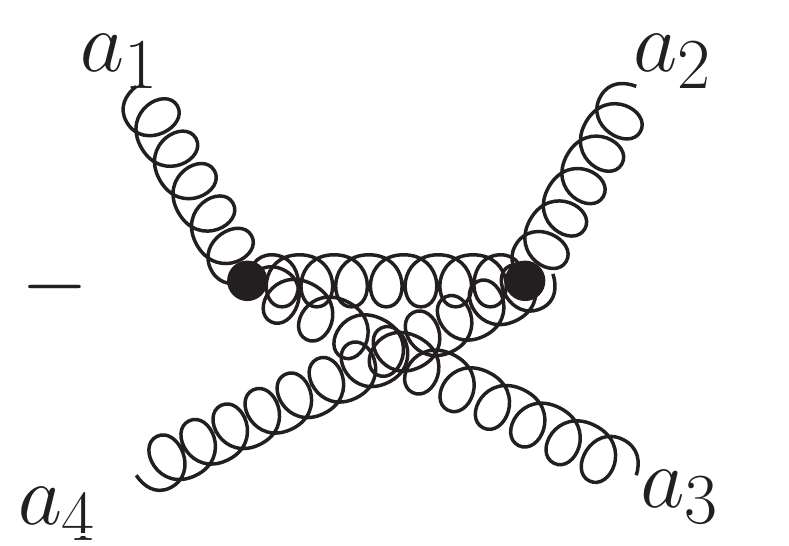}}\quad
    \underbrace{\big(g^{\mu_1\mu_2}g^{\mu_3\mu_4}-g^{\mu_1\mu_4}g^{\mu_3\mu_2}\big)}_{
       \includegraphics[scale=0.275]{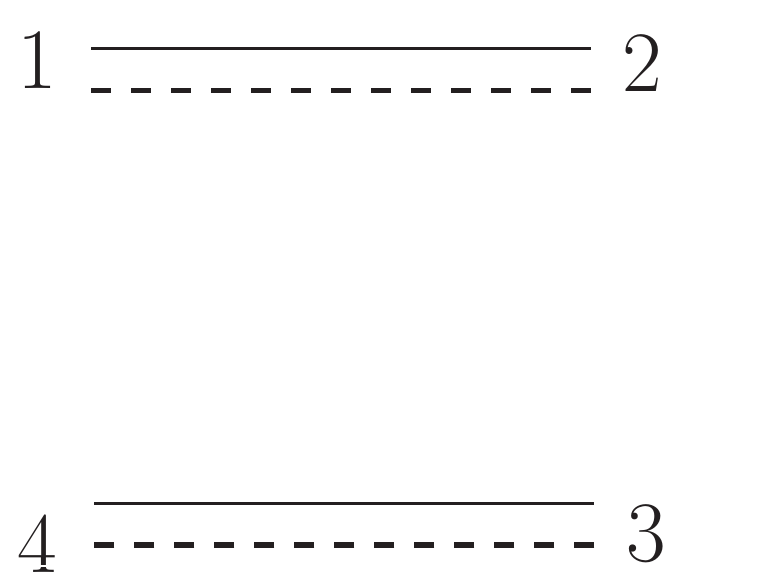}
       \raisebox{+3.5\height}{$-$}\quad
       \includegraphics[scale=0.275]{Jaxodraw/MS/4GluonVertex_Kin14}
    }
  \nonumber\\
   &&+\;\;
  \underbrace{i f^{a_1a_4b}i f^{ba_3a_2}}_{\includegraphics[scale=0.275]{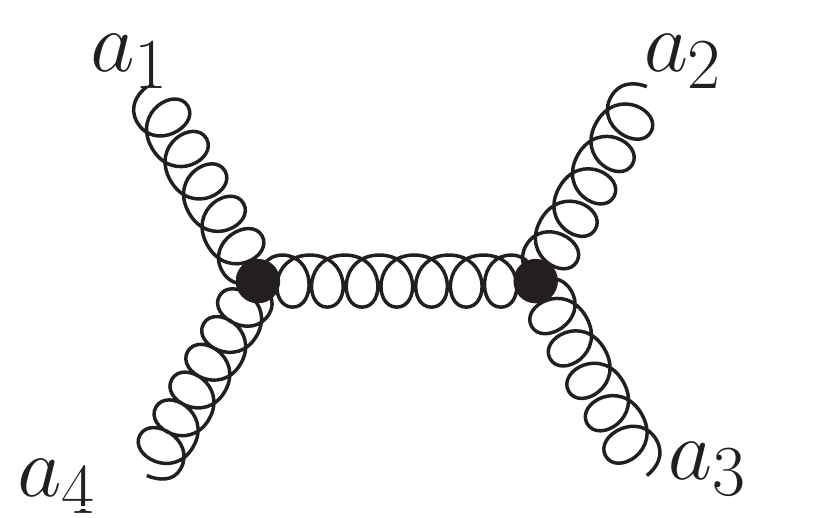}}\quad
     \underbrace{\big(g^{\mu_1\mu_3}g^{\mu_4\mu_2}-g^{\mu_1\mu_2}g^{\mu_4\mu_3}\big)}_{
      \includegraphics[scale=0.275]{Jaxodraw/MS/4GluonVertex_Kin13}
      \raisebox{+3.5\height}{$-$}\quad  
      \includegraphics[scale=0.275]{Jaxodraw/MS/4GluonVertex_Kin12}
    }
  \bigg)\;,
\end{eqnarray}
where we also give the conventional graphical representation for the color factors,
as well as the graphical chirality-flow representation for the kinematic parts. 
Again, the arrow directions in the chirality-flow lines must oppose each other
and be adjusted to match the rest of the diagram.

The four-gluon vertex can also be written in a more condensed form, 
\begin{eqnarray}
  \raisebox{-0.425\height}{\includegraphics[scale=0.325]{Jaxodraw/CR/4gVertex}}
  =\;\;
  \;\;
  i\left(\!\frac{g_s}{\sqrt{2}}\right)^{\!\!2}\!\!
  \;\sum\limits_{Z(2,3,4)}
  \;if^{a_1a_2b}if^{ba_3a_4}
  \;\;\;\Big(
  g^{\mu_1\mu_3}g^{\mu_2\mu_4}
  -g^{\mu_1\mu_4}g^{\mu_2\mu_3}
  \Big)\;,
  \label{eq:four_gluon_vertex0}
\end{eqnarray}
or if we sort by the same metric factors,
\begin{eqnarray}
  \raisebox{-0.425\height}{\includegraphics[scale=0.325]{Jaxodraw/CR/4gVertex}}
  =
  \;\;
  i\left(\!\frac{g_s}{\sqrt{2}}\right)^{\!\!2}\!\!
  \sum\limits_{Z(2,3,4)}
  g^{\mu_1\mu_3}g^{\mu_4\mu_2}
  \;\;\Big(if^{a_1a_2b}if^{ba_3a_4}+if^{a_1a_4b}if^{ba_3a_2}\Big)\;.
  \label{eq:four_gluon_vertex2}
\end{eqnarray}
Translating to the chirality-flow picture, 
using the conventional graphical representation for the color structure,
we note that this corresponds to sorting by chirality flows,
such that
\begin{eqnarray}
  \raisebox{-0.425\height}{\includegraphics[scale=0.325]{Jaxodraw/CR/4gVertex}}
  \rightarrow
  \;\;
  i\left(\!\frac{g_s}{\sqrt{2}}\right)^{\!\!2}\!\!
  \sum\limits_{Z(2,3,4)}
  \raisebox{-0.425\height}{\includegraphics[scale=0.325]{Jaxodraw/MS/4GluonVertex_Kin13}}
  \left(\;
  \raisebox{-0.425\height}{\includegraphics[scale=0.325]{Jaxodraw/MS/4GluonVertex_Color12}}
  +
  \raisebox{-0.425\height}{\includegraphics[scale=0.325]{Jaxodraw/MS/4GluonVertex_Color14}}
  \right)\;.
  \label{eq:four_gluon_vertex3}
\end{eqnarray}
Note that the gluons which are chirality-flow connected are \textit{not}
connected in the structure constants.
Finally, we can use \eqsrefa{eq:color_fierz0}{eq:f_expr}
to translate the color parts of the four-gluon vertex to the color-flow picture,
such that
\begin{eqnarray}
  \hspace{-4.5ex}
  \raisebox{-0.425\height}{\includegraphics[scale=0.325]{Jaxodraw/CR/4gVertex}}
  \!\!\!\!\!  
  \rightarrow
  \;
  i\left(\!\frac{g_s}{\sqrt{2}}\right)^{\!\!2}\!\!
  \!\!\sum\limits_{S(2,3,4)}\!\!
  \underbrace{\delta_{i_1\ibar_2} \delta_{i_2\ibar_3} \delta_{i_3\ibar_4}\delta_{i_4\ibar_1}}_{\;\;\,\includegraphics[scale=0.325]{Jaxodraw/CR/4GluonVertex_ColorFlow}}
  \underbrace{\left(
  \;
  2\!\!\raisebox{-0.4\height}{ \includegraphics[scale=0.325]{./Jaxodraw/MS/4GluonVertex_Kin13}} 
  \!\!\!\!-\raisebox{-0.4\height}{ \includegraphics[scale=0.325]{./Jaxodraw/MS/4GluonVertex_Kin12}} 
  \!\!\!\!-\raisebox{-0.4\height}{ \includegraphics[scale=0.325]{./Jaxodraw/MS/4GluonVertex_Kin14}}
  \right)}_{\equiv V_4^{\mu_1...\mu_4}},
  \label{eq:four_gluon_vertex4}
\end{eqnarray}
where the lines in the lower graph represent the color flows\footnote{
  In the trace basis, 
  and with our conventions in \eqsref{eq:color_fierz0} and (\refd{eq:color_norm}{eq:f_expr}),
  this vertex is \\ $ig_s^2/2\sum_{S(2,3,4)}\trace\big(t^{a_1}t^{a_2}t^{a_3}t^{a_4}\big)V_4^{\mu_1\mu_2\mu_3\mu_4}$.
}.
The lines in the graphs in parentheses represent chirality flows,
and we note that they
correspond to the Lorentz structure of the kinematic part 
$V_4^{\mu_1...\mu_4}=\big(2g^{\mu_1\mu_3}g^{\mu_4\mu_2}-g^{\mu_1\mu_2}g^{\mu_3\mu_4}-g^{\mu_1\mu_4}g^{\mu_2\mu_3}\big)$
in the color-flow decomposition.

\subsection{Propagators}
\label{sec:propagators}

We also need the propagators for fermions and vector bosons.
Starting with the propagator for a massless fermion with momentum $p$ in QED
\begin{eqnarray} \label{eq:fermion_propagator_full}
  \raisebox{-0.1\height}{\includegraphics[scale=0.45]{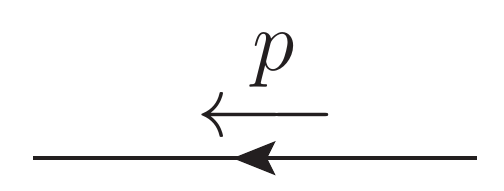}}
  =\frac{ip_{\mu}\gamma^{\mu}}{p^2}
  =\frac{i}{p^2}
  \sqrt{2}
  \begin{pmatrix}
    0 & p_\mu\tau^{\mu}  \\
    p_\mu\taubar^{\mu} & 0
  \end{pmatrix}
  =\frac{i}{p^2}
  \begin{pmatrix}
    0 \;&\; \slashed{p}  \\
    \bar{\slashed{p}} \;&\; 0
  \end{pmatrix}\;,
\end{eqnarray}
we see that
--- like the fermion-photon vertex --- 
the fermion propagator separates into two parts,
\begin{eqnarray}
  \label{eq:fermion_propagator}
  \frac{i}{p^2}\slashed{p}
  &\;\;\rightarrow\;\;&
  \frac{i}{p^2}\sqrt{2}p^{\dal\be}
  =
  \frac{i}{p^2}
  \raisebox{-0.10\height}{\includegraphics[scale=0.55]{./Jaxodraw/FermionPropFlowg}}\;,
  \\
  \frac{i}{p^2}\bar{\slashed{p}}
  &\;\;\rightarrow\;\;&
  \frac{i}{p^2}\sqrt{2}\bar{p}_{\al\dbe}
  =
  \frac{i}{p^2}
  \raisebox{-0.10\height}{\includegraphics[scale=0.55]{./Jaxodraw/FermionPropFlowh}}\;.
  \label{eq:fermion_propagator2}
\end{eqnarray}
Note that since we only consider tree-level diagrams with massless particles, 
the momentum can always be written as a linear combination of external momenta $p_i$, 
with $p_i^2=0$, 
such that we can use \eqsrefd{eq:Mom two-spinors_pi}{eq:p_sum} 
to reduce the corresponding momentum bispinors to outer products, or dyads, of momentum spinors.
In the case that the fermion is a quark we have to multiply by an additional $\delta_{i\jbar}$,
with $i$ and $\jbar$ being the color indices of the two ends of the quark propagator.

The propagator for a photon with momentum $p$ is given by
\begin{equation}
  \raisebox{-0.25\height}{\includegraphics[scale=0.45]{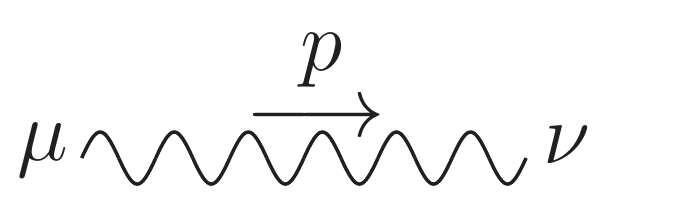}}
  \!\!=\;\; 
  -i\frac{g_{\mu\nu}}{p^2}\;.
\end{equation}
We recall from the previous section that
--- as for the three- and four-gluon vertices --- 
considering a full diagram, 
we can translate $g^{\mu\nu}$ to the chirality-flow picture using \eqref{eq:g_munuPic},
such that
\begin{equation}
  -i\frac{g_{\mu\nu}}{p^2}
  \;\;=\; 
  \raisebox{-0.25\height}{\includegraphics[scale=0.45]{./Jaxodraw/PhotonProp}}
  \rightarrow\;\;\;
  -\frac{i}{p^2}\raisebox{-0.25\height}{\includegraphics[scale=0.45]{./Jaxodraw/PhotonPropFlowa}}
  \quad \mbox{or}\quad
  -\frac{i}{p^2}\raisebox{-0.25\height}{\includegraphics[scale=0.45]{./Jaxodraw/PhotonPropFlowb}}\;,
\label{eq:vbpropagator}
\end{equation}
for which the arrow directions of the dotted and undotted lines in the double line must oppose each other,
but should be adjusted to match the rest of the diagram.
In the case of a gluon propagator we simply multiply by an additional $\delta^{ab}$, 
with $a$ and $b$ being the color indices of the two ends of the gluon propagator\footnote{
  In the color-flow picture, 
  and with our conventions in \eqsrefa{eq:color_fierz0}{eq:color_norm},
  $\delta^{ab}$ is translated to $\delta_{i_a\ibar_b}\delta_{i_b\ibar_a}-1/N\delta_{i_a\ibar_a}\delta_{i_b\ibar_b}$.
}.

To conclude this section, 
we note that we have formulated a set of chirality-flow rules, 
analogous to color-flow rules.
In fact, in a sense, 
the flow picture works even better here than in the case of color, 
since for color we have to bear in mind the $1/N$ suppressed term in \eqref{eq:color_fierz0} for the gluon propagator, 
as well as for external gluons upon squaring 
--- as long as we are not considering purely gluonic processes.
For the Lorentz structure, this complication does not arise.

\subsection{Application}
\label{sec:recipe}

In the following
we give a ``recipe'' for using the chirality-flow Feynman rules.
The Feynman rules are conveniently collected in \tabsref{tab:Flow rules comparison QED} and \ref{tab:Flow rules comparison QCD},
and concrete examples are given in \secref{sec:Examples}.

Consider a Feynman diagram with a given combination of external (outgoing) helicities:
\begin{enumerate}
  \item \label{step:prefactors} Collect all factors of $\pm i$, $\sqrt{2}$ and coupling constants from
    vertices (see \secref{sec:verticesqedqcd}), 
    as well as denominators from propagators 
    (\secref{sec:propagators}) 
    and external polarization vectors 
    (\secref{sec:polarizationvectors}).
  \item \label{step:add lines} Assign chirality-flow lines,
    i.e. dotted and undotted lines.
    Ignore chirality-flow directions in this step.
    When assigning internal momentum labels to momentum-dots, 
    write the corresponding momentum in terms of external momenta,
    directed as usually in Feynman diagrams.
  \begin{itemize}
    \item 
      External fermions with positive or negative helicity are assigned a
      single dotted or undotted line, respectively, and a momentum label 
      (see \eqsrefd{eq:lambdaupper}{eq:lambdatildelower}
      ).
    \item
      External vector bosons are assigned a double line
      and two momentum labels; 
      the line corresponding to the physical helicity 
      (the dotted line for positive helicity, the undotted line for negative helicity) 
      is assigned the physical momentum whereas 
      the other line is assigned the reference momentum
      (see \eqsrefd{eq:epsilon0}{eq:epsilon1} and (\refd{eq:epsilon2}{eq:epsilon3})
      ). 
    \item
      Vector boson propagators are assigned a double line, 
      one dotted and one undotted 
      (see \eqref{eq:vbpropagator}). 
    \item 
      Fermion propagators are assigned a pair of successive lines, 
      turning from dotted to undotted (or vice versa), 
      joined by a momentum-dot, with the corresponding momentum label
      (see \eqsrefd{eq:fermion_propagator}{eq:fermion_propagator2}
      ).
    \item Using the appropriate vertices, all lines are connected in the
      only possible way for the Feynman diagram to form a chirality-flow diagram.
      For photon- and gluon-fermion vertices, use either one of the chirality-flow structures in \eqref{eq:fermion_photon_vertex0} or (\ref{eq:fermion_photon_vertex}) 
      (equivalently \eqref{eq:fermion_gluon_vertex0} or (\ref{eq:fermion_gluon_vertex})),
      for three-gluon vertices use the sum of chirality-flow structures in \eqref{eq:trpgluvrt2}
      (or \eqref{eq:trpgluvrt3} for the color-ordered approach) and 
      for four-gluon vertices use the chirality-flow structures in
      \eqref{eq:four_gluon_vertex3} (or \eqref{eq:four_gluon_vertex4} for
      the color-ordered approach).

  \end{itemize}
  \item \label{step:arrows} Assign chirality-flow directions.
    Start with any external chirality-flow line and assign to it a chirality-flow arrow in an arbitrary direction\footnote{
      In a sum of Feynman diagrams there is no rule where and
      how to set the initial chirality-flow arrow.
      As shown in the previous section,  
      chirality-flow arrow swaps are identity operations,
      and performing them for each Feynman diagram independently
      can therefore not introduce any relative minus signs.
    }.
    Follow the line through the chirality-flow diagram, continuing
    through any potential momentum-dot,
    and assign chirality-flow arrows in the same direction.
    Assign the other arrow directions such that double lines
    from gauge bosons have opposing arrows.
    Note that non-abelian vertices will give rise to
    disconnected pieces (see \secref{sec:QCDremarks}).
    For each such disconnected piece, independently apply
    the above arrow direction rules.
\end{enumerate}
Due to the sums of chirality flows from the non-abelian vertices, 
each Feynman diagram is now turned into a sum of chirality-flow diagrams.
Multiplied by the collected factors in step \ref{step:prefactors},
as well as by potential color factors\footnote{
  Note that in the case of four-gluon vertices the color parts do not trivially factorize.
  Due to this, 
  depending on whether or not we work in a color-ordered approach,
  color factors may or may not be collected globally in front of the Feynman diagram;
  i.e. different chirality flows may have different color factors.
},
we obtain the result of the Feynman diagram without
any non-trivial algebraic manipulation.

If it is desired to obtain the result in conventional form with spinor
brackets, expand the momentum-dots and translate the lines to
spinor inner products.

\section{Examples}

\label{sec:Examples}

In order to demonstrate how the chirality-flow picture is applied,
we give some examples.
We remind the reader that we adopt the convention of counting all particles
in a scattering process as outgoing\footnote{
  We remind the reader that after crossing from incoming to outgoing states, 
  a (left-chiral) negative helicity incoming particle for instance turns into a (left-chiral) positive helicity outgoing anti-particle, 
  etc. 
  }.
The recipe for turning a Feynman diagram into a chirality-flow diagram is given in \secref{sec:recipe}, 
and for convenience the Feynman and chirality-flow rules are collected in 
\tabsref{tab:Flow rules comparison QED} and \ref{tab:Flow rules comparison QCD}.

\subsection{\texorpdfstring{$e^+e^-\to \mu^+ \mu^-$}{e+e- -> mu+mu-}}

We begin with a simple example of electron-positron annihilation,
creating a muon-antimuon pair through photon exchange. 
We follow the recipe in \secref{sec:recipe} and arbitrarily choose one
of the possible helicity configurations for the external fermions,
considering $M(0\rightarrow e_L^- e_R^+ \mu_L^- \mu_R^+ )$.

Step \ref{step:prefactors} asks to collect prefactors and denominators, 
while step \ref{step:add lines} requires assigning dotted lines to positive helicity
particles and undotted lines to negative helicity particles, giving
\begin{align}
M(0\rightarrow e_L^- e_R^+ \mu_L^- \mu_R^+ )
= \raisebox{-0.4\height}{\includegraphics[scale=0.4]{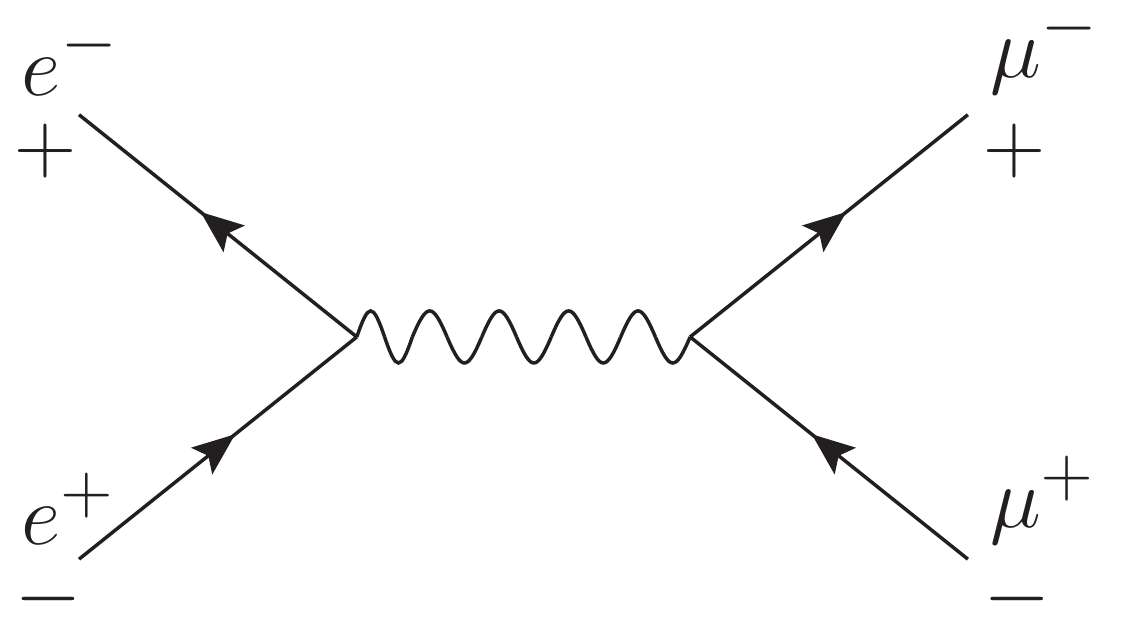}} \sim
\frac{2ie^2}{s_{e^+e^-}}\raisebox{-0.4\height}{\includegraphics[scale=0.4]{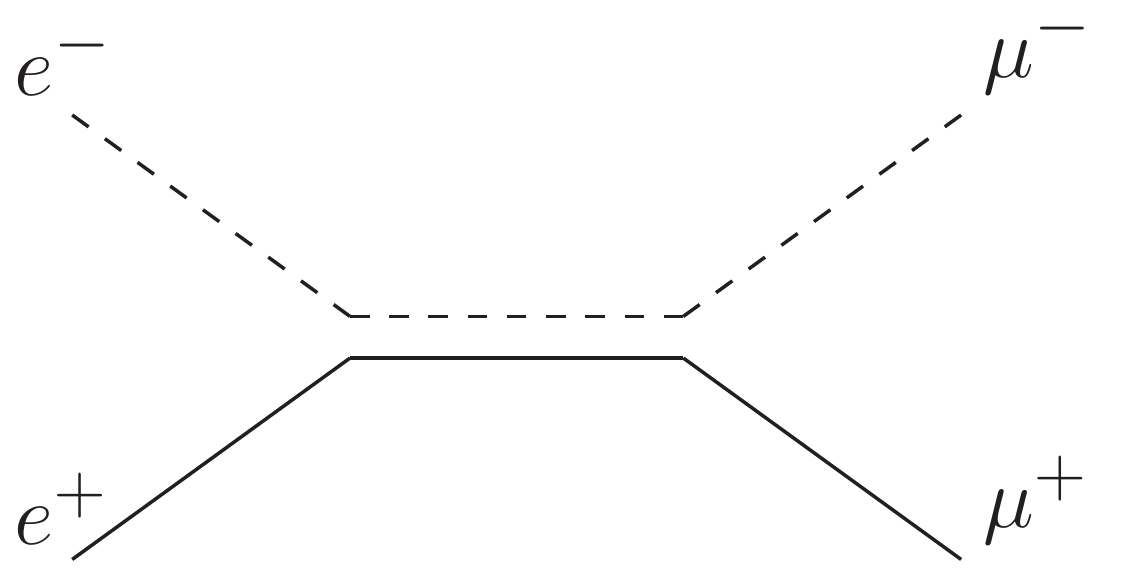}}~,
\end{align}
with $s_{e^+e^-}=(p_{e^+}+p_{e^-})^2=2p_{e^+}\cdot p_{e^-}$.
From the chirality-flow diagram,
we see that we no longer need the helicity labels, 
since the dotted and undotted lines give the same information. 
Step \ref{step:arrows} advises us how to add chirality-flow arrows.
We arbitrarily choose the chirality-flow arrow from the $e^-$ to point inward and follow its line through the chirality-flow diagram to the $\mu^-$, 
which then has its chirality-flow arrow pointing outward. 
The arrow on the solid line is then fixed by the double line from the photon propagator;
it has to be opposite to that of the dotted line,
such that
\begin{align}
M(0\rightarrow  e_L^-  e_R^+ \mu_L^- \mu_R^+ )
= \raisebox{-0.4\height}{\includegraphics[scale=0.4]{./Jaxodraw/eemumuExFeynPla}} = 
\frac{2ie^2}{s_{e^+e^-}}\raisebox{-0.4\height}{\includegraphics[scale=0.4]{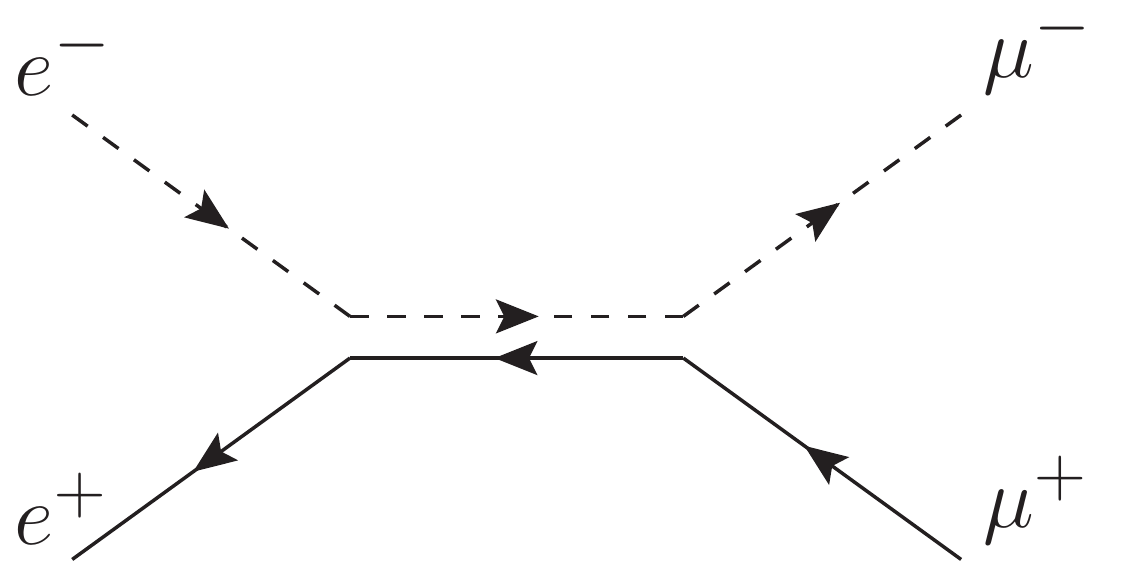}}~, \label{eq:QED ex 1 flow}
\end{align}
which is our final result.
The directed dotted and solid lines are equivalent to the spinor inner products 
which we may, if desired, convert to the more familiar square  and angled brackets
using \eqsrefa{eq:Angle Prod Flow}{eq:Square Prod Flow}
to obtain $M(0\rightarrow e_L^- e_R^+ \mu_L^- \mu_R^+ )=\frac{2ie^2}{s_{e^+e^-}} [e^-\mu^-]\langle \mu^+ e^+ \rangle$.

The only other non-trivial diagram needed for calculating the helicity-summed result is generated by swapping the helicities of the muons,
\begin{align}
M(0\rightarrow e_L^- e_R^+ \mu_R^- \mu_L^+)
= \raisebox{-0.4\height}{\includegraphics[scale=0.4]{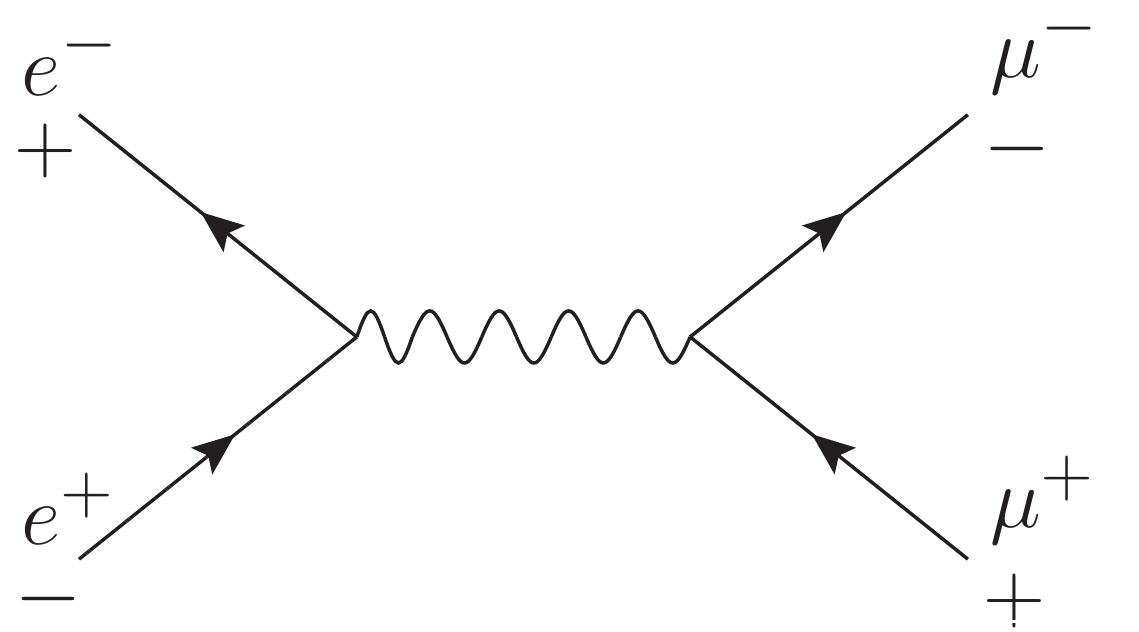}} &= 
\frac{2ie^2}{s_{e^+e^-}}\raisebox{-0.4\height}{\includegraphics[scale=0.4]{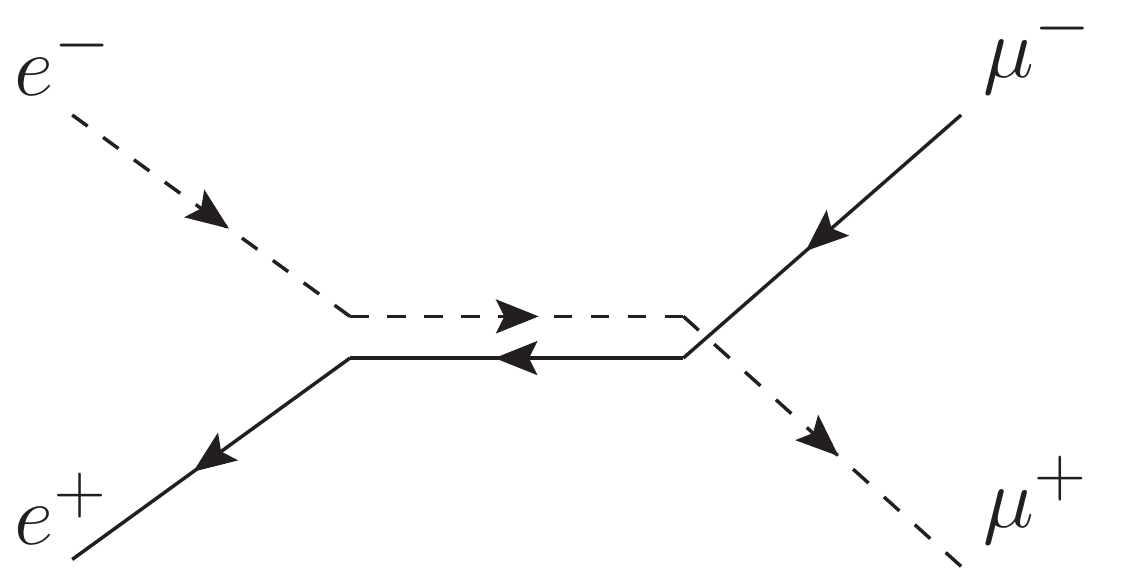}}~. 
\label{eq:QED ex 2 flow}
\end{align}
Using \eqsrefa{eq:Angle Prod Flow}{eq:Square Prod Flow},
we get $M(0\rightarrow  e_L^- e_R^+\mu_R^- \mu_L^+)=\frac{2ie^2}{s_{e^+e^-}} [e^-\mu^+]\langle \mu^- e^+ \rangle$.
The remaining two helicity configurations needed for calculating the helicity-summed result are simply given by exchanging the dotted and solid lines in the chirality-flow diagrams in 
\eqsrefa{eq:QED ex 1 flow}{eq:QED ex 2 flow}. 

While the above is a simple example --- also in the ordinary spinor-helicity
formalism,
we stress that we did not need to perform a single 
algebraic manipulation to arrive at the result.

\subsection{\texorpdfstring{$e^+e^- \to \mu^+\mu^- \gamma$}{e+e- -> mu+mu- photon}} \label{sec:QEDphotonEx}

As our next example we consider the same process,
but with an additional photon radiated externally.
This introduces a fermion propagator and a polarization vector. 
We choose a particular helicity configuration to start with,
considering the amplitude $M(0\rightarrow  e_L^- e_R^+ \mu_L^- \mu_R^+ \gamma_1^+)$. 
One diagram for this process has the photon emitted from the $\mu^-$.
Steps \ref{step:prefactors} and \ref{step:add lines} from the recipe ask to draw
the chirality-flow structure, giving
\begin{align}
 \raisebox{-0.4\height}{\includegraphics[scale=0.4]{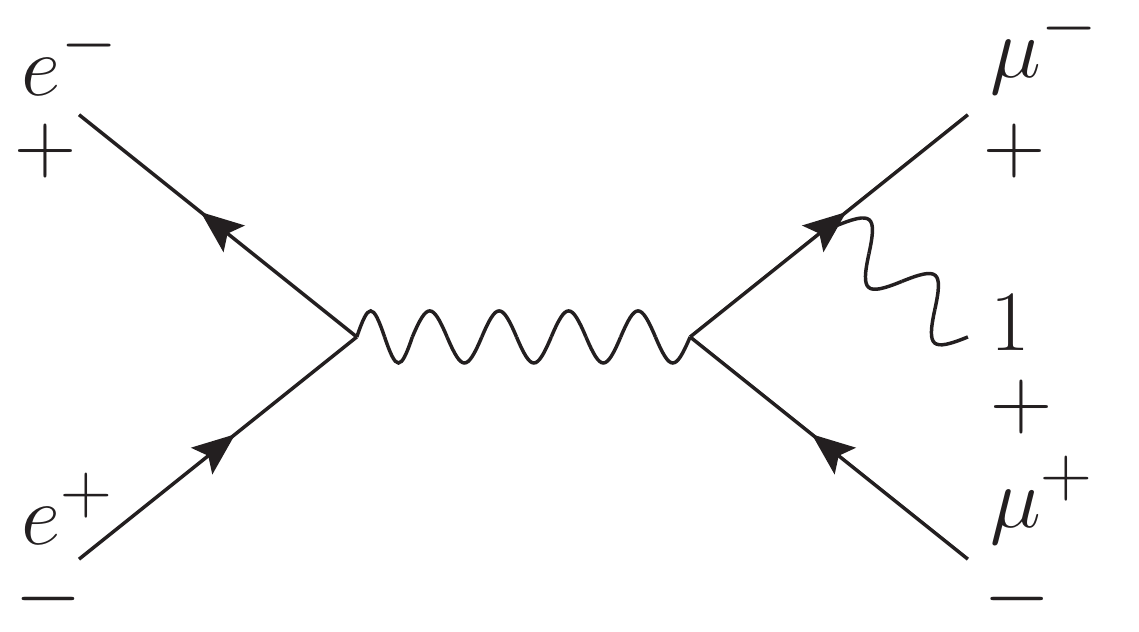}} \sim 
  \frac{-i2\sqrt{2}e^3}{s_{e^+e^-}s_{1\mu^-}\langle r1\rangle}\raisebox{-0.4\height}{\includegraphics[scale=0.4]{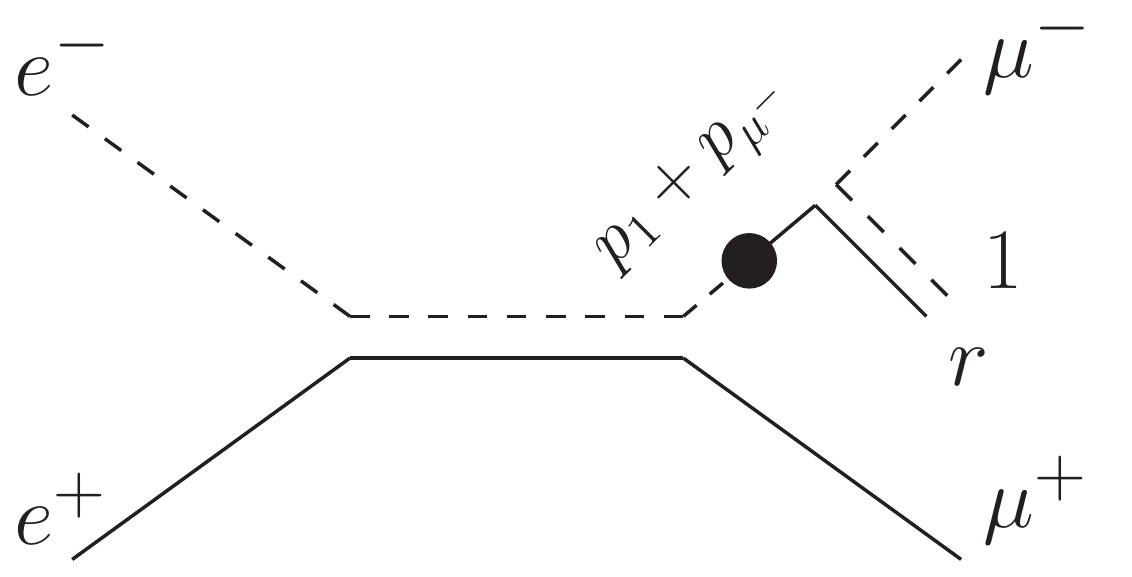}}~,
\end{align}
where $r$ denotes the reference momentum of the external photon, 
the fermion propagator has momentum $p_1+p_{\mu^-}$
and we have used the shorthand notation $p_1 \rightarrow 1$ to denote the photon momentum.
Step \ref{step:arrows} is to assign chirality-flow arrows. 
Following the same procedure as in the previous example, 
remembering to continue the flow through the momentum-dot,
we get
\begin{align}
 \raisebox{-0.4\height}{\includegraphics[scale=0.4]{./Jaxodraw/QEDFlowEx2a}} = 
 \frac{-i2\sqrt{2}e^3}{s_{e^+e^-}s_{1\mu^-}\langle r1\rangle}\raisebox{-0.4\height}{\includegraphics[scale=0.4]{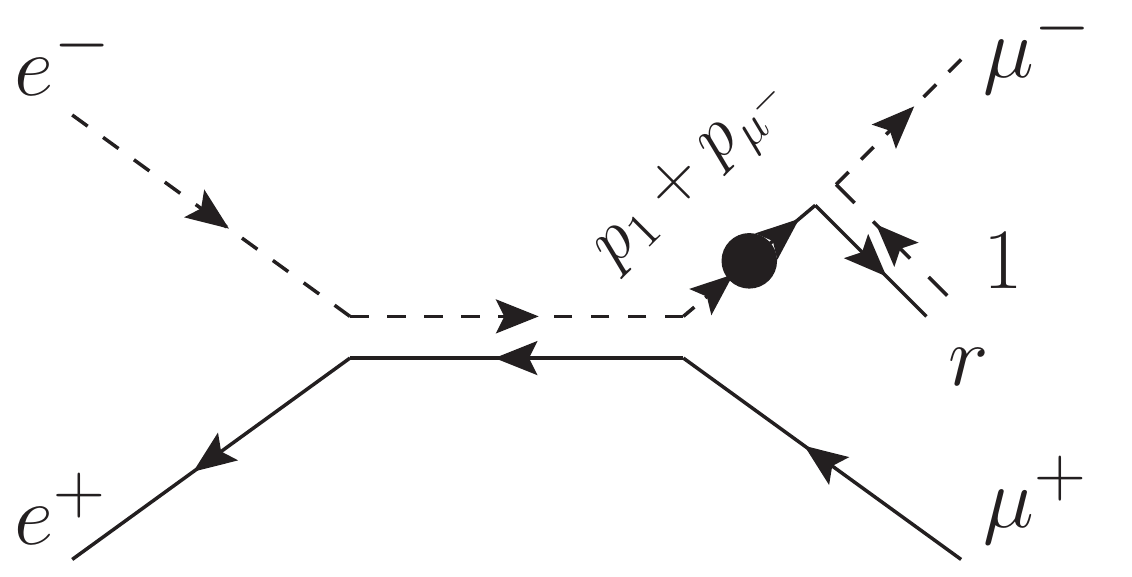}}~.
\end{align}
This is in principle our final result, 
as it contains all the spinor inner products in the diagram.
To write out the inner products as square or angled brackets,
we first expand the momentum-dot using \eqref{eq:p_sum0},
and then use \eqsrefa{eq:Angle Prod Flow}{eq:Square Prod Flow}
to obtain
\begin{align}
\raisebox{-0.4\height}{\includegraphics[scale=0.4]{./Jaxodraw/QEDFlowEx2aDL2}} = 
 \Big([e^-1]\langle 1 r \rangle + [e^-\mu^-]\langle \mu^- r \rangle \Big)[1\mu^-]\langle \mu^+ e^+ \rangle~. 
\end{align}

The diagram with the photon emitted from the $\mu^+$ has a similar structure, 
\begin{align}
\raisebox{-0.4\height}{\includegraphics[scale=0.4]{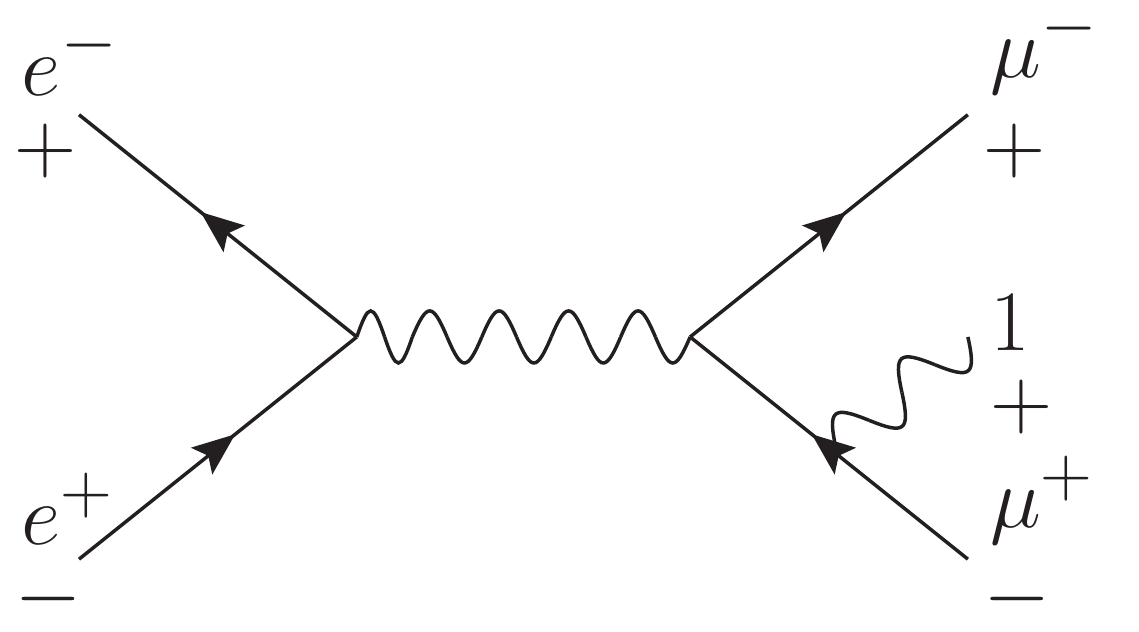}} &= 
\frac{-i2\sqrt{2}e^3}{s_{e^+e^-}s_{1\mu^+}\langle r1\rangle} \raisebox{-0.4\height}{\includegraphics[scale=0.4]{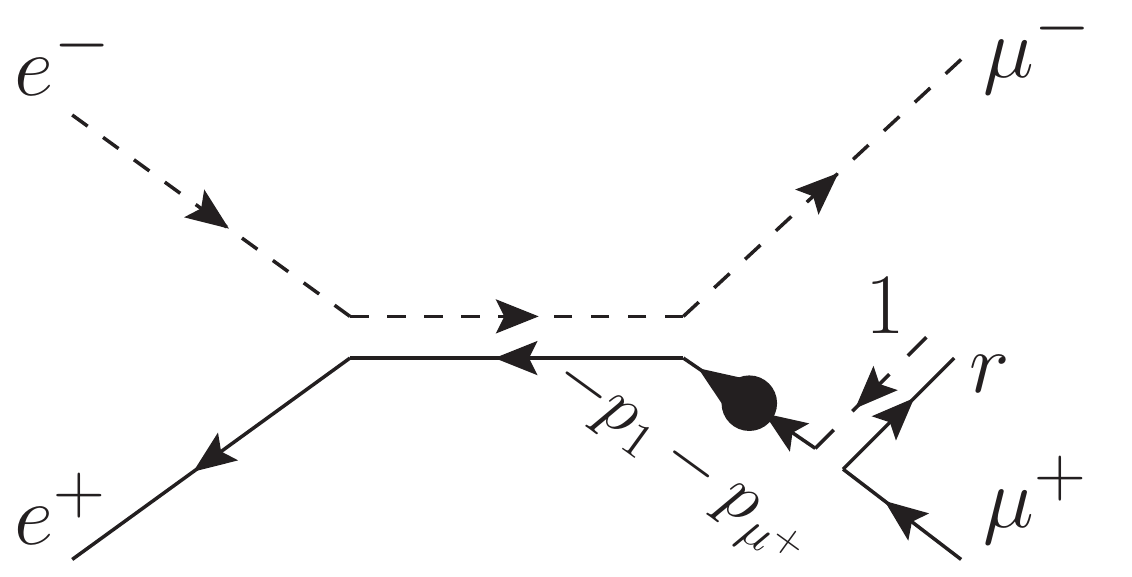}} \nonumber \\
&= \frac{-i2\sqrt{2}e^3}{s_{e^+e^-}s_{1\mu^+}\langle r1\rangle} [e^-\mu^-]\langle \mu^+ r \rangle \Big(-0-[1\mu^+]\langle \mu^+ e^+\rangle \Big)~,\label{eq:QEDphoton2}
\end{align}
where in the last step we have used that $[11]=0$.

Following the same procedure to obtain results for the remaining
two diagrams gives
\begin{eqnarray}
  &&\quad\quad\quad\quad\quad\quad
  \raisebox{-0.4\height}{\includegraphics[scale=0.4]{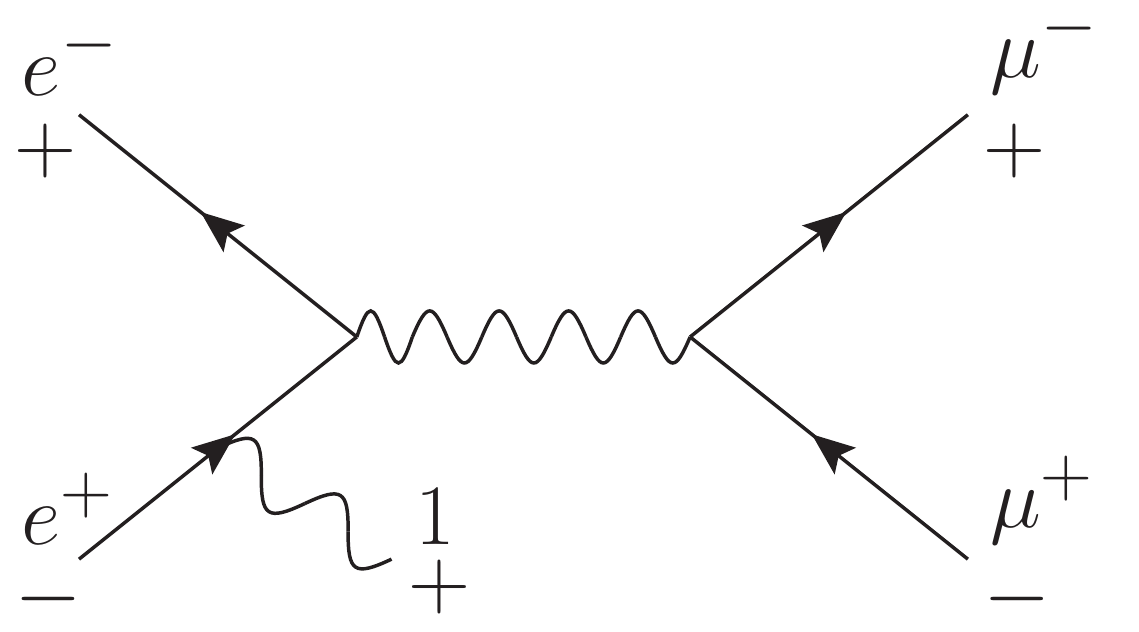}}
  \;\;+ \quad\quad\quad\quad\quad\quad\quad\raisebox{-0.4\height}{\includegraphics[scale=0.4]{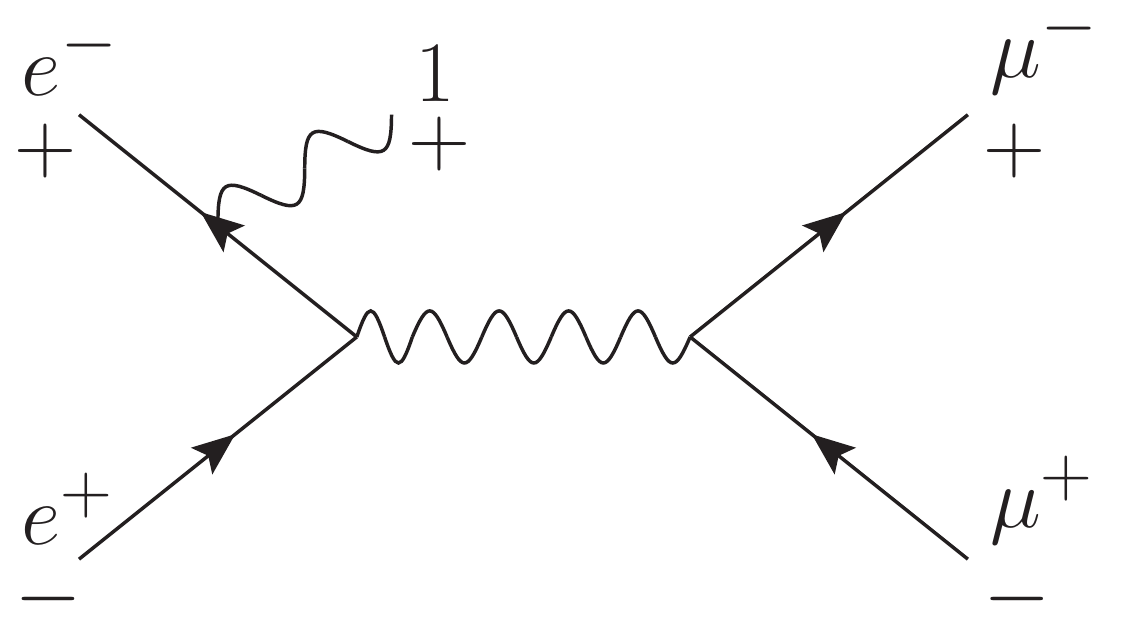}} \label{eq:QEDphoton34}\\
  &=& \frac{-i2\sqrt{2}e^3}{s_{\mu^+\mu^-}s_{1e^+}\langle r1\rangle}\raisebox{-0.4\height}{\includegraphics[scale=0.4]{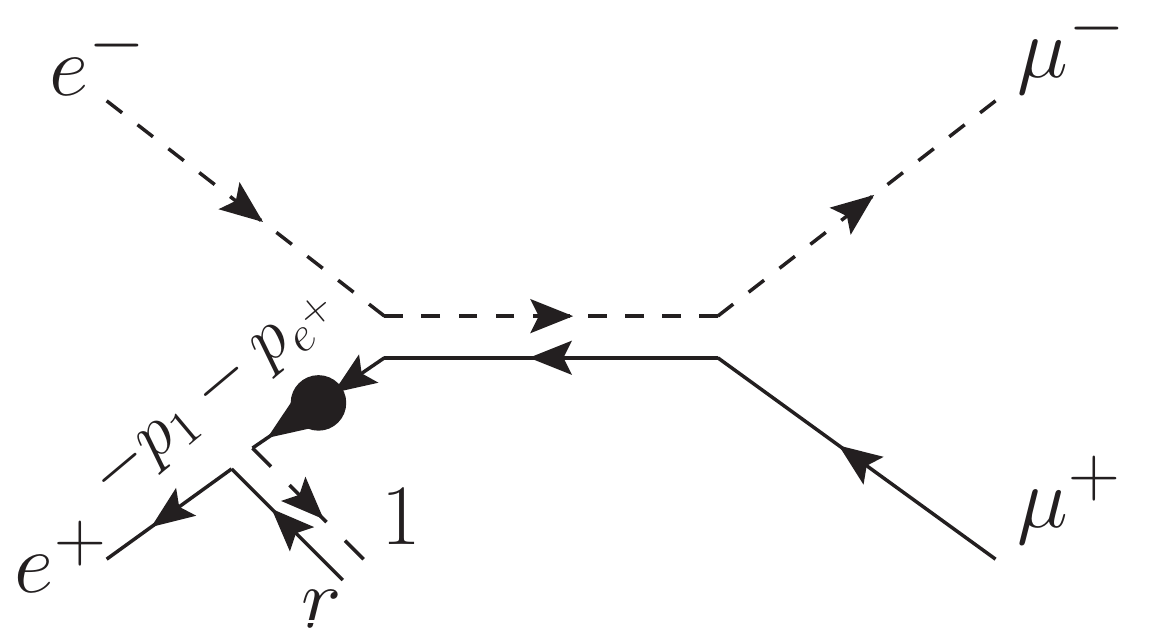}}
  +\;\frac{-i2\sqrt{2}e^3}{s_{\mu^+\mu^-}s_{1e^-}\langle r1\rangle}\raisebox{-0.4\height}{\includegraphics[scale=0.4]{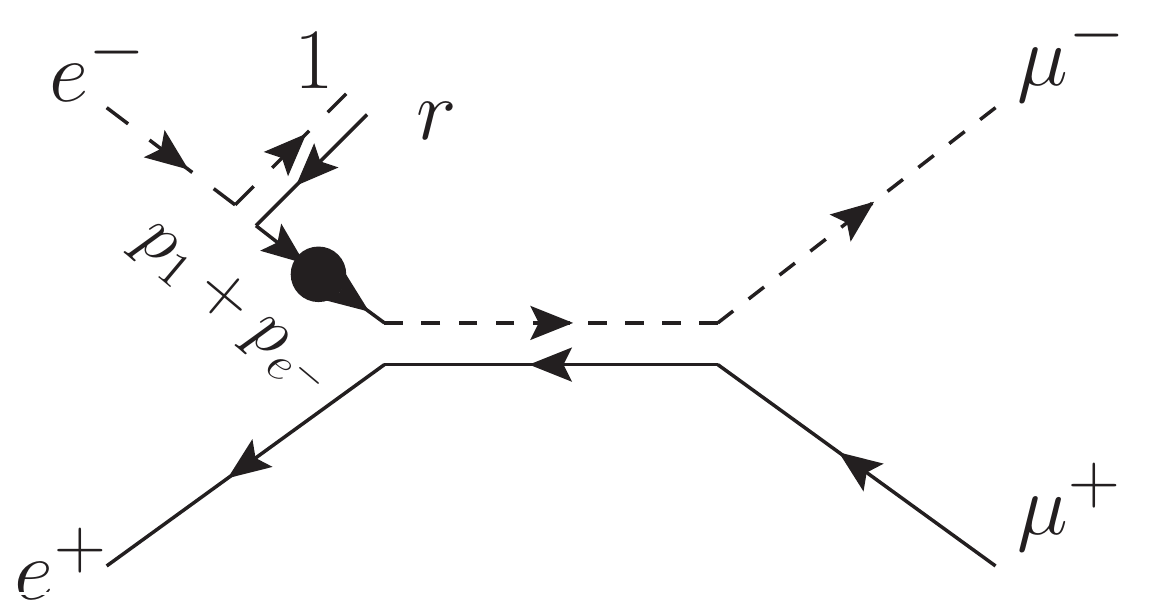}} \nonumber \\
  &=& \frac{i2\sqrt{2}e^3}{s_{\mu^+\mu^-}s_{1e^+}\langle r1\rangle} \;\;[e^-\mu^-]\langle\mu^+ e^+ \rangle [e^+1]\langle re^+\rangle \quad
  +\;\frac{-i2\sqrt{2}e^3}{s_{\mu^+\mu^-}s_{1e^-}\langle r1\rangle}[e^-1]\Big(\langle r1\rangle[1\mu^-] + \langle re^-\rangle[e^-\mu^-]\Big)\langle\mu^+e^+\rangle~.\nonumber
\end{eqnarray}
Note that in the results above,
we have left the reference momentum of the external photon unassigned. 
We may simplify the results by choosing it appropriately\footnote{
  We need to choose the same reference vector for each gauge-invariant subset of diagrams.
},
to generate spinor inner products of the form $\langle ii\rangle=[ii]=0$.
We stress again that we did not need to perform a single 
algebraic manipulation,
other than to expand momentum-dots,
to arrive at the results.

We also remark on the simplicity of the result: In QED
every Feynman diagram gives a single chirality-flow graph, where
every spinor line is contracted with the ``nearest'' (closest
possible, following the chirality flow) spinor of the same kind.

\subsection{\texorpdfstring{$q_1 \bar{q}_1\to q_2 \bar{q}_2 g$}{q1qb1 -> q2qb2g}} \label{sec:4q1gEx}

We now consider our first QCD example,
$q_1 \bar{q}_1\to q_2 \bar{q}_2 g$.
For this example we will --- for comparison --- also go through the
standard spinor-helicity calculation. 
We will also for the first time encounter a disconnected chirality-flow
structure.

\begin{figure}
\centering
\subfloat[\label{fig:4q1g 1}]{\includegraphics[scale=0.4]{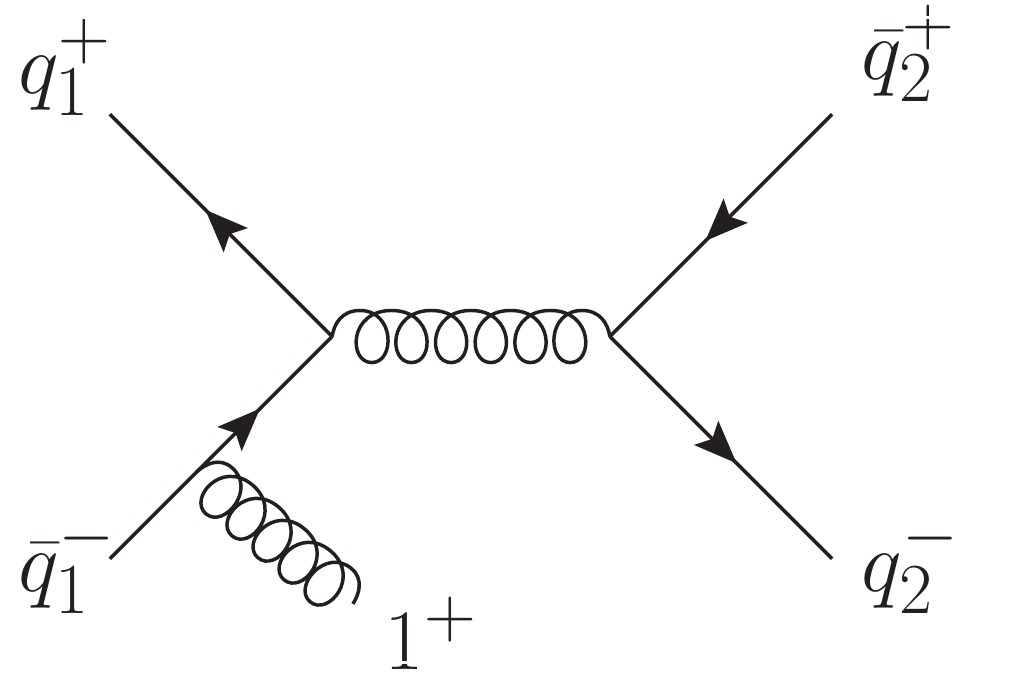}}
\hspace{1cm}
\subfloat[\label{fig:4q1g 2}]{\includegraphics[scale=0.4]{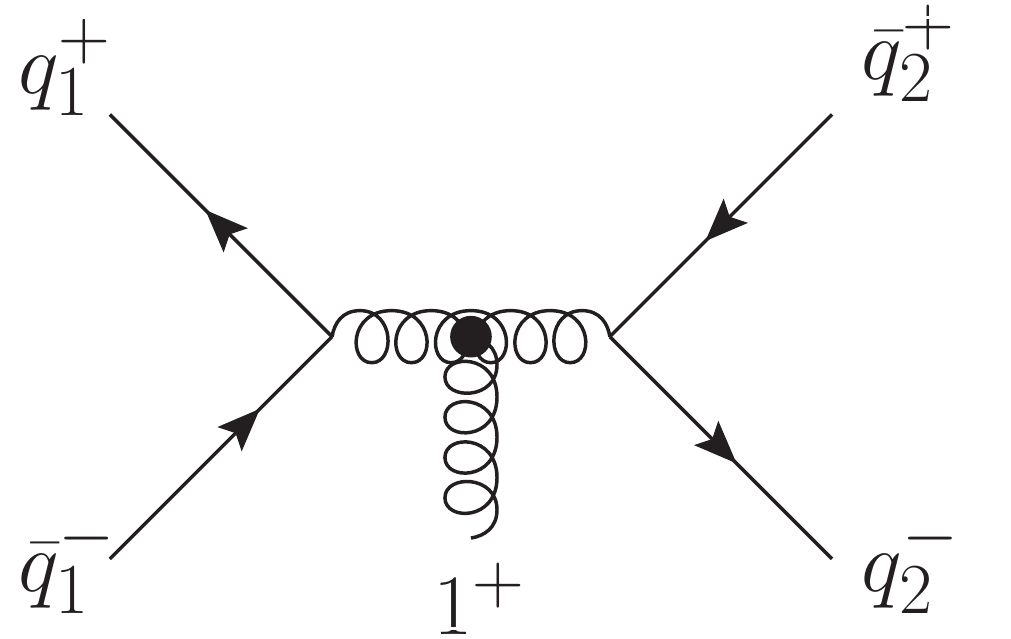}}
\hspace{1cm}
\subfloat[\label{fig:4q1g 3}]{\includegraphics[scale=0.4]{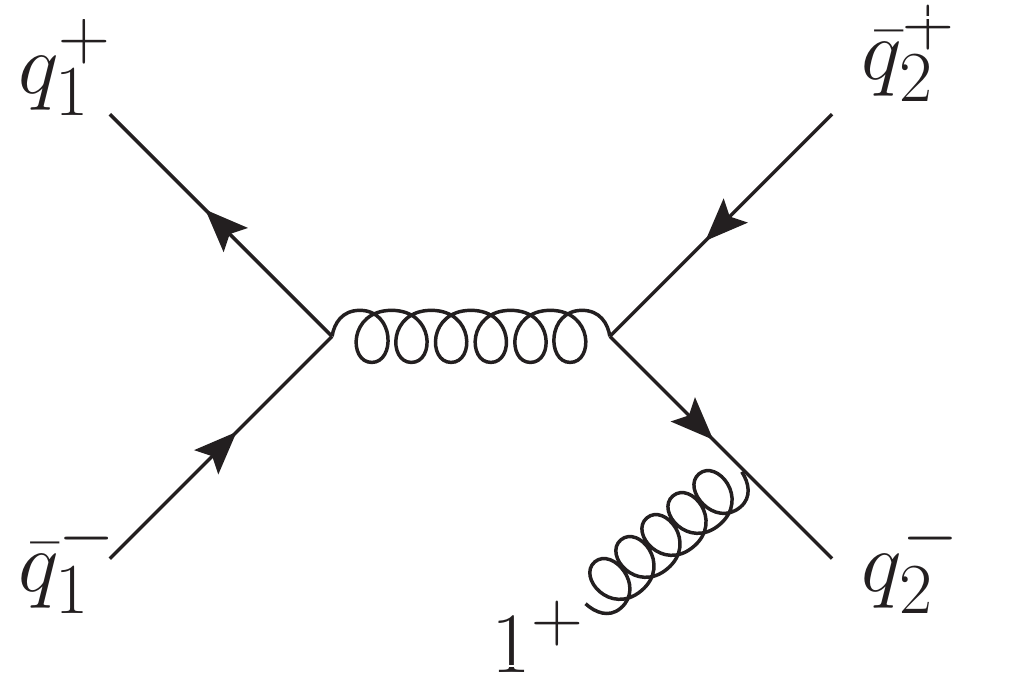}}
\caption{The three diagrams contributing to the 
partial helicity amplitude
$M(0\rightarrow q_2^- 1^+ \bar{q}_1^- q_1^+ \bar{q}_2^+)$
multiplying the color factor 
$t^{a_1}_{q_2\bar{q}_1}\de_{q_1\bar{q}_2}$.
} 
\label{fig:4q1g diagrams}
\end{figure}

There are six Feynman diagrams in total, 
and the color structure can be decomposed into four linearly independent
basis vectors (color factors). 
We will consider the (color-ordered) 
partial helicity amplitude
$M(0\rightarrow q_2^- 1^+ \bar{q}_1^- q_1^+ \bar{q}_2^+)$
multiplying the color factor 
$t^{a_1}_{q_2\bar{q}_1}\de_{q_1\bar{q}_2}$, 
which has contributions from the three diagrams in \figRef{fig:4q1g diagrams}.

Using the standard spinor-helicity method, 
the Lorentz structure of the diagram in \figRef{fig:4q1g 1} is given by
\begin{align}
\raisebox{-0.4\height}{\includegraphics[scale=0.4]{./Jaxodraw/4q1g1Feyn}} &\rightarrow 
\frac{-ig_s^3}{s_{1\bar{q}_1}s_{q_2\bar{q}_2}}[ q_1|\tau^{\mu}|-(\overline{\slashed{p}}_1+\overline{\slashed{p}}_{\bar{q}_1})|\eps_1^+|\bar{q}_1\rangle \langle q_2|\taubar_{\mu}|\bar{q}_2] \nonumber \\
&= \frac{-ig_s^3}{s_{1\bar{q}_1}s_{q_2\bar{q}_2}\langle r 1\rangle}
\Big(-[q_1|\tau^{\mu}|1 \rangle [1 1]\langle r\bar{q}_1\rangle
- [q_1|\tau^{\mu}|\bar{q}_1\rangle [\bar{q}_1 1]\langle r\bar{q}_1\rangle\Big)
\langle q_2|\taubar_{\mu}|\bar{q}_2] \nonumber \\
&= \frac{ig_s^3}{s_{1\bar{q}_1}s_{q_2\bar{q}_2}\langle r 1\rangle}
[q_1 \bar{q}_2 ] \langle q_2 \bar{q}_1\rangle [\bar{q}_1 1] \langle r\bar{q}_1\rangle ~,\label{eq:4q1gCalcFeyn1}
\end{align}
where, in the first line, we collected the prefactors and denominators
from propagators, and wrote down the spinor expression for each fermion line. 
The spinor expressions begin with the quark, 
have a $\tau^{\mu}$ or $\taubar^{\mu}$ for each vertex, 
and a $\slashed{p}$ or $\overline{\slashed{p}}$ for each fermion propagator.
We 
expand out the propagator momentum
and contract polarization vectors with a $\tau$ where possible.
In the second line we 
rewrote the slashed propagator momenta 
and the slashed polarization vector in terms of spinors.
Finally, we used $[11]=0$, 
and utilized the Fierz identity between $\tau$ and $\taubar$ to write the result in terms of spinor inner products.

Within the chirality-flow formalism, we collect scalar factors and
write down the result immediately (first without arrows) 
\begin{align}
\hspace{-2ex}
\raisebox{-0.45\height}{\includegraphics[scale=0.4]{./Jaxodraw/4q1g1Feyn}} 
\!\!\!\!\!\!\!\!\!\!
\rightarrow \frac{-ig_s^3}{s_{1\bar{q}_1}s_{q_2\bar{q}_2}\langle r 1\rangle} 
\!\raisebox{-0.425\height}{\includegraphics[scale=0.4]{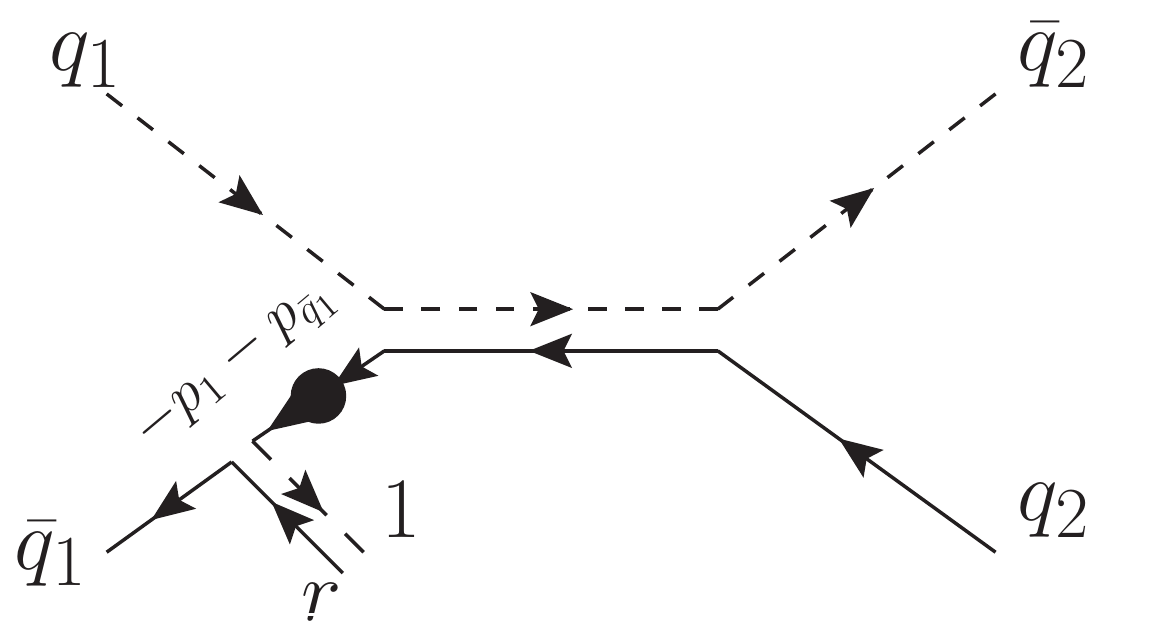}} 
\!\!\!\!\!\!\!\!\!\!
= \frac{ig_s^3}{s_{1\bar{q}_1}s_{q_2\bar{q}_2}\langle r 1\rangle}
[q_1 \bar{q}_2 ] \langle q_2 \bar{q}_1\rangle [\bar{q}_1 1] \langle r\bar{q}_1\rangle ~,
\end{align}
in close resemblance to the first term in \eqref{eq:QEDphoton34}.

Following the standard spinor-helicity procedure, as in \eqref{eq:4q1gCalcFeyn1}, 
the diagram in \figRef{fig:4q1g 3} gives
\begin{align}
\raisebox{-0.4\height}{\includegraphics[scale=0.4]{./Jaxodraw/4q1g3Feyn}} &\rightarrow 
\frac{-ig_s^3}{s_{1q_2}s_{q_1\bar{q}_1}}[ q_1|\tau^{\mu}|\bar{q}_1\rangle \langle q_2|\bar{\eps}_1^+|(\slashed{p}_1+\slashed{p}_{q_2})|\taubar_{\mu}|\bar{q}_2] \nonumber \\
&= \frac{-ig_s^3}{s_{1q_2}s_{q_1\bar{q}_1}\langle r 1\rangle}[ q_1|\tau^{\mu}|\bar{q}_1\rangle 
\Big(\langle q_2 r \rangle [11]\langle 1|\taubar_{\mu}|\bar{q}_2] + \langle q_2 r \rangle[1 q_2]\langle q_2 |\taubar_{\mu}|\bar{q}_2] \Big)\nonumber \\
&= \frac{-ig_s^3}{s_{1q_2}s_{q_1\bar{q}_1}\langle r 1\rangle}
[q_1 \bar{q}_2 ] \langle q_2 \bar{q}_1\rangle [1q_2] \langle q_2r \rangle ~,\label{eq:4q1gCalcFeyn3}
\end{align}
which requires a few steps.
However, in the chirality-flow formalism,
we immediately write this down (cf. \eqref{eq:QEDphoton2})
\begin{align}
\hspace{-2ex}
\raisebox{-0.45\height}{\includegraphics[scale=0.4]{./Jaxodraw/4q1g3Feyn}} 
\!\!\!\!\!\!\!\!\!\!
\rightarrow \frac{-ig_s^3}{s_{1q_2}s_{q_1\bar{q}_1}\langle r 1\rangle} \raisebox{-0.425\height}{\includegraphics[scale=0.4]{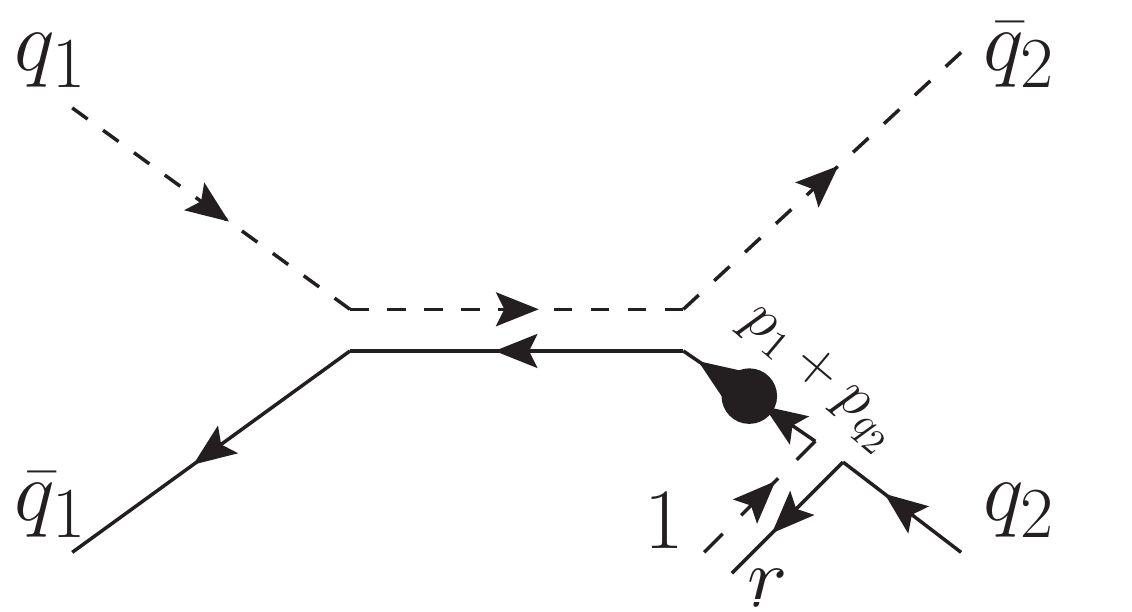}} 
\!\!\!\!\!\!\!\!\!\!
= \frac{-ig_s^3}{s_{1q_2}s_{q_1\bar{q}_1}\langle r 1\rangle}
[q_1 \bar{q}_2 ] \langle q_2r \rangle[1q_2]\langle q_2 \bar{q}_1\rangle   ~.
\end{align}

The new type of diagram in this example is \figRef{fig:4q1g 2}.
Using the standard spinor-helicity method we get
\begin{eqnarray}
  \raisebox{-0.4\height}{\includegraphics[scale=0.4]{./Jaxodraw/4q1g2Feyn}}
  &\rightarrow& 
  \frac{ig_s^3}{\sqrt{2}s_{q_1\bar{q}_1}s_{q_2\bar{q}_2}} [q_1|\tau^{\mu}|\bar{q}_1\rangle \langle q_2|\taubar^{\nu}|\bar{q}_2](\epsilon_1^+)^{\rho} \times \nonumber \\
  &\times& \Big( g_{\mu\nu}\big(p_{q_1}+p_{\bar{q}_1}-(p_{q_2}+p_{\bar{q}_2})\big)_{\rho} + g_{\nu\rho}\big(p_{q_2}+p_{\bar{q}_2}-p_1\big)_{\mu} + g_{\rho\mu}\big(p_1-(p_{q_1}+p_{\bar{q}_1})\big)_{\nu}\Big)\nonumber \\
  &=& \frac{ig_s^3}{\sqrt{2}s_{q_1\bar{q}_1}s_{q_2\bar{q}_2}}\bigg( [q_1|\tau^\mu|\bar{q}_1\rangle\langle q_2|\taubar_{\mu}|\bar{q}_2]
  \epsilon_1^+\cdot \big(2(p_{q_1}+p_{\bar{q}_1})+p_1\big) \nonumber \\
  &&+ \frac{1}{\sqrt{2}}[q_1|(-2\slashed{p}_1-\slashed{p}_{q_1}-\slashed{p}_{\bar{q}_1})|\bar{q}_1\rangle\langle q_2|\bar{\eps}_1^+|\bar{q}_2]
  + \frac{1}{\sqrt{2}} [q_1|\eps_1^+|\bar{q}_1\rangle\langle q_2|(2\bar{\slashed{p}}_1+\bar{\slashed{p}}_{q_2}+\bar{\slashed{p}}_{\bar{q}_2})|\bar{q}_2] \bigg) \nonumber 
\end{eqnarray}
\begin{eqnarray}
  \quad\quad\quad\quad\
   &=& \frac{ig_s^3}{s_{q_1\bar{q}_1}s_{q_2\bar{q}_2}\langle r1 \rangle}\bigg([q_1\bar{q}_2] \langle q_2\bar{q}_1\rangle[1|(\slashed{p}_{q_1}+\slashed{p}_{\bar{q}_1})|r\rangle - [q_1|\slashed{p}_1|\bar{q}_1\rangle \langle q_2r\rangle [1 \bar{q}_2] \nonumber \\
      &&+ [q_1 1] \langle r\bar{q}_1\rangle \langle q_2|\bar{\slashed{p}}_1|\bar{q}_2] \bigg) \nonumber \\
    &=& \frac{ig_s^3}{s_{q_1\bar{q}_1}s_{q_2\bar{q}_2}\langle r1 \rangle}\bigg( [q_1\bar{q}_2] \langle q_2\bar{q}_1\rangle [1q_1]\langle q_1r\rangle + [q_1\bar{q}_2] \langle q_2\bar{q}_1\rangle [1\bar{q}_1]\langle \bar{q}_1 r\rangle \nonumber \\
    &&- [q_1 1] \langle 1\bar{q}_1\rangle \langle q_2 r \rangle [1 \bar{q}_2]
    + [q_11] \langle r\bar{q}_1\rangle \langle q_2 1\rangle [1\bar{q}_2]\bigg)~,\label{eq:4q1gCalcFeyn2}
\end{eqnarray}
where in the second step we used momentum conservation within a triple gluon vertex, 
$p_a-p_b = 2p_a + p_c = -p_c - 2p_a$.
This was done to simplify the result, 
since in the third step we used the Dirac equation to remove the terms containing $p_c$. 
In this step we also used the Fierz identity between $\tau$ and $\taubar$
and rewrote the polarization vector in terms of spinors. Finally, in
the last step, we expanded the slashed propagator momenta in terms of spinors.

To calculate \figref{fig:4q1g 2} using the chirality-flow method is again simpler.
For the triple-gluon vertex,
we start with the term in which the two quark pairs are contracted with the metric, 
and then add the cyclic permutations.
Performing steps \ref{step:prefactors} and \ref{step:add lines} gives
\begin{align}
\raisebox{-0.4\height}{\includegraphics[scale=0.4]{./Jaxodraw/4q1g2Feyn}} \!\!\!\!\!\!\!\!\!\!\rightarrow
\frac{ig_s^3}{2s_{q_1\bar{q}_1}s_{q_2\bar{q}_2}\langle r 1\rangle} \left( \raisebox{-0.425\height}{\includegraphics[scale=0.385]{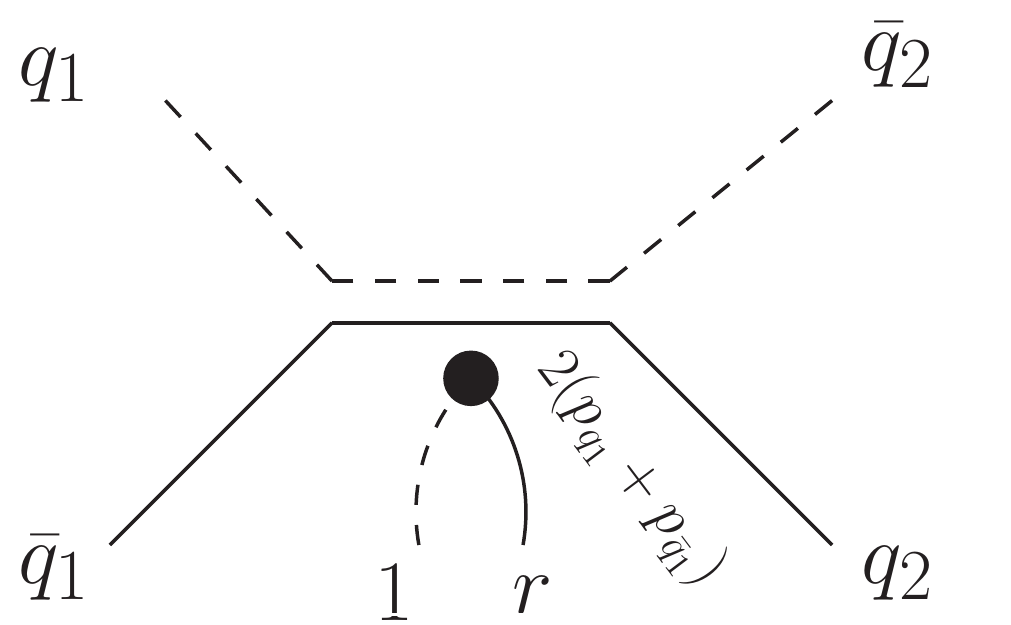}}  
\!\!\!\!\!+\! \raisebox{-0.4\height}{\includegraphics[scale=0.4]{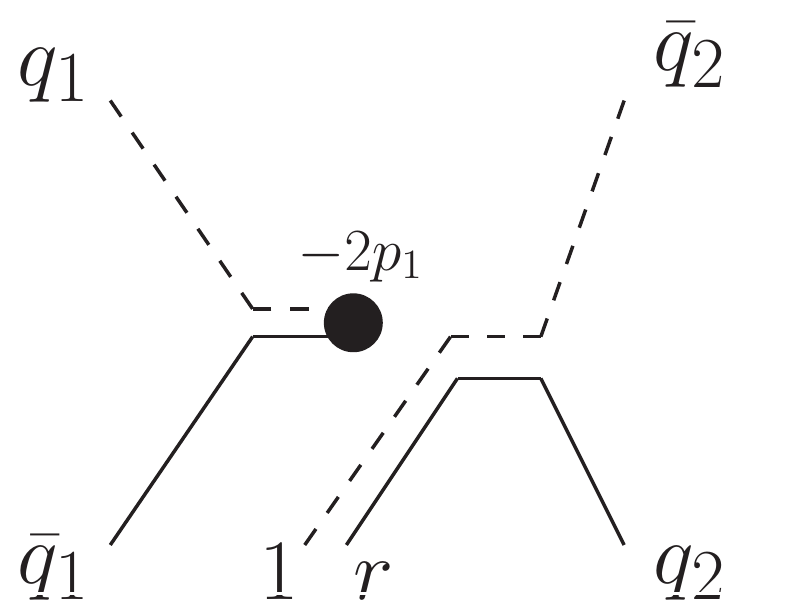}} 
\!\!\!\!\!+\! \raisebox{-0.375\height}{\includegraphics[scale=0.4]{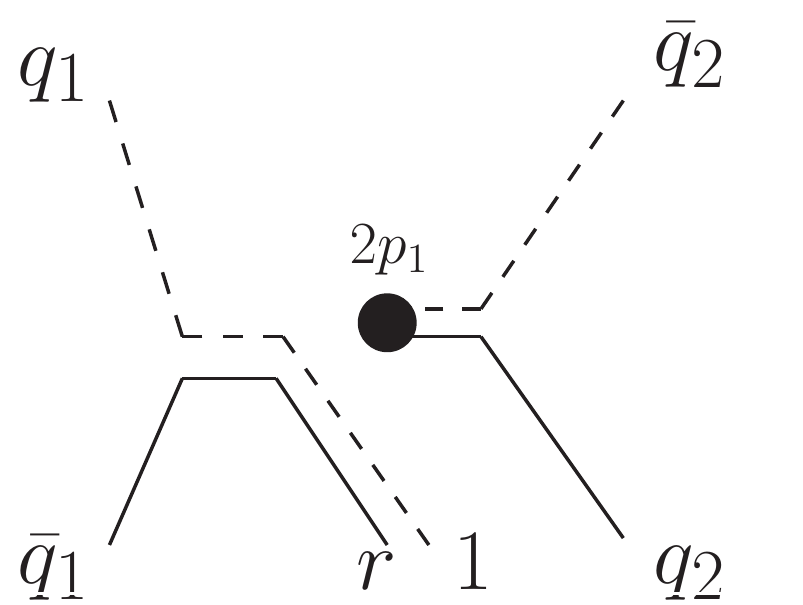}} \!\right)~, \label{eq:4q1gTripGlueNoArrows}
\end{align}
where, as in \eqref{eq:4q1gCalcFeyn2}, 
we used momentum conservation and antisymmetry of the spinor inner product to simplify the slashed momenta 
(i.e.\ the labels of the momentum-dots).

Step \ref{step:arrows} is to assign the arrows. 
We begin with the first chirality-flow diagram, 
and choose the arrow of $q_1$ to point inward.
As before, 
this uniquely fixes the arrows of the dotted line to point from $q_1 \rightarrow \bar{q}_2$,
and the solid line from $q_2 \rightarrow \bar{q}_1$.
However, the spinor line from $1 \leftrightarrow r$ is unaffected by this arrow choice.
In this sense, we say that the chirality-flow diagram has two disconnected pieces,
and we are free to choose the arrow to e.g.\ point from $1 \rightarrow r$.
Similarly, in the second and third terms, 
the arrow directions of $q_1$ and $\bar{q}_2$ can be chosen independently of each other.
An appropriate set of arrow choices gives the spinor structure
\begin{align}
\raisebox{-0.4\height}{\includegraphics[scale=0.4]{./Jaxodraw/4q1g2Feyn}} \!\!\!\!\!\!\!\!\!\!&\rightarrow 
\frac{ig_s^3}{2s_{q_1\bar{q}_1}s_{q_2\bar{q}_2}\langle r 1\rangle} \left( \raisebox{-0.425\height}{\includegraphics[scale=0.385]{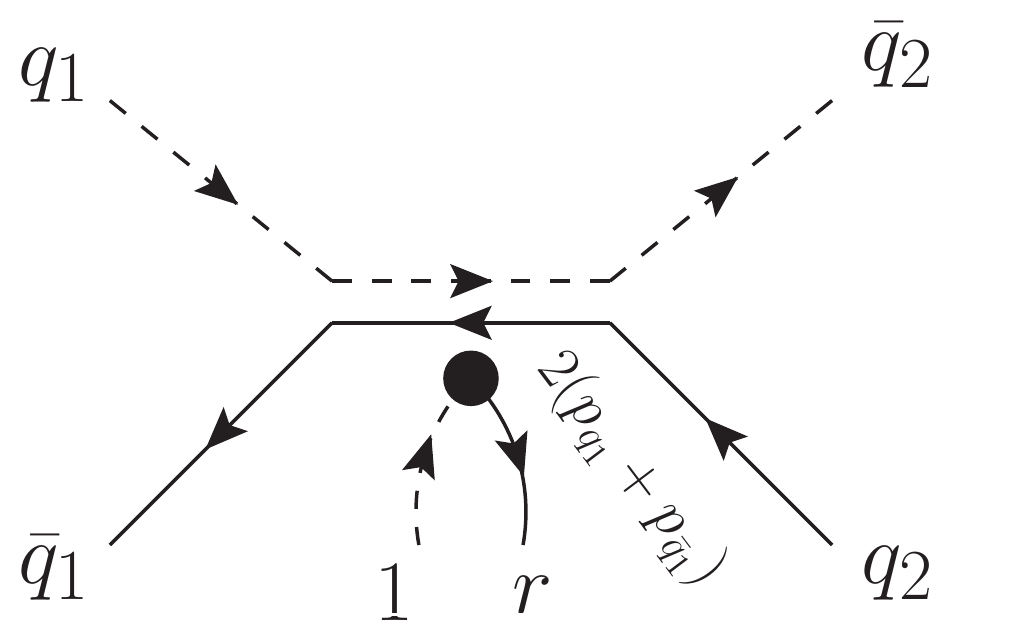}}  
\!\!\!\!\!+\! \raisebox{-0.4\height}{\includegraphics[scale=0.4]{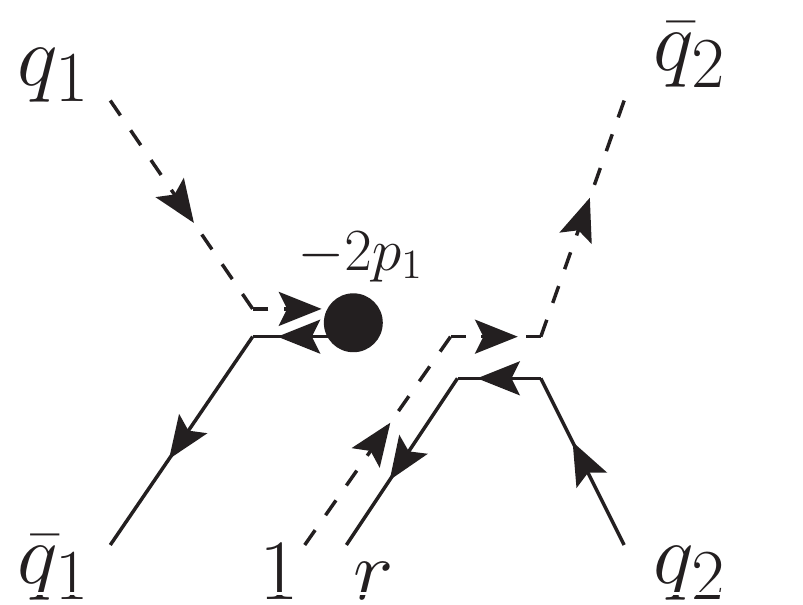}} 
\!\!\!\!\!+\! \raisebox{-0.375\height}{\includegraphics[scale=0.4]{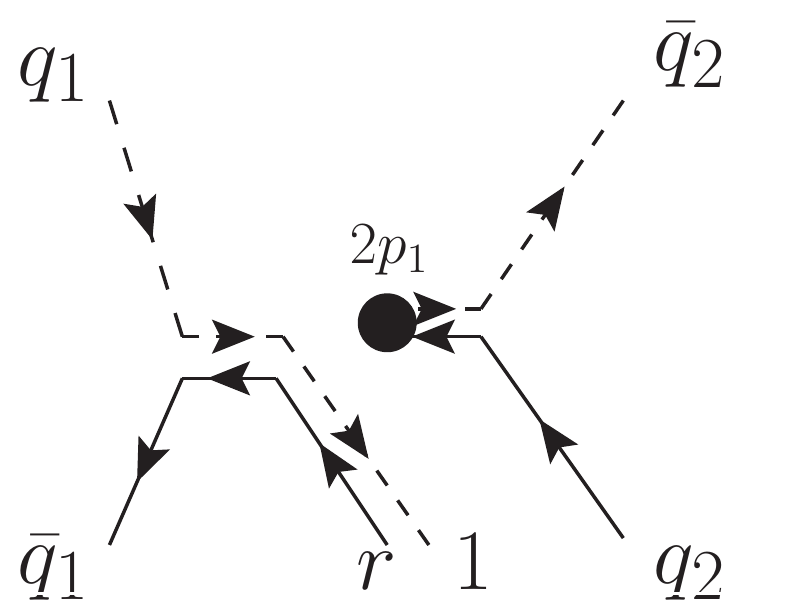}} \!\right) \nonumber \\
\quad\quad\quad\quad&= \frac{ig_s^3}{s_{q_1\bar{q}_1}s_{q_2\bar{q}_2}\langle r 1\rangle} \bigg( [q_1\bar{q}_2] \langle q_2\bar{q}_1\rangle \Big([1q_1]\langle q_1r\rangle + [1\bar{q}_1]\langle \bar{q}_1 r\rangle\Big) \nonumber \\
&   \hspace{5cm}
- [q_1 1] \langle 1\bar{q}_1\rangle \langle q_2 r \rangle [1 \bar{q}_2]
+ [q_11] \langle r\bar{q}_1\rangle \langle q_2 1\rangle [1\bar{q}_2]\bigg)~, \label{eq:4q1gTripGlue}
\end{align}
as in \eqref{eq:4q1gCalcFeyn2}. 

Comparing the two methods, 
we see that 
the chirality-flow method simplifies the calculation in two ways.
First, there is no need to explicitly write all objects as spinor expressions,
and then use the Fierz identity to remove the remaining Lorentz indices.
Second, 
it is much more transparent which spinor inner products occur.

Finally, we stress that once one is familiar with the flow formalism, 
it is possible to write down an amplitude like this in one step,
\begin{align}
M&(0\rightarrow q_2^- 1^+ \bar{q}_1^- q_1^+ \bar{q}_2^+)
= -\frac{ig_s^3}{\langle r 1\rangle} \left( \frac{1}{s_{1\bar{q}_1}s_{q_2\bar{q}_2}} \raisebox{-0.45\height}{\includegraphics[scale=0.35]{./Jaxodraw/4q1g1Flow}} \!\! + \;\; \frac{1}{s_{q_1\bar{q}_1}s_{1q_2}} \raisebox{-0.45\height}{\includegraphics[scale=0.35]{./Jaxodraw/4q1g3Flow}}\right. \nonumber \\
&\quad\quad\quad\quad\quad
 \left. - \frac{1}{2s_{q_1\bar{q}_1}s_{q_2\bar{q}_2}} \left[ \raisebox{-0.45\height}{\includegraphics[scale=0.35]{./Jaxodraw/4q1g2Flowa}} \!\! + \raisebox{-0.45\height}{\includegraphics[scale=0.35]{./Jaxodraw/4q1g2FlowbStr}} \!\! + \raisebox{-0.45\height}{\includegraphics[scale=0.35]{./Jaxodraw/4q1g2FlowcStr}} \right] \quad \right) ~, \label{eq:4q1gOneStep}
\end{align}
where the first two terms correspond to the diagrams in \figsRef{fig:4q1g 1} and \ref{fig:4q1g 3} respectively, 
and the last three terms to the diagram in \figRef{fig:4q1g 2}. 
We again emphasize that the above already contains the sum of all spinor inner products,
and that we may simplify the result by choosing the reference momentum appropriately to generate spinor inner products of the form $\langle ii\rangle$ or $[ii]$.

\subsection{\texorpdfstring{$q\bar{q}\to gg$}{qqb -> gg}}

\begin{figure}
\centering
\subfloat[\label{fig:qqBgg 3Glue}]{\includegraphics[scale=0.4]{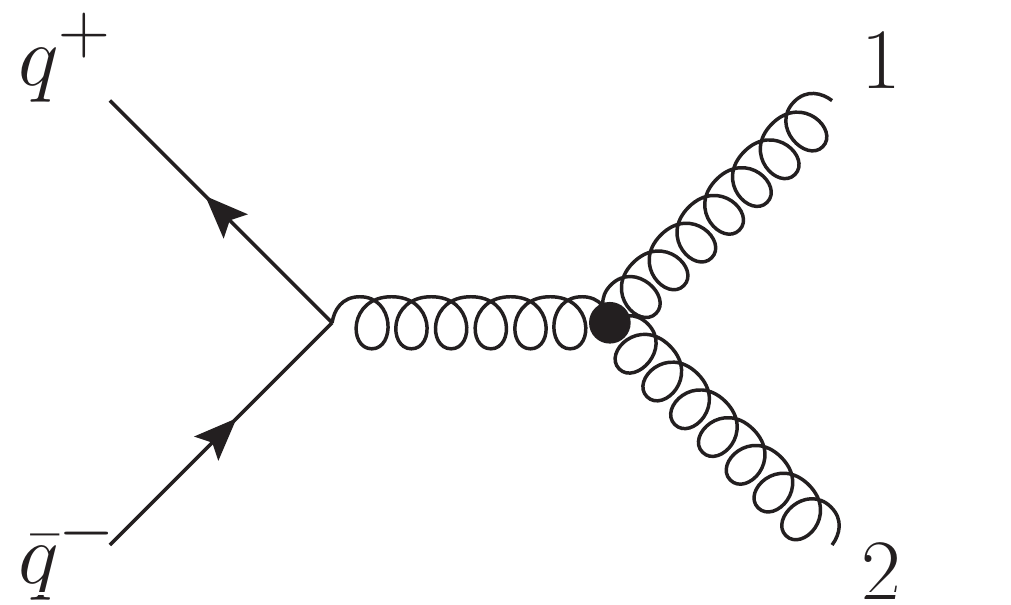}}
\hspace{2cm}
\subfloat[\label{fig:qqBgg QED}]{\includegraphics[scale=0.4]{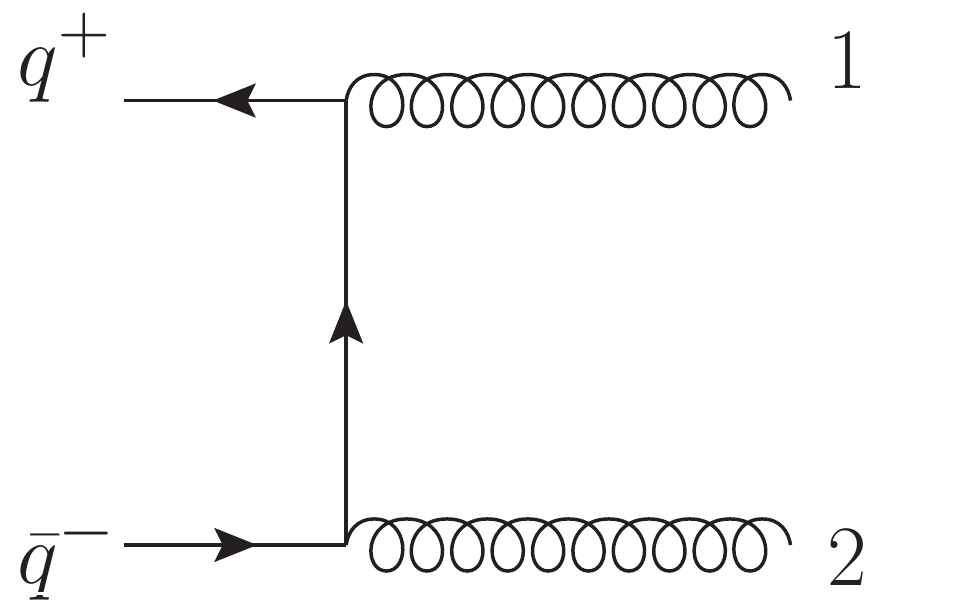}}
\caption{The two diagrams contributing to 
the partial helicity amplitude $M(0\to q^+ 1 \, 2 \, \bar{q}^-)$, multiplying the color factor $t^1_{qi}t^2_{i\bar{q}}$.} 
\label{fig:qqBgg diagrams}
\end{figure}

For this example we will leave the helicity of the vector bosons unassigned. 
We call this the helicity-agnostic case.
This can be done in the traditional spinor-helicity method as well,
but its consequences are more obvious in the chirality-flow picture, 
as the helicity and chirality structures of the diagrams are more transparent.
To this end, 
we introduce a common notation for the bispinor representations of the negative- and positive-helicity polarization vectors in
\eqsrefd{eq:epsilon0}{eq:epsilon1} 
and \eqsrefd{eq:epsilon2}{eq:epsilon3}, 
\begin{align}
\eps_{h}^{\dbe\al}(p_i,r)
=
\frac{|i_p]\langle i_m|}{f_h(i_p,i_m)}
=
\frac{1}{f_h(i_p,i_m)}
\raisebox{-0.25\height}{\includegraphics[scale=0.5]{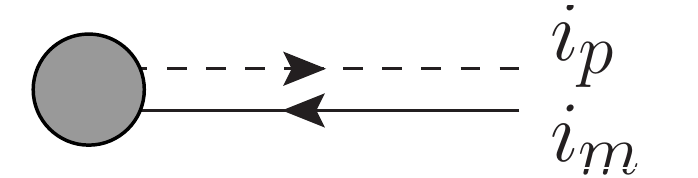}}\;,  \nonumber \\
\bar{\eps}_{h,\be\dal}(p_i,r)
=
\frac{|i_m\rangle[i_p|}{f_h(i_p,i_m)}
=
\frac{1}{f_h(i_p,i_m)}
\raisebox{-0.25\height}{\includegraphics[scale=0.5]{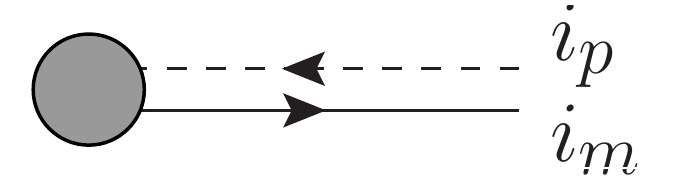}}\;,
\end{align}
where $h=\mp$,
\begin{align}
i_p\text{ and }i_m
=
\left\{
\begin{matrix}
r\text{ and }i\,,\;h=-\\
i\text{ and }r\,,\;h=+
\end{matrix}
\right.
\;,
\qquad \quad
f_h(i_p,i_m)
=
\left\{
\begin{matrix}
[i_mi_p]=[ir]\,,\ \ h=-\\
\langle i_mi_p\rangle=\langle ri\rangle\,,\;h=+
\end{matrix}
\right.
\;,
\end{align}
and where $i_p$ ($i_m$) denotes the positive- (negative-)helicity spinor in gluon $i$.

For this example we consider the partial helicity amplitude $M(0\to q^+ 1\, 2\, \bar{q}^-)$, multiplying the color factor $t^1_{qi}t^2_{i\bar{q}}$. 
This means that we only consider the two diagrams in \figRef{fig:qqBgg diagrams}. 
For the Lorentz structure of the diagram in \figRef{fig:qqBgg QED} we get
\begin{align}
  \raisebox{-0.45\height}{\includegraphics[scale=0.4]{./Jaxodraw/qqBggQEDFeyn}}
  &\rightarrow \frac{-ig_s^2}{s_{q1}f_{h,1}f_{h,2}} \quad \raisebox{-0.45\height}{\includegraphics[scale=0.4]{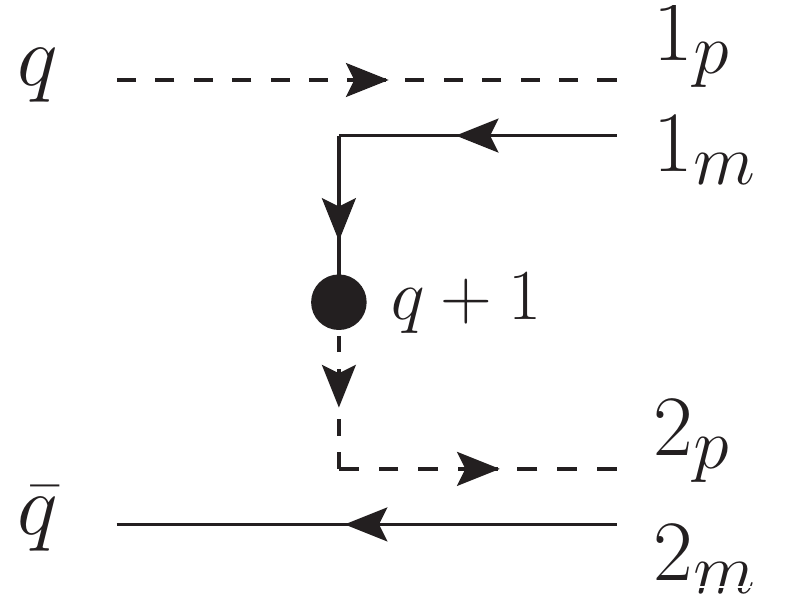}} \quad \nonumber \\
 &=  \frac{-ig_s^2}{s_{q1}f_{h,1}f_{h,2}} 
 \bigg([q 1_p] \Big(\langle1_m q\rangle [q 2_p] + \langle1_m 1\rangle [1 2_p]\Big)\langle 2_m \bar{q} \rangle\bigg)~, \label{eq:qqgg QED flow}
\end{align}
while for the diagram in \figRef{fig:qqBgg 3Glue} we obtain
\begin{align}
  \label{eq:qqgg 3 glue flow}
  \raisebox{-0.45\height}{\includegraphics[scale=0.4]{./Jaxodraw/qqBggTripGlueFeyn}}
  &\rightarrow \frac{ig_s^2}{2s_{12}f_{h,1}f_{h,2}} \left( \quad \raisebox{-0.45\height}{\includegraphics[scale=0.4]{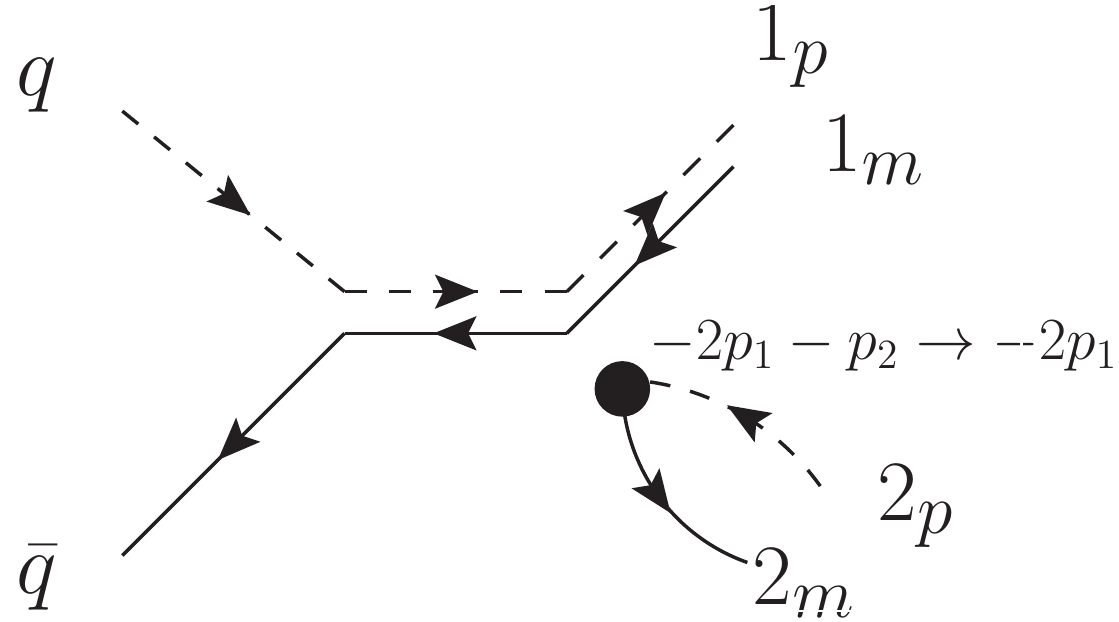}} \right.  \\
  & \hspace{2.5cm}
  +  \raisebox{-0.45\height}{\includegraphics[scale=0.4]{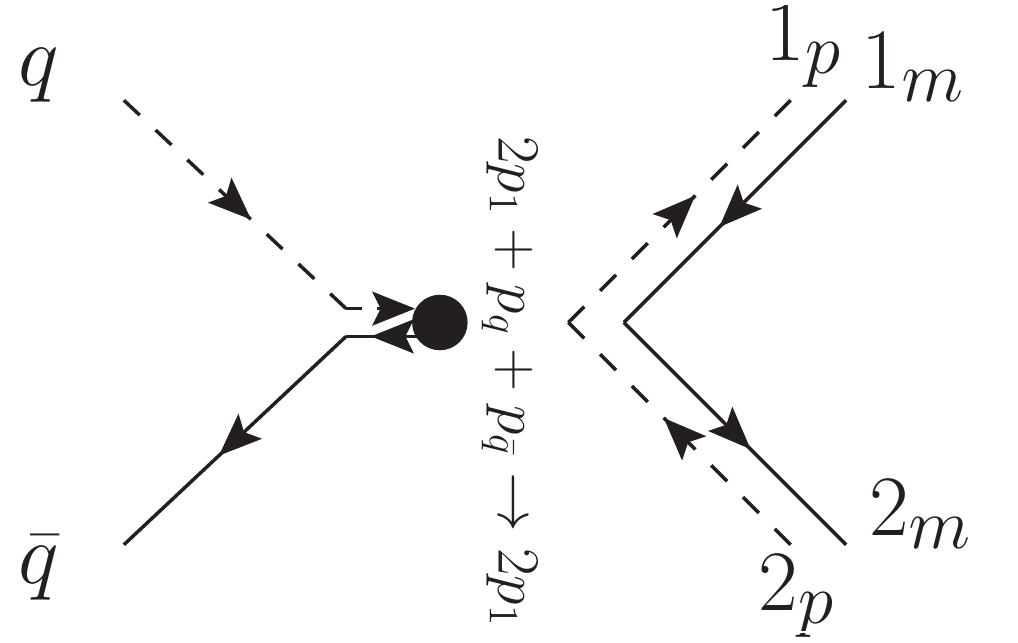}} \quad \left.
  +  \raisebox{-0.45\height}{\includegraphics[scale=0.4]{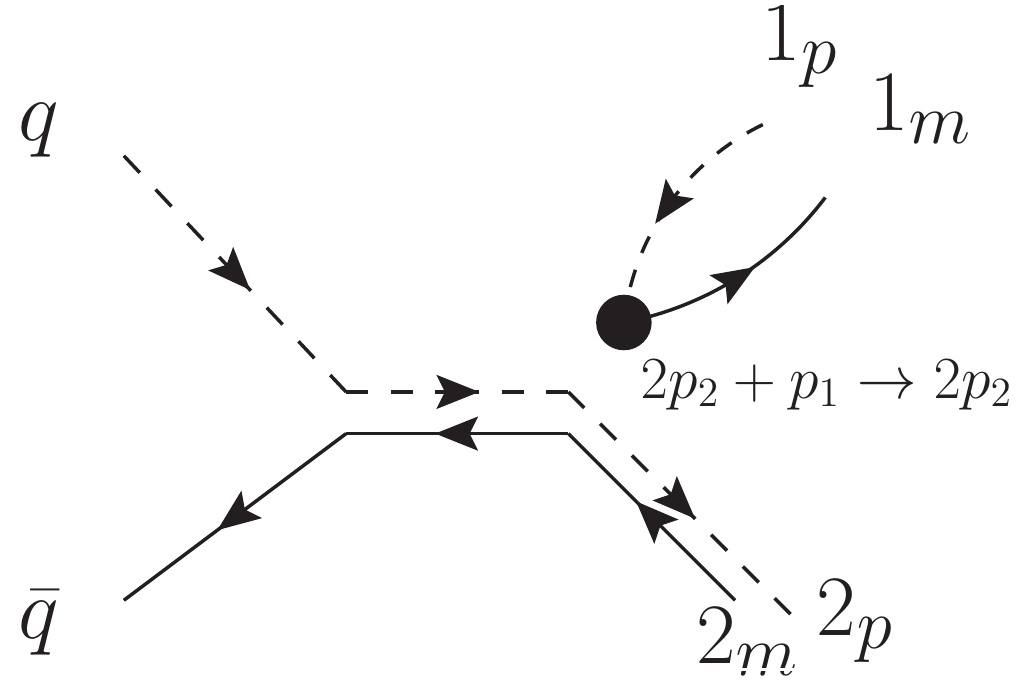}} \right) \nonumber \\
 &=  \frac{ig_s^2}{s_{12}f_{h,1}f_{h,2}} \bigg( -[q 1_p] \langle 1_m \bar{q} \rangle [2_p 1] \langle1 2_m\rangle 
 + \langle 1_m 2_m \rangle [2_p 1_p] [q 1]\langle 1 \bar{q} \rangle \nonumber \\
 & \hspace{2.5cm} + [q 2_p] \langle 2_m \bar{q} \rangle  [1_p 2]\langle 2 1_m \rangle \bigg)\,, \nonumber   
\end{align}
where we again used momentum conservation to rewrite all momenta in the triple-gluon vertex 
$p_a-p_b = 2p_a +p_c = -2p_b - p_c$.
Since $p_c=p_2$ in the first flow diagram, 
and $p_c=p_1$ in the last diagram,
and in each of these either $i_m = i$ or $i_p = i$ for $i=1,2$, 
we could again remove the term with $p_c$.
The full result for a gluon of either helicity is then
\begin{align}
M(0\to q^+ 1\, 2\, \bar{q}^-)
&=\frac{ig_s^2}{f_{h,1}f_{h,2}}\Bigg(\frac{1}{s_{12}} \bigg[ -[q 1_p] \langle 1_m \bar{q} \rangle [2_p 1] \langle 1 2_m\rangle + \langle 1_m 2_m \rangle [2_p 1_p] [q 1]\langle 1 \bar{q} \rangle  \\
  & \hspace{2.9cm} + [q 2_p] \langle 2_m \bar{q} \rangle [1_p 2]\langle 2 1_m \rangle  \bigg]- \frac{1}{s_{q1}} [q 1_p] \Big(\langle1_m q\rangle [q 2_p] + \langle1_m 1\rangle [1 2_p]\Big)\langle 2_m \bar{q} \rangle \Bigg) ~. \nonumber
\end{align}
If both gluons have positive helicity, 
then $(1_m,1_p) = (r_1,1)$ and $(2_m,2_p) = (r_2,2)$ such that
choosing $1_m = 2_m = \bar{q}$ and using the antisymmetry 
of the spinor inner products we see that the amplitude vanishes. 
If both gluons have negative helicity, 
choosing $1_p=2_p=q$ we see that the amplitude vanishes. 
Looking at the chirality-flow diagrams in 
\eqsrefa{eq:qqgg QED flow}{eq:qqgg 3 glue flow}, 
this is easy to see from an early stage,
since e.g.\ $q$ always flows to/from either $1_p$ or $2_p$, 
except for when it flows to/from $\bar{q}$,
in which case $2_p$ flows to/from $1_p$.  

The only non-zero case is the MHV case, 
where e.g.\ $h_1 = -$, $h_2 = +$,  $(1_m,1_p) = (1,r_1)$ and $(2_m,2_p) = (r_2,2)$. 
We choose $1_p = q$ and $2_m = \bar{q}$, 
such that the reference momentum of each gluon is equal to the momentum of the quark with opposite helicity,
giving
\begin{align}
M(0\to q^+ 1^- 2^+ \bar{q}^-)
&=\frac{ig_s^2}{[1q]\langle \bar{q} 2 \rangle} \Bigg[ \frac{1}{s_{12}}\bigg[ -0 + \langle 1 \bar{q} \rangle[2q][q1]\langle 1 \bar{q} \rangle + 0 \bigg] - \frac{1}{s_{q1}}(0)\bigg] \nonumber \\
&
= -\frac{ig_s^2 \langle  \bar{q} 1 \rangle^2[2q]}{\langle \bar{q} 2 \rangle\langle 1 2 \rangle [21]} \frac{\langle \bar{q} q \rangle \langle q 1 \rangle}{\langle \bar{q} q \rangle \langle q 1 \rangle}  
= -\frac{ig_s^2 \langle  \bar{q} 1 \rangle^2 \langle q 1 \rangle (-[21]\langle 1 \bar{q} \rangle) }{\langle q 1 \rangle \langle 1 2 \rangle\langle 2 \bar{q} \rangle  \langle \bar{q} q \rangle  [21]}  \nonumber \\
&
= -\frac{ig_s^2 \langle  \bar{q} 1 \rangle^3 \langle q 1 \rangle }{\langle q 1 \rangle \langle 1 2 \rangle\langle 2 \bar{q} \rangle  \langle \bar{q} q \rangle }~,
\end{align}
which is the standard MHV formula\footnote{
  The minus sign here is often omitted in the literature \cite{Dixon:1996wi},
  such that this MHV QCD amplitude resembles the relevant gluino-gluon amplitude,
  and therefore obeys the supersymmetric Ward identities without additional minus signs
  \cite{Grisaru:1977px,Parke:1985pn,Schwinn:2006ca}.
}.
To obtain this, in the second line, we multiplied by one, 
expanded out $s_{12} = \langle 12 \rangle [21]$ (see \eqref{eq:sij})
and canceled the $[q1]$ brackets, 
before using that $[2q]\langle q \bar{q} \rangle = -[21]\langle 1 \bar{q} \rangle $, due to momentum conservation
(see \eqref{eq:mom conservation}).

\subsection{\texorpdfstring{$gg\to gg$}{gg -> gg}}

\begin{figure}
\centering
\subfloat[\label{fig:4g s}]{\includegraphics[scale=0.4]{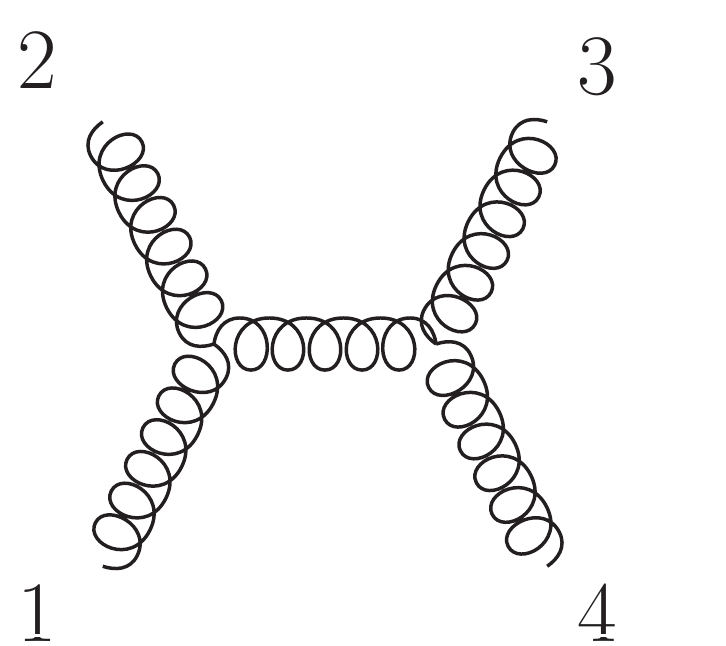}}
\hspace{1cm}
\subfloat[\label{fig:4g t}]{\includegraphics[scale=0.4]{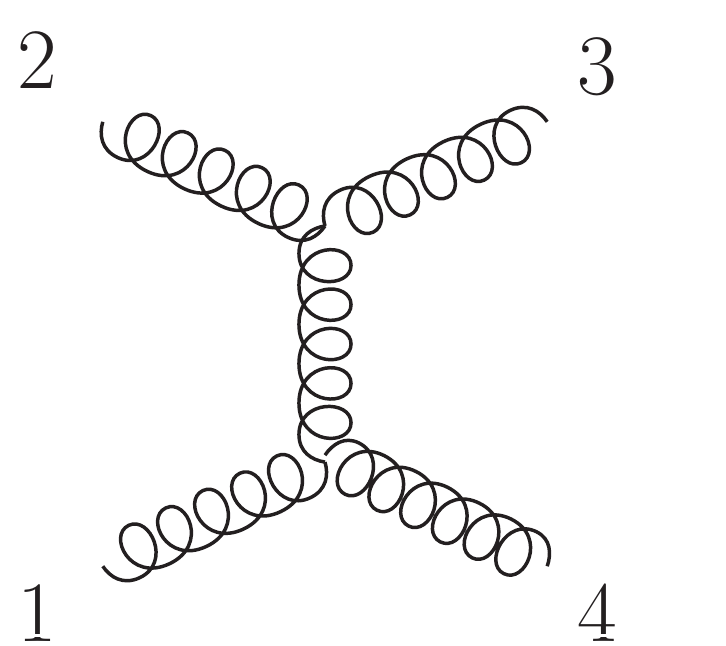}}
\hspace{1cm}
\subfloat[\label{fig:4g contact}]{\includegraphics[scale=0.4]{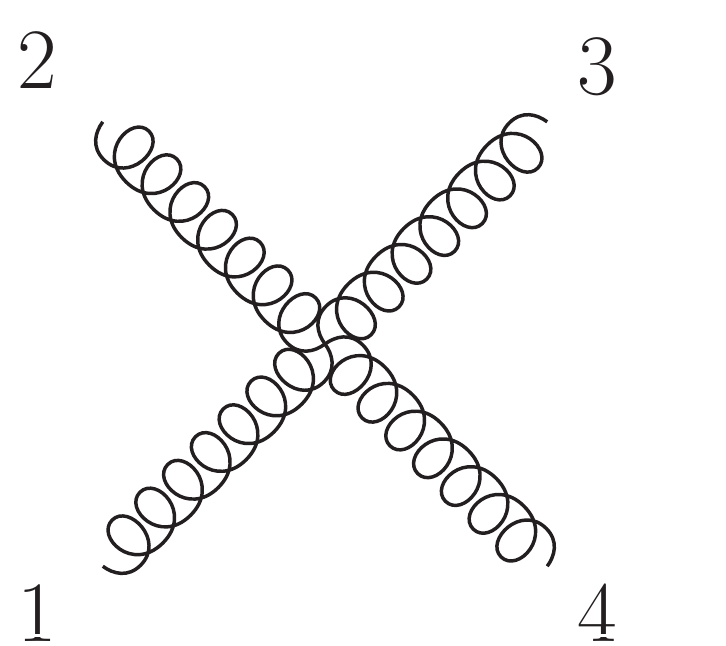}}
\caption{The three diagrams contributing to the 
color-partial helicity amplitude
$M(0\to1^{h_1}2^{h_2}3^{h_3}4^{h_4})$. } 
\label{fig:4g diagrams}
\end{figure}

\noindent In this example, we explore the four-gluon vertex. 
We consider the three diagrams contributing to the partial helicity amplitude 
$M(0\to1^{h_1}2^{h_2}3^{h_3}4^{h_4})$, multiplying the color factor
$\trace\big(t^1t^2t^3t^4\big)$.
For now, we leave the helicities of each particle unassigned.
Steps \ref{step:prefactors} and \ref{step:add lines} give for the s-channel diagram
\begin{align}
\raisebox{-0.45\height}{\includegraphics[scale=0.4]{./Jaxodraw/4gExs}} 
& \rightarrow 
\frac{i g_s^2}{4s_{12}\prod_if_{h,i}} 
\left\lbrace 
\raisebox{-0.45\height}{\includegraphics[scale=0.4]{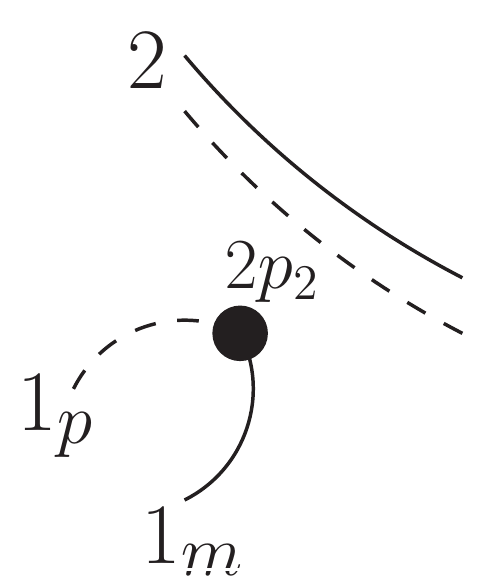}} 
\quad + \quad
\raisebox{-0.45\height}{\includegraphics[scale=0.4]{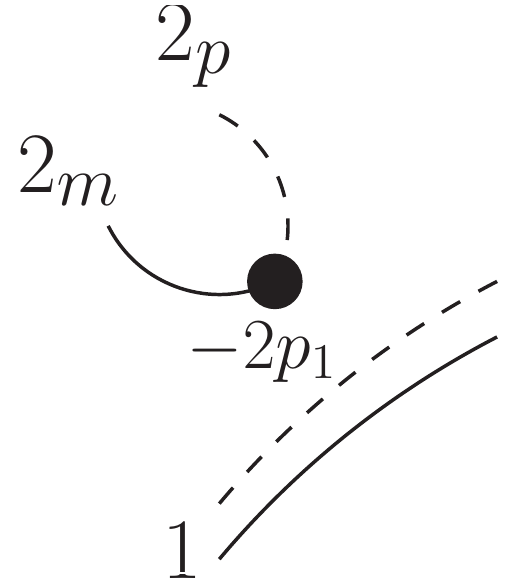}} 
\quad + \quad
\raisebox{-0.45\height}{\includegraphics[scale=0.4]{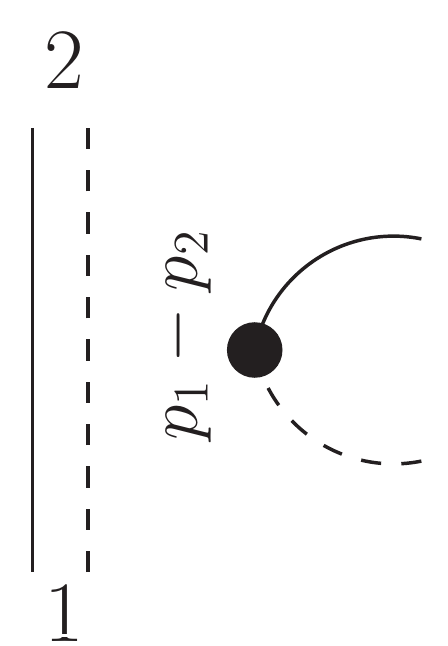}} 
\right\rbrace \nonumber\\
& \quad\quad \;\;\times \quad 
\left\lbrace
\raisebox{-0.45\height}{\includegraphics[scale=0.4]{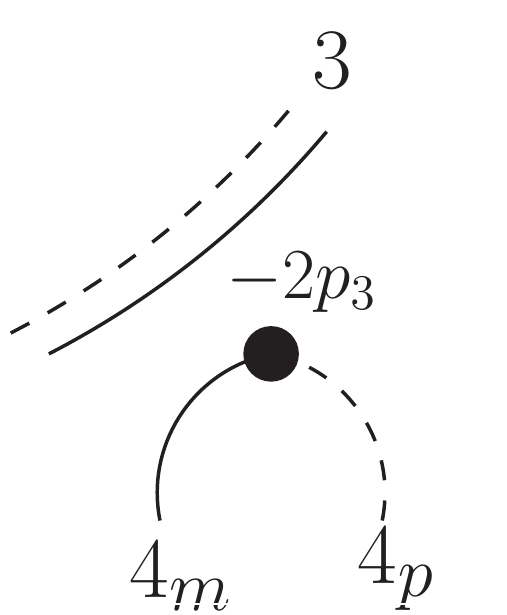}} 
\quad + \quad
\raisebox{-0.45\height}{\includegraphics[scale=0.4]{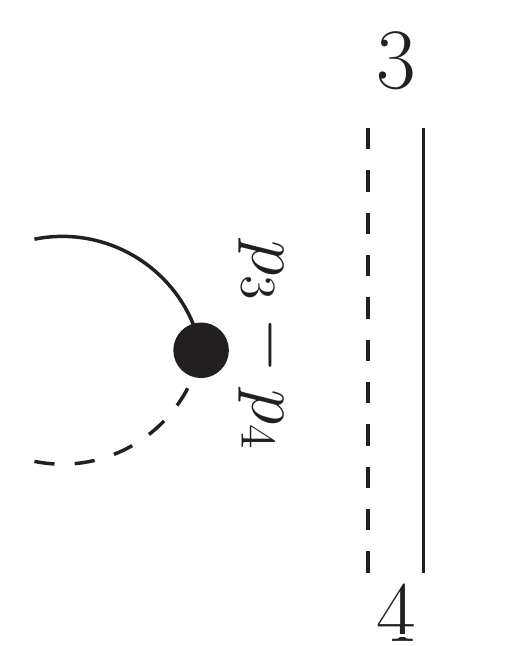}}  
\quad + \quad
\raisebox{-0.45\height}{\includegraphics[scale=0.4]{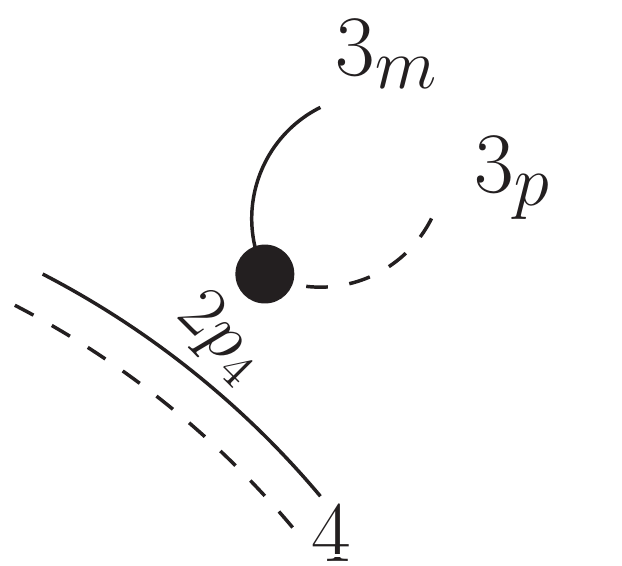}}  
\right\rbrace~,
\end{align}
where we have factorized the two triple-gluon vertices, used the shorthand notation $i=(i_m,i_p)$,
and written the gluon propagator as a double line without a label. 
Stitching together the two factorized terms and assigning arrows results in
\begin{eqnarray}
  \label{eq:4gs}
  \raisebox{-0.45\height}{\includegraphics[scale=0.4]{./Jaxodraw/4gExs}}
  &&\rightarrow
  \frac{i g_s^2}{4s_{12}\prod_if_{h,i}}
  \left(
  \begin{array}{c}
     \\
     \\
     \\
     \\
  \end{array}\right.\\
  & &
  \raisebox{-0.45\height}{\includegraphics[scale=0.4]{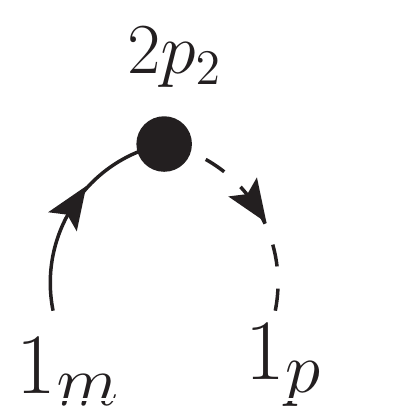}} \left[
    \raisebox{-0.45\height}{\includegraphics[scale=0.4]{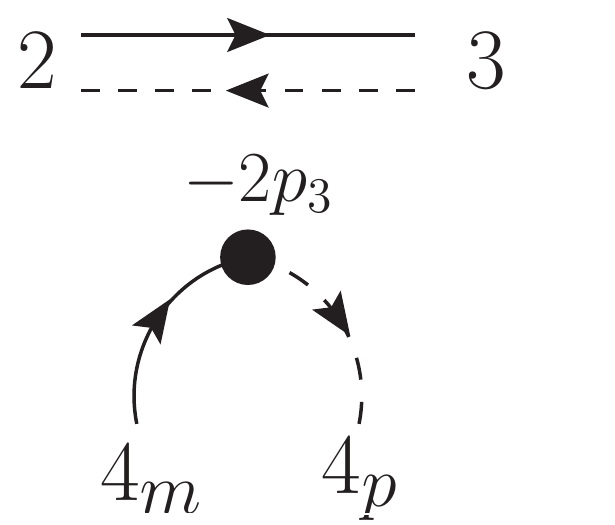}} \quad + \quad
    \raisebox{-0.45\height}{\includegraphics[scale=0.4]{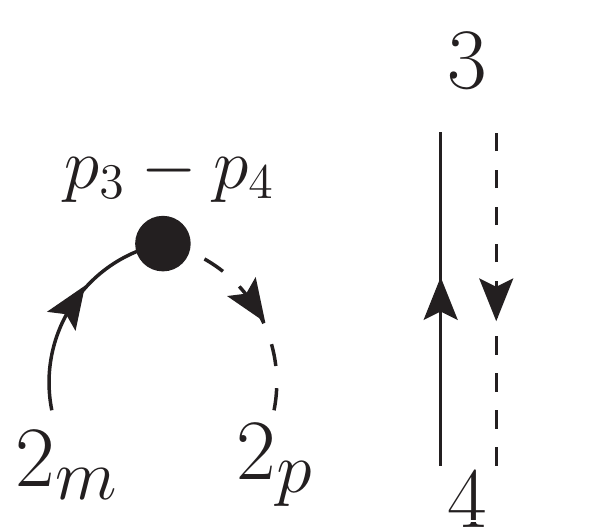}} \quad + \quad
    \raisebox{-0.45\height}{\includegraphics[scale=0.4]{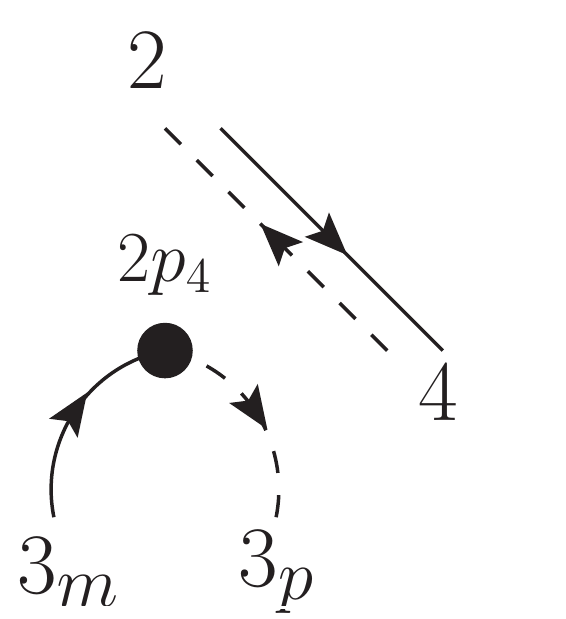}} \right] \nonumber
\end{eqnarray}
\begin{eqnarray}
  \quad\quad\quad\quad\quad\quad\quad\quad\quad\quad\;\;
  & + &
  \raisebox{-0.45\height}{\includegraphics[scale=0.4]{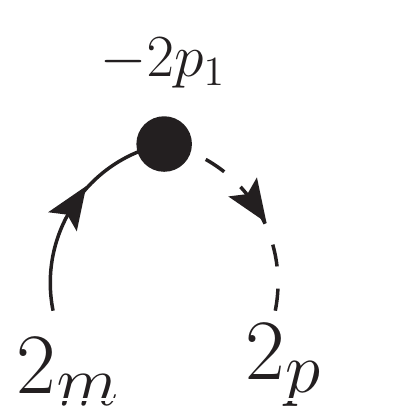}} \left[
    \raisebox{-0.45\height}{\includegraphics[scale=0.4]{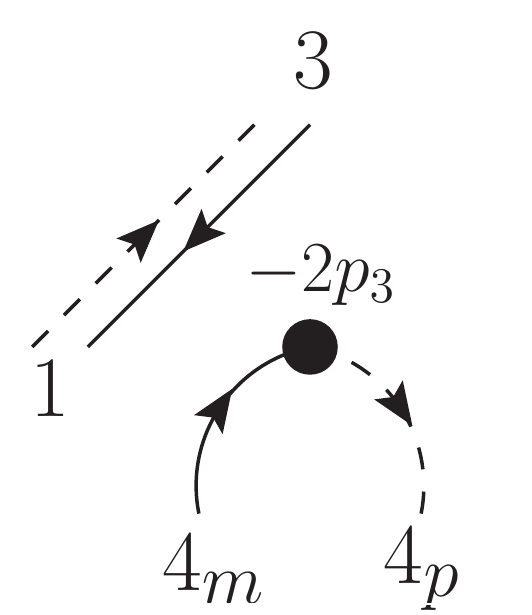}} \quad + \quad
    \raisebox{-0.45\height}{\includegraphics[scale=0.4]{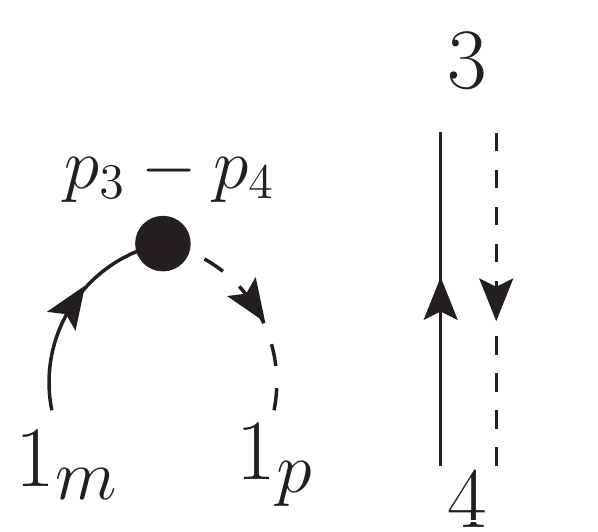}} \quad + \quad
    \raisebox{-0.45\height}{\includegraphics[scale=0.4]{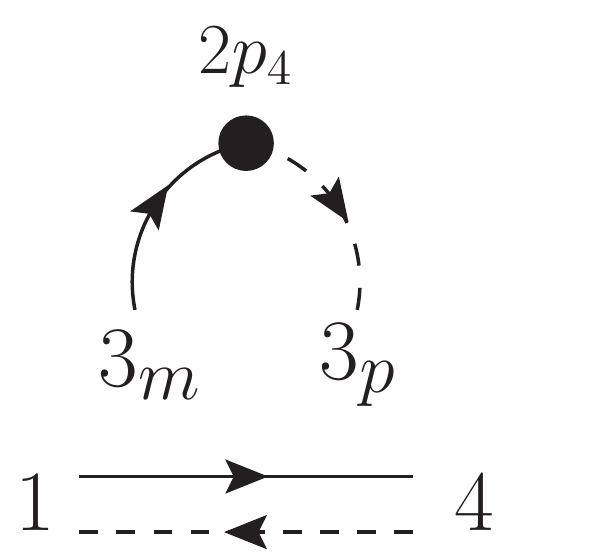}} \right] \nonumber \\
  & + &
  \quad
  \raisebox{-0.45\height}{\includegraphics[scale=0.4]{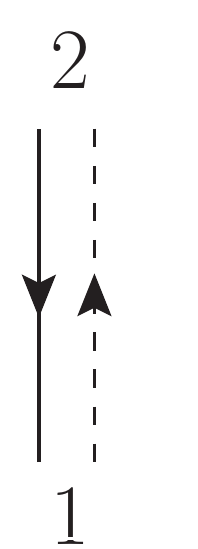}}\quad \left[
    \raisebox{-0.45\height}{\includegraphics[scale=0.4]{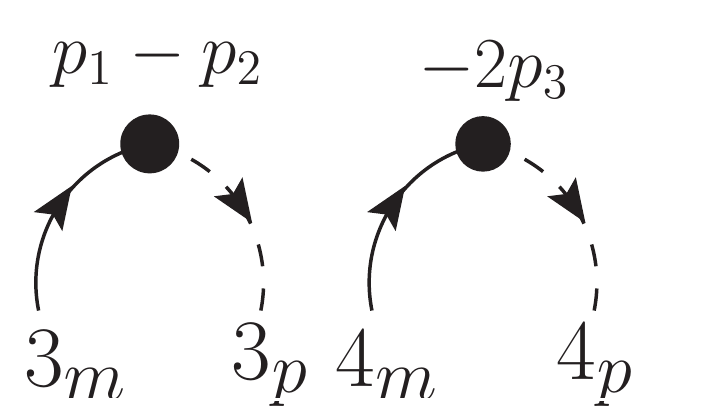}} \quad + \quad
    \raisebox{-0.45\height}{\includegraphics[scale=0.4]{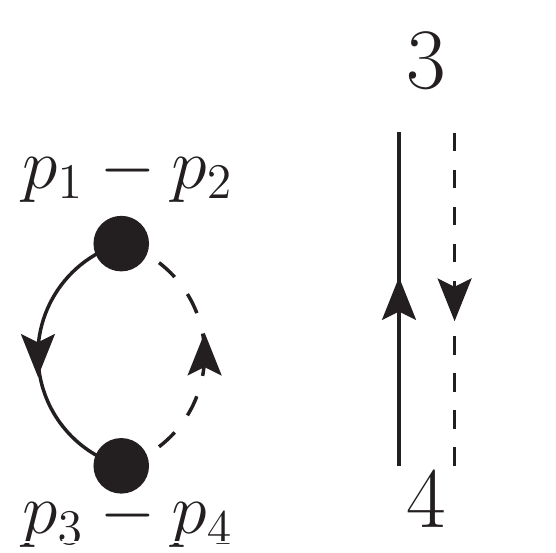}} \quad + \quad
    \raisebox{-0.45\height}{\includegraphics[scale=0.4]{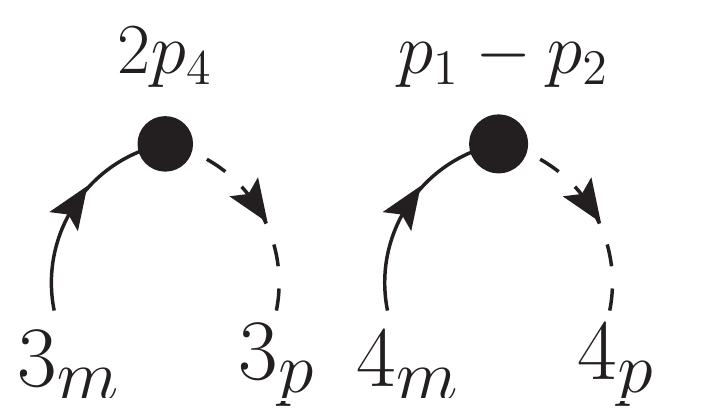}} \right]
  \left.
  \begin{array}{c}
    \\
    \\
    \\
    \\
  \end{array}\right),\nonumber
\end{eqnarray}
where we see a contraction of momenta
\begin{align}
\raisebox{-0.45\height}{\includegraphics[scale=0.4]{./Jaxodraw/pTrace}} 
=2p_i\cdot p_j \overset{p_i^2=p_j^2=0}{=} \langle i j \rangle [ji]~, \label{eq:pi.pj ex}
\end{align}
for the first time.

The t-channel diagram \figRef{fig:4g t} has the same form as the s-channel diagram, 
and can be found by permuting the labels $(1,2,3,4)$ in the cyclic direction, $1\rightarrow 2,\, 2\rightarrow 3$, etc.
What remains to calculate is therefore only the contact diagram, \figRef{fig:4g contact}, for which we directly write
\begin{align}
\raisebox{-0.45\height}{\includegraphics[scale=0.4]{./Jaxodraw/4gExContact}}\rightarrow
\frac{i g_s^2}{2\prod_if_{h,i}}\left[2\raisebox{-0.45\height}{\includegraphics[scale=0.4]{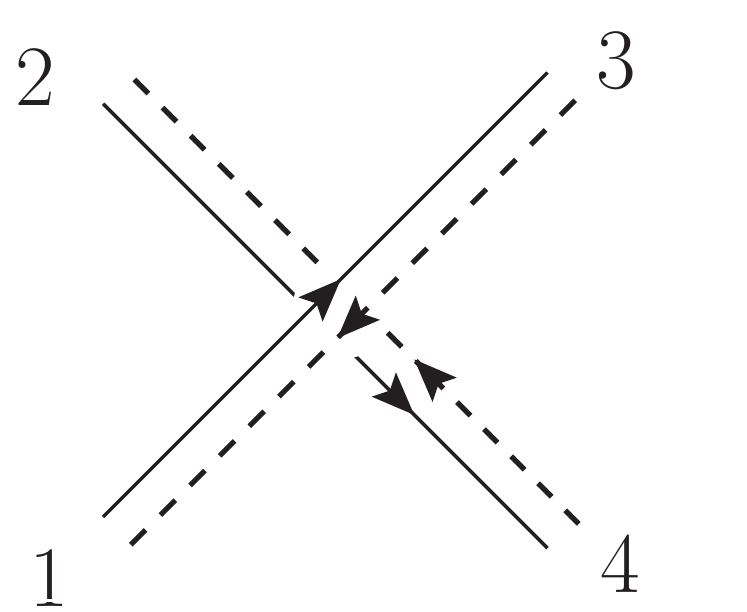}} - \quad
\raisebox{-0.45\height}{\includegraphics[scale=0.4]{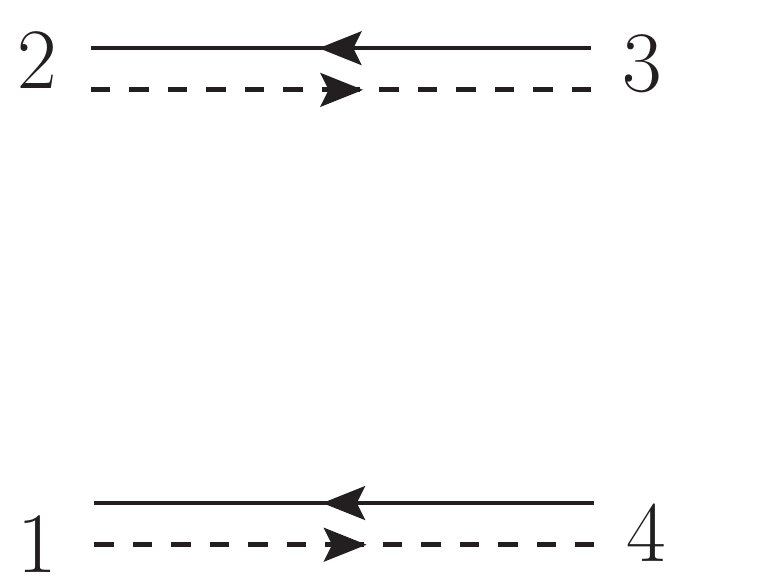}}  - \quad
\raisebox{-0.45\height}{\includegraphics[scale=0.4]{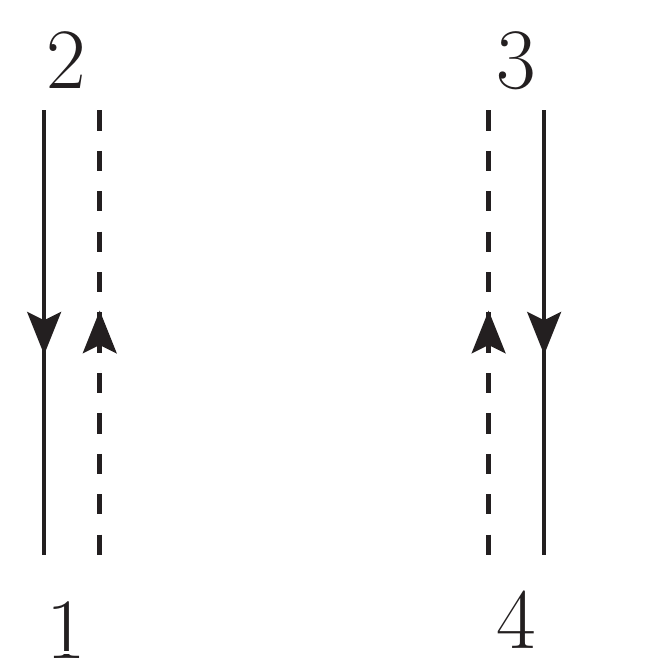}} \right]~.
\end{align}
While the sum of the diagrams still appears complicated,
it contains all of the information needed to obtain the four-gluon scattering
amplitude for any combination of gluon helicities.
For example, if we choose $h_1=h_2=-h_3=-h_4=1$, $r_1 = r_2 = p_4$ and $r_3 = r_4 = p_1$, 
only a single term, the first term in \eqref{eq:4gs},
remains.
The resulting amplitude is then
\begin{align}
M(1^+,2^+,3^-,4^-)&=-ig_s^2\frac{\langle 3 4 \rangle^2[21]^2}{s_{12}\langle 1 4 \rangle [41]} 
= -i g_s^2 \frac{\langle 3 4 \rangle^2[21]}{\langle 1 2 \rangle\langle 1 4 \rangle [41]} \frac{\langle 3 4 \rangle^2}{\langle 3 4 \rangle^2} \nonumber \\
&=i g_s^2 \frac{\langle 3 4 \rangle^4}{\langle 1 2 \rangle  \langle 3 4 \rangle \langle 4 1 \rangle} \frac{[21]}{\langle 3 4 \rangle [41]}
= i g_s^2 \frac{\langle 3 4 \rangle^4}{\langle 1 2 \rangle  \langle 3 4 \rangle \langle 4 1 \rangle} \frac{[21]}{-\langle 3 2 \rangle [21]}\nonumber \\
&= i g_s^2 \frac{\langle 3 4 \rangle^4}{\langle 1 2 \rangle \langle 2 3 \rangle  \langle 3 4 \rangle \langle 4 1 \rangle} ~,
\end{align}
as expected \cite{Parke:1986gb,Berends:1987me}.

\section{Conclusion and outlook}
\label{sec:Conclusion}

In this paper we have presented a new graphical formalism for
calculating massless QED and QCD Feynman diagrams.

In the spinor-helicity formalism, Dirac matrices are abandoned
in favor of the smaller Pauli matrices, and in the Weyl-van-der-Waerden
formalism 
further simplification is obtained by
recasting Feynman rules and diagrams to expressions without Pauli matrices,
which instead depend on the antisymmetric $\epsilon$-tensor, i.e., the SL(2,C)
invariant object.

In this work, we take this one step further, and argue that we can
directly write diagrams, and hence amplitudes, as a combination of spinor inner products.
Using charge conjugation, we argue that we can
formulate a set of chirality-flow Feynman rules for massless QED and
QCD diagrams at tree-level.
After this, there are no algebraic manipulations or matrix equations involved
in the process of finding the spinor inner products appearing in a
scattering amplitude.

Conceptually our method is similar to the idea of a flow of color
in QCD, in the sense that after stripping off the external color/spinor
wave functions, what remains can be thought of as Kronecker contractions
connecting the external color/spinor structures.
We therefore dub our method the chirality-flow formalism. We remark,
however, that the flow picture for color and the flow picture
for chirality can be used completely independently, and in fact
we have given both ordinary Feynman rules and color-flow
Feynman rules for the chirality-flow method.

We also note that the methods differ by using one su(3)-algebra
for color and two (complexified) su(2)-algebras for chirality,
hence we need two different types of lines --- dotted and undotted ---
which can never be contracted with each other, since the
corresponding object would not be Lorentz invariant.

On the more practical side we note that, aside from being more
transparent, the chirality-flow method also shortens actual
calculations by removing a few steps compared to the ordinary
spinor-helicity method. This complete avoidance of matrix structure
may turn out beneficial in automated tools for calculating
amplitudes,
particularly those relying on Feynman diagrams,
e.g. \cite{Honeywell:2018fcl,Cullen:2014yla,Alwall:2014hca}.

Finally, we remark that while the present paper deals only with
massless particles in QED and QCD, the spinor-helicity formalism
for massive particles is well known, and work towards completing
the chirality-flow method for the full standard model is ongoing.

\acknowledgments 
We thank Johan Bijnens and Rikkert Frederix for constructive feedback
on the manuscript.
This work was supported by 
the Swedish Research Council (contract numbers 2012-02744 
and 2016-05996),
as well as 
the European Union's Horizon 2020 research and innovation programme (grant agreement No 668679).
This work has also received funding from the European Union's Horizon 2020 research and innovation programme as part of the Marie Sklodowska-Curie Innovative Training Network MCnetITN3 (grant agreement no. 722104).

\appendix

\section{Conventions and identities}
\label{sec:conventions}

In this appendix, we give some conventions and collect some additional identities,
as well as explicit representations for spinors and four-vectors in the spinor representation.
For convenience,
some of the previous definitions and identities will be repeated in this appendix.

\subsection{Pauli matrices}
\label{sec:apptaumatrices}

We define the Dirac matrices in the chiral, 
or Weyl, 
representation as
\begin{align}
    \gamma^{\mu} 
    = 
    \begin{pmatrix}
      0 & \sigma^{\mu}  \\
      \Bar{\sigma}^{\mu} & 0
    \end{pmatrix}
    =
    \begin{pmatrix}
      0 & \sqrt{2}\tau^{\mu}  \\
      \sqrt{2}\Bar{\tau}^{\mu} & 0
    \end{pmatrix}\;,
\end{align}
and the Pauli matrices as
\begin{align}
& \sigma^\mu
  = (\sigma^0,\vec{\sigma})
  = (\sigma^0,\sigma^1,\sigma^2,\sigma^3)
  =
  \left(
  \begin{pmatrix}
    1 & 0\\
    0 & 1 
  \end{pmatrix}\,,\,
  \begin{pmatrix}
    0 & 1 \\
    1 & 0 
  \end{pmatrix}\,,\,
  \begin{pmatrix}
    0 & -i \\
    i & 0 
  \end{pmatrix}\,,\,
  \begin{pmatrix}
    1 & 0 \\
    0 & -1 
  \end{pmatrix}\right)\;,
  \label{eq:sigma}\\
& \bar{\sigma}^\mu
  = (\sigma^0,-\vec{\sigma})\;,
  \label{eq:sigmabar}
\end{align}
or equivalently the Infeld-van-der-Waerden matrices, 
or $\tau$ matrices, 
as $\tau^\mu=\frac{1}{\sqrt{2}}\sigma^\mu$ and $\taubar^\mu=\frac{1}{\sqrt{2}}\sigmabar^\mu$. 
The elements of the Pauli or $\tau$ matrices are denoted by $\tau^{\mu,\da\be}=\frac{1}{\sqrt{2}}\sigma^{\mu,\da\be}$
and $\taubar^{\mu}_{\al\dbe}=\frac{1}{\sqrt{2}}\sigmabar^{\mu}_{\al\dbe}$. 
The Pauli matrices are Hermitian $2\times2$-matrices, 
i.e. $\big(\sigma^\mu\big)^\dagger=\sigma^\mu$,
and we further have
$
\big(\sigma^\mu\big)^2
=
\big(
\begin{smallmatrix}
1&0\\
0&1 
\end{smallmatrix}
\big)
$, 
for $\mu=0,1,2,3$.
Consequently,
$\big(\tau^\mu\big)^\dagger=\tau^\mu$
and 
$
\big(\tau^\mu\big)^2
=
\frac{1}{2}
\big(
\begin{smallmatrix}
1&0\\
0&1 
\end{smallmatrix}
\big)
$,
for $\mu=0,1,2,3$.
The normalization of the $\tau$ matrices is chosen such that no unnecessary powers of $2$ are carried around in the algebraic relations,
\begin{align}
\label{eq:tautrace0}
\trace\big(\tau^{\mu}\taubar^\nu\big)
&=g^{\mu\nu}\,,\\
\label{eq:tautrace1}
\trace\big(\taubar^\mu\taubar^\nu\big)
&=\delta^{\mu\nu}\,,\\
\label{eq:tautrace2}
\trace\big(\tau^\mu\tau^\nu\big)
&=\delta^{\mu\nu}\,,\\
\label{eq:taufierz0}
\tau_\mu^{\dal\be}\taubar^\mu_{\ga\deta}
&=\delta_{\ga}^{\ \be}\delta_{\ \deta}^{\dal}\,,\\
\label{eq:taufierz1}
\taubar^\mu_{\al\dbe}\taubar_{\mu,\ga\deta}
&=\eps_{\al\ga}\eps_{\dbe\deta}\,,\\
\label{eq:taufierz2}
\tau^{\mu,\dal\be}\tau_{\mu}^{\dga\eta}
&=\eps^{\dal\dga}\eps^{\be\eta}\,,
\end{align}
where 
$g^{\mu\nu}=\mathrm{diag}(1,-1,-1,-1)$ 
denotes the Minkowski metric and $\eps$ the Levi-Civita tensor
(see \eqref{eq:epsilon}).

\subsection{Spinors and spinor inner products}
\label{sec:appspinors}
We recall from \secref{sec:spinors} that the Weyl spinors,
i.e. the two-component spinors in the chiral or Weyl representation,
are solutions to the Weyl equations,
i.e. the decoupled equations for left- and right-chiral two-component spinors that follow from the Dirac equation in the massless case.
We denote the Weyl spinors of massless outgoing 
left-chiral (positive-helicity) fermions of momentum $p$ by $\tla_{\dal}(p)=\tla_{p,\dal}$
and outgoing left-chiral (positive-helicity) antifermions $\tla^{\dal}(p)=\tla_p^{\dal}$,
while right-chiral (negative-helicity) Weyl spinors of massless outgoing 
fermions and antifermions of momentum $p$ are denoted by $\la^{\al}(p)=\la_p^{\al}$
and $\la_{\al}(p)=\la_{p,\al}$ respectively. 

Explicit representations of the (massless) Weyl spinors are for example,
\begin{align}
\label{eq:la}
\la_p^{\al} 
\leftrightarrow
\langle p|
= 
\frac{1}{\sqrt{p^+}}
\begin{pmatrix}
  p^\perp \,,& 
  -p^+ 
\end{pmatrix}
~,\quad\quad 
\la_{p,\al} 
\leftrightarrow
|p\rangle
= 
\frac{1}{\sqrt{p^+}}
\begin{pmatrix}
  p^+ \\
  p^\perp
\end{pmatrix}
~,\\ 
\label{eq:tla}
\tla_{p,\dal} 
\leftrightarrow
[p|
= 
\frac{1}{\sqrt{p^+}}
\begin{pmatrix}
  p^+ \,,&
  p^{\perp^*}
\end{pmatrix}
~,\quad\quad 
\tla_p^{\da} 
\leftrightarrow
|p]
= 
\frac{1}{\sqrt{p^+}}
\begin{pmatrix}
  p^{\perp^*} \\
  -p^+
\end{pmatrix}
~,     
\end{align}
where we used light-cone coordinates\footnote{
  We have $p^+p^-=(p^0+p^3)(p^0-p^3)=(p^0)^2-(p^3)^2$
  and $p^\perp p^{\perp^*}=(p^1+ip^2)(p^1-ip^2)=(p^1)^2+(p^2)^2$.
  Furthermore,
  $p^2=p^+p^--p^\perp p^{\perp^*}$,
  i.e. for $p^2=0$ we have $p^+p^-=p^\perp p^{\perp^*}$. 

},
\begin{equation}
    p^{\pm} = p^0\pm p^3\;, \quad
    p^{\perp} = p^1+ip^2\;, \quad
    p^{\perp^*} 
    = p^1-ip^2\;.
    \label{eq:lightcoord_app}
\end{equation}

The Weyl spinors are related to each other by Hermitian conjugation,
\begin{equation}
  |p\rangle^{\dagger}
  =[p|
  \quad\mbox{and}\quad
  |p]^{\dagger}
  =\langle p|\;,
  \label{eq:spinorhermitian}
\end{equation}
or in case of the components $(\la_{p,\al})^*=(\tla_{p,\dal})$ and $(\tla_p^{\da})^*=(\la_p^{\al})$ for $\al=\dal=1,2$,
which is easily confirmed for the explicit representations in 
\eqsrefa{eq:la}{eq:tla}.

We recall from \eqref{eq:epsilon_action2} that spinor indices of two-component Weyl spinors are raised and lowered by the Levi-Civita tensor\footnote{
  We recall from \eqref{eq:epsilon} that the Levi-Civita or $\epsilon$ tensor is defined as
  $\epsilon^{12}=-\epsilon^{21}=\epsilon_{21}=-\epsilon_{12}=1$,
  or in matrix notation as
  $\eps^{\al\be}=\eps^{\dal\dbe}\leftrightarrow \big(\begin{smallmatrix}0&1\\-1&0\end{smallmatrix}\big)=i\sigma^2$ and $\eps_{\al\be}=\eps_{\dal\dbe}\leftrightarrow\big(\begin{smallmatrix}0&-1\\1&0\end{smallmatrix}\big)=-i\sigma^2$.
  With our definition of the $\eps$-tensor we have that 
  $\eps_{\al\be}\eps^{\be\ga}\!=\!\delta_{\al}^{\ \ga}$ and 
  $\eps^{\dal\dbe}\eps_{\dbe\dga}\!=\!\delta^{\dal}_{\ \dga}$,
  and that e.g. 
  $\epsilon_{\al\be}\la_p^{\be}=-\epsilon_{\be\al}\la_p^{\be}=-\la_p^{\be}\epsilon_{\be\al}$, etc.
},
\begin{equation}
\la_{p,\al} = \epsilon_{\al\be}\la_p^{\be}~, 
\quad 
\tla_{p,\dal} = \epsilon_{\dal\dbe}\tla_p^{\dbe}~,
\quad
\la_p^{\al} = \epsilon^{\al\be}\la_{p,\be}~, 
\quad 
\tla_p^{\dal} = \epsilon^{\dal\dbe}\tla_{p,\dbe}~,
\label{eq:epsilon_action_app}  
\end{equation}
as is easily seen for the representations in
\eqsrefa{eq:la}{eq:tla} using \eqref{eq:epsilon}.

We also recall the antisymmetric Lorentz invariant spinor inner products,
\eqsrefa{eq:spinorproduct1}{eq:spinorproduct2}, 
\begin{align}
  \langle i j \rangle &
  = \la_i^{\al}\la_{j,\al}
  = \epsilon^{\al \be}\la_{i,\be}\la_{j,\al}
  = -\epsilon^{\be \al}\la_{i,\be}\la_{j,\al} = -\la_{i,\be}\la_j^{\be} = -\langle j i \rangle~, 
  \label{eq:lambdaproduct3app}
  \\
  [ i j ] &
  = \tla_{i,\da}\tla_j^{\da}
  = \epsilon_{\da \db}\tla_i^{\db}\tla_j^{\da}
  = -\epsilon_{\db \da}\tla_i^{\db}\tla_j^{\da}
  = - \tla_i^{\db}\tla_{j,\db} = -[ji]~.
  \label{eq:lambdaproduct4app}
\end{align}
Using the explicit representations of the Weyl spinors in
\eqsrefa{eq:la}{eq:tla},
we have
\begin{equation}
\langle i j \rangle 
= 
\la_i^{\al}\la_{j,\al} 
= 
\frac{1}{\sqrt{p_i^+p_j^+}}\big(p_i^\perp p_j^+ - p_j^\perp p_i^+\big)
~~,\quad 
[i j] 
= 
\tla_{i\da}\tla_j^{\da} 
= 
\frac{1}{\sqrt{p_i^+p_j^+}}\big(p_i^+ p_j^{\perp^*} - p_j^+p_i^{\perp^*}\big)~~,
\label{eq:explicitproducts}
\end{equation}
consistent with
$[ii]=\langle jj\rangle=0$ 
and 
$\langle ij\rangle^*=-[ij]=[ji]$.

We also state the Schouten identities 
\begin{align} 
\label{eq:Schouten1}
&
\langle ij\rangle\langle kl\rangle 
+
\langle ik\rangle\langle lj\rangle 
+ 
\langle il\rangle\langle jk\rangle 
=
\langle i|
\Big(
|j\rangle\langle kl\rangle
+
|k\rangle\langle lj\rangle
+
|l\rangle\langle jk\rangle
\Big)
=
0~,
\\ 
\label{eq:Schouten2}
&
\,
[ij]\,[kl]
\,+\,
[ik]\,[lj] 
\,+\, 
[il]\,[jk] 
\,=
[i|
\Big(
|j]\,[kl]
\,+
|k]\,[lj] 
\,+ 
|l]\,[jk]
\,
\Big)
=
0~,
\end{align}
which are a consequence of the fact that any three two-component spinors are linearly dependent.

For the spinor transformations, we use the common notation $(j_L,j_R)$ for
representations of the Lorentz algebra, corresponding to the representations of the mutually commuting generator combinations
\begin{eqnarray}
  \vec{N}^L&=&\frac{1}{2}(\vec{J}-i\vec{K})\;, \\
  \vec{N}^R&=&\frac{1}{2}(\vec{J}+i\vec{K})\;,
\end{eqnarray}
where $\vec{J}$ and $\vec{K}$ denote the generators of rotations and boosts respectively and
where
\begin{eqnarray}
  [ \vec{N}^{L,i} ,\vec{N}^{R,j}]=0\;.
\end{eqnarray}
For left-chiral spinors $\tla_p^{\da}$ in the $(j_L,j_R)=(1/2,0)$ representation, 
we have $\vec{J}=-i\vec{K}=\vec{\sigma}/2$,
i.e. $\vec{N}^L=\vec\sigma/2$ and $\vec{N}^R=0$.
Under Lorentz transformations they transform as
\begin{equation}
  \tla_p^{\da} \rightarrow (\Lambda_L)^{\da}_{~\db}\tla_p^{\db}
  \quad \text{with}\quad
  \Lambda_L=e^{(-i\vec\theta+\vec\eta)\cdot\vec\sigma/2}~,
\end{equation}
where  $\vec\theta$ and $\vec\eta$ denote rotation angles and boost parameters.
For right-chiral spinors $\la_{p,\be}$ in the $(j_L,j_R)=(0,1/2)$ representation, we similarly have $\vec{J}=i\vec{K}=\vec{\sigma}/2$,
  i.e. $\vec{N}^L=0$ and $\vec{N}^R=\vec\sigma/2$.
  Under Lorentz transformations they change as
\begin{equation}  
  \la_{p,\be}\rightarrow (\Lambda_R)^{~\al}_{\be} \la_{p,\al} 
  \quad \text{with} \quad
  \Lambda_R=e^{(-i\vec\theta-\vec\eta)\cdot\vec\sigma/2}.
\end{equation}

\subsection{Four-vectors and bispinors}
\label{sec:appfourvectors}

We recall from \secref{sec:fourvectors},
that using the $\tau$ matrices, any four-vector $p_\mu$ can be mapped to Hermitian $2\!\times\!2$-matrices, 
or bispinors,
\begin{align}
  \label{eq:p_contr2_app}
  p_{\mu}\tau^{\mu,\dal\be} 
  =
  p^{\dal\be}
  \quad\leftrightarrow\quad
  p_{\mu}\tau^{\mu} 
  =
  \frac{1}{\sqrt{2}}
  \slashed{p}
  &=
  \frac{1}{\sqrt{2}}
  \begin{pmatrix}
  p^0-p^3  & ip^2-p^1 \\
  -ip^2-p^1 & p^0+p^3
  \end{pmatrix}
  =
  \frac{1}{\sqrt{2}}
  \begin{pmatrix}
  p^-  & -p^{\perp^*} \\
  -p^{\perp} & p^+
  \end{pmatrix}~,
  \\ 
  p_{\mu} \Bar{\tau}^{\mu}_{\al\dbe} 
  = 
  \bar{p}_{\al\dbe}
  \quad\leftrightarrow\quad
  p_{\mu} \Bar{\tau}^{\mu} 
  =
  \frac{1}{\sqrt{2}}
  \bar{\slashed{p}}
  &=
  \frac{1}{\sqrt{2}}
  \begin{pmatrix}
  p^0+p^3   & p^1-ip^2 \\
  p^1+ip^2 & p^0-p^3
  \end{pmatrix}
  =
  \frac{1}{\sqrt{2}}
  \begin{pmatrix}
  p^+   & p^{\perp^*} \\
  p^{\perp} & p^-
  \end{pmatrix}~,
  \label{eq:p_contr3_app}
\end{align}
for which we have again used the light-cone coordinates in \eqref{eq:lightcoord_app}\footnote{
  We repeat, 
  the slash notation here is not to be confused with the Feynman slash notation,
  denoting contractions with Dirac matrices. 
}.

Raising and lowering spinor indices on the $\tau$ matrices is done by
\begin{align}
  \taubar^\mu_{\al\dbe} 
  = 
  \epsilon_{\al\ga}\epsilon_{\dbe\deta}\tau^{\mu,\deta\ga}~,
  \quad
  \tau^{\mu,\dal\be} 
  = 
  \epsilon^{\dal\dga}\epsilon^{\be\eta}\taubar^\mu_{\eta\dga}~,
  \label{eq:tau taubar}
\end{align}
which is easily confirmed for the explicit matrix representations in 
\eqsrefa{eq:sigma}{eq:sigmabar}.
The raising and lowering of spinor indices on a four-vector $p_\mu$ in the bispinor representation 
follows by contracting the above with $p_\mu$,
\begin{align}
  \Bar{p}_{\al\dbe} 
  = 
  \epsilon_{\al\ga}\epsilon_{\dbe\deta}p^{\deta\ga}~,
  \quad
  p^{\dal\be} 
  = 
  \epsilon^{\dal\dga}\epsilon^{\be\eta}\Bar{p}_{\eta\dga}~,
\end{align}
which may easily be confirmed for the explicit representations in 
\eqsrefa{eq:p_contr2_app}{eq:p_contr3_app}.

We recall from \secref{sec:fourvectors}
that if a momentum $p$ is massless,
i.e. light-like, the corresponding momentum bispinors can be expressed as
outer products, or dyads, of Weyl spinors,
\begin{align}
    \slashed{p} 
    = 
    |p]\langle p| 
    &\quad\mathrm{or}\quad
    \sqrt{2}p^{\da\be} 
    =
    \tla_p^{\da}\la_p^{\be}~,
    \quad p^2=0~,
    \label{eq:slashed p app}
    \\
    \bar{\slashed{p}} 
    = 
    |p\rangle [p|
    &\quad\mathrm{or}\quad
    \sqrt{2}\bar{p}_{\al\db} 
    = 
    \la_{p,\al}\tla_{p,\db}~,
    \quad p^2=0~.
    \label{eq:slashed pbar app}
\end{align}
This is easily confirmed for the explicit representations of the Weyl spinors in 
\eqsrefa{eq:la}{eq:tla},
comparing to the explicit representation of the momentum bispinors in 
\eqsrefa{eq:p_contr2_app}{eq:p_contr3_app},
and using the conditions for the light-cone coordinates in the massless case,
i.e. $p^+p^-=p^\perp p^{\perp^*}$ if $p^2=0$.
If a momentum $p$ is expressed as a linear combination of light-like momenta $p_i$,
then
\begin{align}
    \slashed{p} 
    = 
    \sum\limits_i
    c_i
    |p_i]\langle p_i| 
    &\quad\mathrm{or}\quad
    \sqrt{2}p^{\da\be} 
    =
    \sum\limits_i
    c_i
    \tla_{p_i}^{\da}\la_{p_i}^{\be}~
    \quad \mathrm{for}\quad p=\sum_ic_ip_i~
    \;\;\text{and}
    \quad p_i^2=0~,
    \\
    \bar{\slashed{p}} 
    = 
    \sum\limits_i
    c_i
    |p_i\rangle [p_i|
    &\quad\mathrm{or}\quad
    \sqrt{2}\bar{p}_{\al\db} 
    = 
    \sum\limits_i
    c_i
    \la_{p_i,\al}\tla_{p_i,\db}~
    \quad \mathrm{for} \quad p=\sum_ic_ip_i~
    \;\;\text{and}
    \quad p_i^2=0~.
\end{align}

With the above,
and the properties of the spinor inner products,
it is easy to see that the Weyl spinors obey the Weyl equations,
\begin{align}
  p^{\mu}\sigma_{\mu}^{\da\be}\la_{p,\be} 
  = 
  \sqrt{2}
  p^{\da\be}\la_{p,\be} 
  \overset{\scriptscriptstyle p^2=0}{=}
  \tla_p^\da\la_p^\be\la_{p,\be} 
  = 0~,
  \quad&\quad
  \la_{p}^{\al}p^{\mu}\sigmabar_{\mu,\al\dbe} 
  =
  \sqrt{2}
  \la_{p}^{\al}\bar{p}_{\al\dbe}
  \overset{\scriptscriptstyle p^2=0}{=}
  \la_{p}^{\al}\la_{p,\al}\tla_{p,\dbe}
  = 0~,
  \\
  p^{\mu}\sigmabar_{\mu,\al\dbe}\tla_{p}^{\dbe} 
  = 
  \sqrt{2}
  \bar{p}_{\al\dbe}\tla_{p}^{\dbe} 
  \overset{\scriptscriptstyle p^2=0}{=}
  \la_{p,\al}\tla_{p,\dbe}\tla_{p}^{\dbe} 
  = 0~,
  \quad&\quad   
  \tla_{p,\da}p^{\mu}\sigma_{\mu}^{\da\be}
  = 
  \sqrt{2}
  \tla_{p,\da}p^{\da\be}
  \overset{\scriptscriptstyle p^2=0}{=}
  \tla_{p,\da}\tla_p^\da\la_p^\be
  = 0~,
\end{align}
or
\begin{align}
  \slashed{p}|p\rangle
  \overset{\scriptscriptstyle p^2=0}{=}
  \big(|p]\langle p|\big)|p\rangle
  = |p]\langle pp\rangle
  = 0~,
  \quad&\quad
  \langle p|\bar{\slashed{p}} 
  \overset{\scriptscriptstyle p^2=0}{=}
  \langle p|\big(|p\rangle [p|\big)
  = \langle pp\rangle [p|
  = 0~,
  \\
  \bar{\slashed{p}}|p] 
  \overset{\scriptscriptstyle p^2=0}{=}
  \big(|p\rangle [p|\big)|p]
  = |p\rangle [pp]
  = 0~,
  \quad&\quad   
  [p|\slashed{p}
  \overset{\scriptscriptstyle p^2=0}{=}
  [p|\big(|p]\langle p|\big)
  = [pp]\langle p|
  = 0~,
\end{align}
and that for massless particles $i$ and $j$, 
with $s_{ij}=(p_i+p_j)^2=2p_i.p_j$,
the spinor inner products satisfy 
\begin{align}
s_{ij}
=2p_i.p_j
=2p_{i,\mu}p_{j,\nu}\trace\big(\tau^\mu\taubar^\nu\big)
=\sqrt{2}p_i^{\dal\be}\sqrt{2}\bar{p}_{j,\be\dal}
=\tla_i^{\dal}\la_i^\beta\la_{j,\beta}\tla_{j,\dal}
=\la_i^\beta\la_{j,\beta}\tla_{j,\dal}\tla_i^{\dal}
=\langle ij\rangle[ji]~.
\label{eq:sij}
\end{align}

Using 
\eqsrefa{eq:slashed p app}{eq:slashed pbar app}, 
we have
\begin{equation} 
[k | \slashed{p}_i | l\rangle 
= [k i] \langle i l \rangle ~,
\quad\quad 
\langle k | \bar{\slashed{p}}_i | l] 
= \langle k i \rangle [i l]  
\quad\text{for}\quad 
p_i^2=0 ~.
\end{equation}

Assuming a set of outgoing massless external momenta $p_i$,
and using momentum conservation,
i.e. $\sum_i p_i=0$, 
gives
\begin{equation} \label{eq:mom conservation}
\sum_i [j i] \langle i k \rangle  = \sum_{i \neq j,k} [j i] \langle i k \rangle = 0~, \quad \quad \sum_i \langle j i \rangle [i k] = \sum_{i \neq j,k} \langle j i \rangle [i k] = 0~.
\end{equation}
Using $\la_i^{\al}\taubar^{\mu}_{\al\dbe}\tla_j^{\dbe} = \tla_{j,\dde}\tau^{\mu,\dde\ga}\la_{i,\ga}$ (see \eqref{eq:charge conj current indices}),
or equivalently $\langle i|\taubar^{\mu}|j]=[j|\tau^{\mu}|i\rangle$ (see \eqref{eq:charge conj current matrix}),
and $\tau_\mu^{\dal\be}\taubar^\mu_{\ga\deta}=\delta_{\ga}^{\ \be}\delta_{\ \deta}^{\dal}$ (see \eqref{eq:Fierz Rearrangement Tau}), 
this implies
\begin{align} \label{eq:contract spinor strings}
[i | \tau^{\mu} | j  \rangle [k | \tau_{\mu} | l \rangle  =  [k i] \langle j l \rangle ~, 
\quad \quad \langle i | \taubar^{\mu} | j] \langle k | \taubar_{\mu} | l]  =   [j l] \langle k i \rangle~,
 \quad \quad   [i | \tau^{\mu} | j  \rangle \langle k | \taubar_{\mu} | l] = [il]\langle k j\rangle ~.
\end{align}

For completeness, 
we also repeat the expressions for the polarization vectors of massless vector bosons from \secref{sec:polarizationvectors}.
The polarization vectors can be written in terms of Weyl spinors and $\tau$ matrices,
\begin{align}
  \epsilon_-^{\mu}(p_i,r)
  = 
  \frac{\la_i^{\al}\taubar^{\mu}_{\al\db}\tla_r^{\db}}{\tla_{i,\dga} \tla_{r}^{\dga}}
  = 
  \frac{\langle i|\taubar^{\mu}|r]}{[ir]}\;\;,~~
  &~~
  \epsilon_+^{\mu}(p_i,r)
  = 
  \frac{\la_r^{\al}\taubar^{\mu}_{\al\db}\tla_i^{\db}}{\la_r^{\ga} \la_{i,\ga}}
  = 
  \frac{\langle r|\taubar^{\mu}|i]}{\langle ri\rangle}\;\;,
  \label{eq:apppolvec1}\\
  \epsilon_-^{\mu}(p_i,r)
  = 
  \frac{\tla_{r,\da}\tau^{\mu,\da\be}\la_{i,\be}}{\tla_{i,\dga} \tla_r^{\dga}}
  = 
  \frac{[r|\tau^{\mu}|i\rangle}{[ir]}\;\;,~~
  &~~
  \epsilon_+^{\mu}(p_i,r)
  = 
  \frac{\tla_{i,\da}\tau^{\mu,\da\be}\la_{r,\be}}{\la_r^\ga \la_{i,\ga}}
  = \frac{[i|\tau^{\mu}|r\rangle}{\langle ri\rangle}\;\;,
  \label{eq:apppolvec2}
\end{align}
where $p_i$ is the vector boson momentum and $r$ is an arbitrary, 
light-like reference momentum satisfying $p_i\cdot r \neq 0$.
To get from \eqref{eq:apppolvec1} to (\ref{eq:apppolvec2}) we have used 
$\la_i^{\al}\taubar^{\mu}_{\al\dbe}\tla_j^{\dbe} = \tla_{j,\dde}\tau^{\mu,\dde\ga}\la_{i,\ga}$ (see \eqref{eq:charge conj current indices}).
With the above, and the properties of massless spinors and bispinors,
as stated before, 
it is easily confirmed that $\big(\epsilon_-^{\mu}(p_i,r)\big)^*=\epsilon_+^{\mu}(p_i,r)\,$,
as well as $\eps_\pm^\mu(p_i,r)p_{i,\mu}=0$ and $\eps_\pm^\mu(p_i,r)r_\mu=0$.
Contracting \eqref{eq:apppolvec1} with $\tau$, 
\eqref{eq:apppolvec2} with $\taubar$, 
and using \eqref{eq:taufierz0},
we may write the polarization vectors in the bispinor representation
in terms of outer products, 
or dyads, 
of Weyl spinors\footnote{
  We could have equally well contracted \eqref{eq:apppolvec1} with $\taubar$ and \eqref{eq:apppolvec2} with $\tau$,
  using \eqsrefa{eq:taufierz1}{eq:taufierz2}, respectively.
},
\begin{align}
  \eps_{-}^{\dbe\al}(p_i,r)
  =
  \eps_{-}^\mu(p_i,r)\tau_\mu^{\dbe\al} 
  =
  \frac{\tla_r^{\db}\la_i^{\al}}{\tla_{i,\dga} \tla_{r}^{\dga}}
  = 
  \frac{|r]\langle i|}{[ir]}\;\,,&\,\;\;
  \eps_{+}^{\dbe\al}(p_i,r)
  =
  \eps_{+}^\mu(p_i,r)\tau_\mu^{\dbe\al} 
  =
  \frac{\tla_i^{\db}\la_r^{\al}}{\la_r^{\ga} \la_{i,\ga}}
  =
  \frac{|i]\langle r|}{\langle ri\rangle}\;,\\
  \bar{\eps}_{-,\be\dal}(p_i,r)
  =
  \eps_{-}^\mu(p_i,r)\taubar_{\mu,\be\dal} 
  =
  \frac{\la_{i,\be}\tla_{r,\da}}{\tla_{i,\dga} \tla_{r}^{\dga}}
  = 
  \frac{|i\rangle[r|}{[ir]}\;\,,&\,\;\;
  \bar{\eps}_{+,\be\dal}(p_i,r)
  =
  \eps_{+}^\mu(p_i,r)\taubar_{\mu,\be\dal} 
  =
  \frac{\la_{r,\be}\tla_{i,\da}}{\la_r^{\ga} \la_{i,\ga}}
  =
  \frac{|r\rangle[i|}{\langle ri\rangle}\;.
\end{align}

\subsection{Tables with QED and QCD conventions and Feynman rules}
\label{sec:Rosetta stone}

We collect here spinor notation conventions, as well as chirality-flow
Feynman rules for QED and QCD.

\afterpage{
    \clearpage
    \thispagestyle{empty}
    \begin{landscape}
    \begin{table}
       \centering 
       \hspace*{-3.75cm}
       \begin{tabular}{c|c|c|c|c|c}
        \textbf{(Outgoing) Species}  & \textbf{Dirac} & \textbf{Index} & \textbf{Bra-ket} & \textbf{Feynman} & \textbf{Chirality-flow}\\ \hline 
\rule{0cm}{0.8cm}        right-chiral fermion &  $\bar{u}(p_i) P_R$ = $\begin{pmatrix} 0\,, & \big(u_L(p_i)\big)^{\dagger} \end{pmatrix}$ & $\lambda_i^{\alpha}$ & 
        $\langle i|$  & 
        \raisebox{-0.2\height}{\includegraphics[scale=0.4]{./Jaxodraw/FermionExtMi}} &  \includegraphics[scale=0.40]{./Jaxodraw/ExtSpinorSolidi}  \\
        right-chiral anti-fermion & $P_Rv(p_j)$ = $\begin{pmatrix} 0 \\ v_R(p_j) \end{pmatrix}$  & $\la_{j,\alpha}$ & 
        $| j \rangle$  & 
        \raisebox{-0.2\height}{\includegraphics[scale=0.4]{./Jaxodraw/FermionAntiExtMi}} & \includegraphics[scale=0.40]{./Jaxodraw/ExtSpinorSolidj}  \\ 
        left-chiral fermion & $\bar{u}(p_i) P_L$ = $\begin{pmatrix} \big(u_R(p_i)\big)^{\dagger}\,, & 0 \end{pmatrix}$  & $\tilde{\lambda}_{i,\da}$ & 
        $[i|$  & 
        \raisebox{-0.2\height}{\includegraphics[scale=0.4]{./Jaxodraw/FermionExtPl}} & \includegraphics[scale=0.40]{./Jaxodraw/ExtSpinorDottedi}  \\
        left-chiral anti-fermion & $P_Lv(p_j)$ = $\begin{pmatrix} v_L(p_j) \\ 0 \end{pmatrix}$  & $\tilde{\lambda}_j^{\dot{\alpha}}$ & 
        $|j]$  & 
        \raisebox{-0.2\height}{\includegraphics[scale=0.4]{./Jaxodraw/FermionAntiExtPl}} & \includegraphics[scale=0.4]{./Jaxodraw/ExtSpinorDottedj}  \\  
       negative-helicity photon & $\epsilon_-^{\mu}(p_i,r)$ & $\frac{\la_i^{\al}\taubar^{\mu}_{\al\db}\tla_r^{\db}}{[ir]}$ or $\frac{\tla_{r,\db}\tau^{\mu,\db\al}\la_{i,\al}}{[ir]}$ & $\frac{\langle i|\taubar^{\mu}|r]}{[ir]}$ or $\frac{[r|\tau^{\mu}|i\rangle}{[ir]}$ & \raisebox{-0.2\height}{\includegraphics[scale=0.4]{./Jaxodraw/PhotonExtMi}} & $\,\;\frac{1}{[ i r ]}$\raisebox{-0.20\height}{\includegraphics[scale=0.4]{./Jaxodraw/PhotonExtFlowMia}}\hspace{-2ex} or $\,\;\;\frac{1}{[ i r ]}$\raisebox{-0.20\height}{\includegraphics[scale=0.4]{./Jaxodraw/PhotonExtFlowMiaRev}}\hspace{-2ex} \\
        positive-helicity photon & $\epsilon_+^{\mu}(p_i,r)$ & $\frac{\la_r^\be\taubar^\mu_{\be\da}\tla_i^{\da}}{\langle ri\rangle}$ or $\frac{\tla_{i,\da}\tau^{\mu,\da\be}\la_{r,\be}}{\langle ri\rangle}$ & $\frac{\langle r|\taubar^{\mu}|i]}{\langle ri\rangle}$ or $\frac{[i|\tau^{\mu}|r\rangle}{\langle ri\rangle}$ & \raisebox{-0.2\height}{\includegraphics[scale=0.4]{./Jaxodraw/PhotonExtPl}} & $\;\;\frac{1}{\langle r i \rangle}$\raisebox{-0.2\height}{\includegraphics[scale=0.4]{./Jaxodraw/PhotonExtFlowPlaRev}}\hspace{-2ex} or $\;\;\frac{1}{\langle r i \rangle}$\raisebox{-0.2\height}{\includegraphics[scale=0.4]{./Jaxodraw/PhotonExtFlowPla}}\hspace{-2ex} \\ \hline
        \textbf{Vertices} & \textbf{Dirac} & \textbf{Index} & \textbf{Bra-ket} & \textbf{Feynman} & \textbf{Chirality-flow}\\ \hline
 fermion-photon & $ieQ_f\gamma^{\mu}$ & $ieQ_f\sqrt{2}\tau^{\mu,\da\be}$ &  $ieQ_f\sqrt{2}\tau^{\mu,\da\be}$ &        \rule{0cm}{1.2cm}\raisebox{-0.35\height}{\includegraphics[scale=0.4]{./Jaxodraw/qPhotonVertexRighta}} &  $ieQ_f\sqrt{2}$\raisebox{-0.3\height}{\includegraphics[scale=0.35]{./Jaxodraw/TauVertexHighb}} \\
        fermion-photon & $ieQ_f\gamma^{\mu}$  & $ieQ_f\sqrt{2}\taubar^{\mu}_{\al\db}$ &  $ieQ_f\sqrt{2}\taubar^{\mu}_{\al\db}$ & \raisebox{-0.35\height}{\includegraphics[scale=0.4]{./Jaxodraw/qPhotonVertexRightb}} &  $ieQ_f\sqrt{2}$\raisebox{-0.3\height}{\includegraphics[scale=0.35]{./Jaxodraw/TauVertexLowa}} \\ \hline
        \textbf{Propagators}& \textbf{Dirac} & \textbf{Index} & \textbf{Bra-ket} & \textbf{Feynman} & \textbf{Chirality-flow}\\ \hline
       \rule{0cm}{1cm}        fermion & $\frac{i}{p_{\mu}\gamma^{\mu}} = \frac{ip_{\mu}\gamma^{\mu}}{p^2}$  & $ i \frac{\tla_p^{\da}\la_p^{\be}}{p^2}$ & $i\frac{|p] \langle p|}{p^2}$ & \raisebox{-0.1\height}{\includegraphics[scale=0.4]{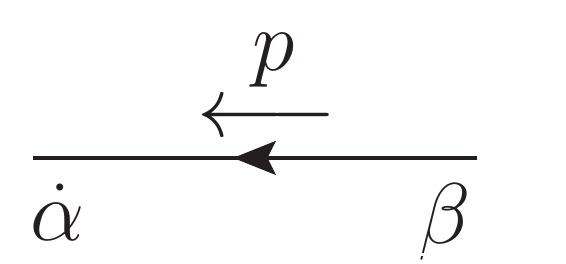}} & $\frac{i}{p^2}\raisebox{-0.15\height}{\includegraphics[scale=0.45]{./Jaxodraw/FermionPropFlowg}}$ \\
     fermion & $\frac{i}{p_{\mu}\gamma^{\mu}} = \frac{ip_{\mu}\gamma^{\mu}}{p^2}$  & $i \frac{\la_{p,\al}\tla_{p,\db}}{p^2}$ & $i \frac{|p\rangle [p|}{p^2}$ & \raisebox{-0.1\height}{\includegraphics[scale=0.4]{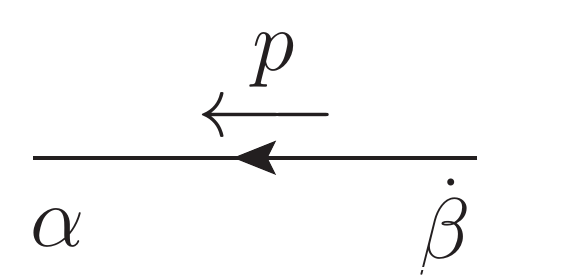}} & $\frac{i}{p^2}\raisebox{-0.15\height}{\includegraphics[scale=0.45]{./Jaxodraw/FermionPropFlowh}}$ \\

        photon & $-i\frac{g_{\mu\nu}}{p^2}$ & $-i\frac{g_{\mu\nu}}{p^2}$ & $-i\frac{g_{\mu\nu}}{p^2}$ & \raisebox{-0.2\height}{\includegraphics[scale=0.385]{./Jaxodraw/PhotonProp}} & $ - \frac{i}{p^2}\raisebox{-0.25\height}{\includegraphics[scale=0.4]{./Jaxodraw/PhotonPropFlowa}}$ \ or \ $ - \frac{i}{p^2}\raisebox{-0.25\height}{\includegraphics[scale=0.4]{./Jaxodraw/PhotonPropFlowb}}$ 
    \end{tabular}
   \captionof{table}{The QED ``Rosetta Stone" translating the chirality-flow notation to widely-used spinor-helicity notations.
   For more information see \secsref{sec:spinors} (external fermions), \ref{sec:polarizationvectors} (external vector bosons), \ref{sec:verticesqedqcd} (vertices) and \ref{sec:propagators} (propagators).
   } 
    \label{tab:Flow rules comparison QED}
    \end{table}
    \end{landscape}
    \clearpage
}

\afterpage{
    \clearpage
    \thispagestyle{empty}
    \begin{landscape}
    \begin{table}
        \centering 
        \hspace*{-2.5cm} 
        \begin{tabular}{c|c|c|c|c|c}
        \textbf{Vertices} & \textbf{Dirac} & \textbf{Index} & \textbf{Bra-ket} & \textbf{Feynman} & \textbf{Chirality-flow}\\ \hline
         \rule{0cm}{1cm}      quark-gluon & $i\frac{g_s}{\sqrt{2}}t^a_{i\jbar}\gamma^{\mu}$ & $ig_st^a_{i\jbar}\tau^{\mu,\da\be}$ &  $ig_s t^a_{i\jbar}\tau^{\mu,\da\be}$ & \raisebox{-0.45\height}{\includegraphics[scale=0.4]{./Jaxodraw/qGluonVertexRighta}} &  $ig_s t^a_{i\jbar}$\raisebox{-0.45\height}{\includegraphics[scale=0.35]{./Jaxodraw/TauVertexHighb}} \\
        quark-gluon & $i\frac{g_s}{\sqrt{2}}t^a_{i\jbar}\gamma^{\mu}$ & $ig_st^a_{i\jbar}\taubar^{\mu}_{\al\db}$ &  $ig_st^a_{i\jbar}\taubar^{\mu}_{\al\db}$ & \raisebox{-0.45\height}{\includegraphics[scale=0.4]{./Jaxodraw/qGluonVertexRightb}} &  $ig_s t^a_{i\jbar}$\raisebox{-0.45\height}{\includegraphics[scale=0.35]{./Jaxodraw/TauVertexLowa}} \\
        three-gluon & $V_3 \equiv i\frac{g_s}{\sqrt{2}}(if^{a_1a_2a_3})V_3^{\scriptscriptstyle\mu_1\mu_2\mu_2}$ & $V_3$ & $V_3$ &
        \raisebox{-0.5\height}{ \includegraphics[scale=0.35]{./Jaxodraw/CR/3GluonVertex_Full}} & 
        $i\frac{g_s}{\sqrt{2}}(if^{a_1a_2a_3})\frac{1}{\sqrt{2}}\!\left( \raisebox{-0.35\height}{ \includegraphics[scale=0.275]{./Jaxodraw/MS/3GluonVertex_HelFlow1}}
        \!\!\!+\!\!\!
        \raisebox{-0.45\height}{ \includegraphics[scale=0.275]{./Jaxodraw/MS/3GluonVertex_HelFlow2}}
        \!\!\!+\!\!\!
        \raisebox{-0.4\height}{ \includegraphics[scale=0.275]{./Jaxodraw/MS/3GluonVertex_HelFlow3}} \right)$  \\ 
        four-gluon & 
        \footnotesize{$\!\!V_4\equiv$}\parbox[t]{4.2cm}{\footnotesize$\ i\!\left(\!\frac{g_s}{\sqrt{2}}\!\right)^{\!2}\!\!\!\!\!\sum\limits_{Z(2,3,4)}\!\!\!\!(if^{a_1a_2b})(if^{ba_3a_4})\times$\\$\times(g^{\mu_1\mu_3}g^{\mu_2\mu_4}-g^{\mu_1\mu_4}g^{\mu_2\mu_3})$} & $V_4$  & $V_4$ &  
        \raisebox{-0.5\height}{ \includegraphics[scale=0.30]{./Jaxodraw/CR/4gVertex}} &  
        $i\left(\!\frac{g_s}{\sqrt{2}}\right)^{\!2}\!\!\!\sum\limits_{Z(2,3,4)}\!\!
(if^{a_1a_2b})(if^{ba_3a_4})\;\left(\raisebox{-0.4\height}{ \includegraphics[scale=0.2]{./Jaxodraw/MS/4GluonVertex_Kin13}} \!\!-\, \raisebox{-0.4\height}{ \includegraphics[scale=0.2]{./Jaxodraw/MS/4GluonVertex_Kin14}}\right)\!\!$ \\ 
        &  &  &  & 
        &  
        $ \begin{matrix} \hspace{-3.75cm} = i\left(\!\frac{g_s}{\sqrt{2}}\right)^{\!2}\!\!\!\sum\limits_{S(2,3,4)}\!\!\!\trace\big(t^{a_1}t^{a_2}t^{a_3}t^{a_4}\big) \times \\
       \hspace{2cm}\times \left( 2 \!\!\raisebox{-0.4\height}{ \includegraphics[scale=0.2]{./Jaxodraw/MS/4GluonVertex_Kin13}} -\; \raisebox{-0.4\height}{ \includegraphics[scale=0.2]{./Jaxodraw/MS/4GluonVertex_Kin12}} - \;\;\raisebox{-0.4\height}{ \includegraphics[scale=0.2]{./Jaxodraw/MS/4GluonVertex_Kin14}} \right)
        \end{matrix}$  \\  \hline
        \textbf{Propagators} & \textbf{Dirac} & \textbf{Index} & \textbf{Bra-ket} & \textbf{Feynman} & \textbf{Chirality-flow}\\ \hline
       \rule{0cm}{1cm}
        fermion & $\frac{i\delta_{i\jbar}}{p_{\mu}\gamma^{\mu}} = \frac{i\delta_{i\jbar}\, p_{\mu}\gamma^{\mu}}{p^2}$  & $ i \frac{\delta_{i\jbar}\tla_p^{\da}\la_p^{\be}}{p^2}$ & $i\frac{\delta_{i\jbar}|p] \langle p|}{p^2}$ & \ \ \raisebox{-0.1\height}{\includegraphics[scale=0.4]{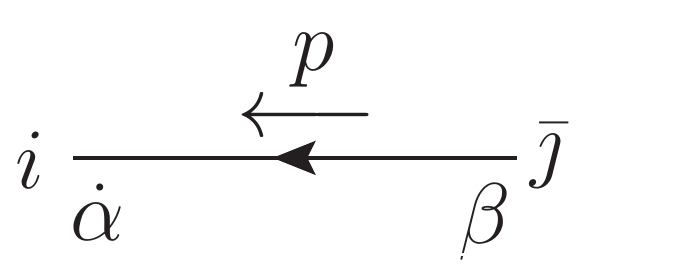}} & $\frac{i\delta_{i\jbar}}{p^2}\raisebox{-0.15\height}{\includegraphics[scale=0.45]{./Jaxodraw/FermionPropFlowg}}$ \\
        fermion & $\frac{i\delta_{i\jbar}}{p_{\mu}\gamma^{\mu}} = \frac{i\delta_{i\jbar}\,p_{\mu}\gamma^{\mu}}{p^2}$  & $i \frac{\delta_{i\jbar}\la_{p,\al}\tla_{p,\db}}{p^2}$ & $i \frac{\delta_{i\jbar}|p\rangle [p|}{p^2}$ & \ \ \raisebox{-0.1\height}{\includegraphics[scale=0.4]{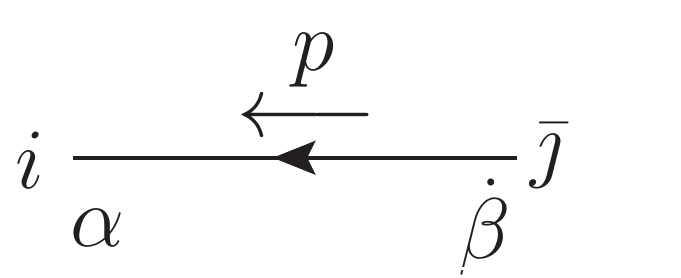}} & $\frac{i\delta_{i\jbar}}{p^2}\raisebox{-0.15\height}{\includegraphics[scale=0.45]{./Jaxodraw/FermionPropFlowh}}$ \\
        gluon & $-i\frac{\delta^{ab}g_{\mu\nu}}{p^2}$ & $-i\frac{\delta^{ab}g_{\mu\nu}}{p^2}$ & $-i\frac{\delta^{ab}g_{\mu\nu}}{p^2}$ & \hspace{-0.5ex}\raisebox{-0.2\height}{\includegraphics[scale=0.385]{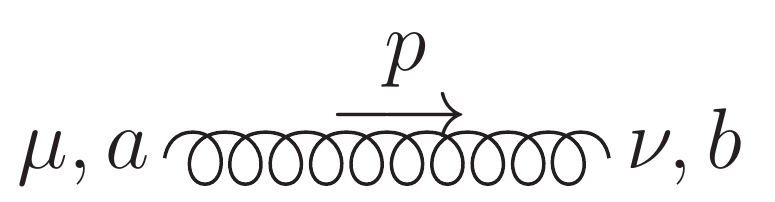}} & $ - \frac{i\delta^{ab}}{p^2}\raisebox{-0.25\height}{\includegraphics[scale=0.4]{./Jaxodraw/PhotonPropFlowa}}$ \ or \ $ - \frac{i\delta^{ab}}{p^2}\raisebox{-0.25\height}{\includegraphics[scale=0.4]{./Jaxodraw/PhotonPropFlowb}}$ 
    \end{tabular}
    \captionof{table}{The QCD ``Rosetta Stone" translating the chirality-flow notation to widely-used spinor-helicity notations. 
    $V_3^{\mu_1\mu_2\mu_2}\equiv V_3^{\mu_1\mu_2\mu_2}(p_1,p_2,p_3)$ is given in \eqref{eq:trpgluvrt}.
    For more information, in particular on the various forms of the four-gluon vertex, see \secref{sec:verticesqedqcd} (vertices) and \ref{sec:propagators} (propagators).
  }
    \label{tab:Flow rules comparison QCD}
    \end{table}
    \end{landscape}
    \clearpage
}

\newpage

\bibliographystyle{JHEP}  
\bibliography{HelFlowBib} 

\end{document}